\documentclass[a4paper,11pt]{article}
\pdfoutput=1 % to allow pdflatex compilation in JCAP
\usepackage{jcappub}
\usepackage{amsmath}
\usepackage{graphicx}
\usepackage{latexsym}
\usepackage{xspace}
\usepackage{color}
\usepackage{hyperref} 
\usepackage{bm}
\usepackage{relsize}
\usepackage{tabularx}
\usepackage{multirow}
\usepackage{amssymb}
\usepackage[table]{xcolor}
\usepackage{slashbox}
\usepackage{braket}
\usepackage{comment}
\usepackage[utf8]{inputenc}
\usepackage{slashed}
\usepackage{empheq}
\usepackage[T1]{fontenc} % if needed
\usepackage{soul}

\allowdisplaybreaks

\makeatletter
\gdef\@fpheader{}
\g@addto@macro\bfseries{\boldmath}
\makeatother

\makeatletter
\renewcommand*\env@matrix[1][\arraystretch]{%
  \edef\arraystretch{#1}%
  \hskip -\arraycolsep
  \let\@ifnextchar\new@ifnextchar
  \array{*\c@MaxMatrixCols c}}
\makeatother

\newcommand{\ds}{\displaystyle}

\newcommand{\ie}{\textsl{i.e.~}}

\newcommand{\eg}{\textsl{e.g.~}}

\newcommand{\dd}{\mathrm{d}}
\newcommand{\sss}[1]{{\scriptscriptstyle{#1}}}

\newcommand{\uPl}{\mathrm{Pl}}

\newcommand{\usssPl}{\sss{\uPl}}

\newcommand{\Mp}{M_\usssPl}

\newcommand{\beq}{\begin{equation}}
\newcommand{\eeq}{\end{equation}}
\newcommand{\bea}{\begin{eqnarray}}
\newcommand{\eea}{\end{eqnarray}}

\newlength{\wsingfig}
\newlength{\wdblefig}
\newlength{\wquadfig}
\newlength{\wtriplefig}
\newcommand{\Eq}[1]{Eq.~(\ref{#1})}

\newcommand{\MM}{M_{pl}}
\newcommand{\tsr}{_{\mu\nu}}                %tensor
\newcommand{\utsr}{^{\mu\nu}}               %up tensor
\newcommand{\stsr}{_{ij}}                   %spatial tensor
\newcommand{\ustsr}{^{ij}}                  %up spatial tensor
\newcommand{\Tsr}{_{IJ}}                    %tensor in phase space
\newcommand{\uTsr}{^{IJ}}                   %up tensor in phase space
\subheader{}

\title{Separate universe in multifield inflation: a phase-space approach}

\author[a]{Julien Grain,}
\emailAdd{julien.grain@cnrs.fr}

\author[a]{Hugo Holland}
\emailAdd{hugo.holland@universite-paris-saclay.fr}

\affiliation[a]{Universit\'e Paris-Saclay, CNRS, Institut d'Astrophysique Spatiale, 91405, Orsay, France}

\date{today}

\begin{document}
\sloppy

\abstract{In this article we extend a study of the validity conditions of the separate-universe approach of cosmological perturbations to models of inflation with multiple fields. The separate-universe approach consists in describing the universe as a collection of homogeneous and isotropic patches, giving us an effective description of perturbation theory at large scales through phase-space reduction. This approximation is a necessary step in stochastic inflation, an effective theory of coarse-grained, super-Hubble, scalar fields fluctuations. One needs a stochastic inflation description in the context of primordial black hole productions since it needs enhancements of the curvature power spectrum. It is easily achievable in multifield inflation models but necessarily comes with strong diffusive effects. We study and compare cosmological perturbation theory and the separate-universe approach in said non-linear sigma models as a typical framework of multifield inflation and employ the Hamiltonian formalism to keep track of the complete phase space (or the reduced isotropic phase space in the separate-universe approach). We find that the separate-universe approach adequately describes the cosmological perturbation theory provided the wavelength of the modes considered is greater that several lower bounds that depend on the cosmological horizon and the inverse of the effective Hamiltonian masses of the fields; the latter being fixed by the coupling potential and the field-space geometry. We also compare gauge-invariant variables and several gauge fixing procedures in both approaches. For instance, we showed that the uniform-expansion gauge is nicely described by the separate-universe picture, hence justifying its use in stochastic inflation as commonly done.}

\keywords{cosmological perturbation theory, inflation}

\maketitle

\flushbottom

\section{Introduction}

Inflation is the most favoured model to produce inhomogeneities in the early universe which later seed the anisotropies of the cosmic microwave background and large-scale structures. They are predicted to be adiabatic and Gaussian distributed with an almost scale-invariant power spectrum. This is confirmed by current observations \cite{Planck:2018jri} and these properties can be explained considering that inflation is driven by a single scalar field. However, current observations probe a small part of the 40-to-60 e-folds of the inflationary phase which leaves a wide berth for models departing from the single-field setup. Indeed, high-energy constructions of inflationary models typically lead to multiple scalar fields with derivative couplings \cite{Lyth:1998xn,Cremonini:2010ua,Renaux_Petel_2015,Carrasco:2015uma,Hetz:2016ics,Achucarro:2018ngj,Krajewski:2018moi,Christodoulidis:2018qdw,Linde:2018hmx,Achucarro:2017ing,Achucarro:2018vey,Bjorkmo:2019aev,Mizuno:2019pcm,Bravo:2019xdo,Aragam:2019khr,Assassi:2013gxa,Pinol:2024arz}. Such extra-fields lead to a rich phenomenology including primordial non-gaussianities \cite{Bjorkmo:2019qno,Achucarro:2019lgo,achucarro2019orbitalinflationinflatingangular,Ach_carro_2020,Tzavara:2013wxa,SGrootNibbelink_2002,Garcia_Saenz_2018,Fumagalli_2019,Garcia_Saenz_2020,Pinol:2021aun}, gravitational waves \cite{Fumagalli:2020nvq,Fumagalli:2021cel} and primordial black holes \cite{Fumagalli:2020adf,Geller:2022nkr,Qin:2023lgo}. 

Primordial inhomogeneities are small deviations from the homogeneous and isotropic background which are described by means of perturbative techniques. In that context, an important simplification may occur if the perturbed universe can be described at large scales by a collection of independent patches which are locally homogeneous and isotropic. In practice, it means that cosmic inhomogeneities are treated by solving for an homogeneous and isotropic problem with different initial conditions. This is the so-called separate-universe picture \cite{Wands_2000,Tanaka_2021,PhysRevD.68.103515,PhysRevD.68.123518,LIFSHITZ1992493,article,PhysRevD.49.2759,I_M_Khalatnikov_2002} which offers a simple method to compute cosmic inhomogeneities at large scales including their non-linear evolution. It leads to the $\delta\mathcal{N}$-formalism with many applications to predict inflationary observables \cite{PhysRevD.42.3936,Sasaki_1996,Lyth_2005,PhysRevLett.95.121302,Sugiyama_2013,Tanaka:2024mzw,Artigas:2024xhc}, including multifield models (see \eg \cite{Tada_2017,achucarro2019orbitalinflationinflatingangular}). The conditions for applying this approximation scheme has been derived for single-field inflation in Ref. \cite{Nambu_2005,doi:10.1142/10953,Pattison:2019hef,Artigas_2022,Cruces_2023}. They are shown to depend on both the Hubble radius and the effective mass of the inflaton. Yet, these conditions still need to be assessed when multiple fields are present which is the main goal of this paper.

Another motivation for establishing these conditions is stochastic inflation \cite{STAROBINSKY1982175,10.1007/3-540-16452-9_6,10.1143/PTP.80.1041}. It heavily relies on the separate-universe picture and it has been recently extended to inflationary models with multiple fields and derivative couplings \cite{Pinol_2019,Pinol:2020cdp}. Stochastic inflation is an effective theory aimed at describing the evolution of the fields coarse-grained at a fixed physical radius typically greater than the Hubble one. In this setup, the dynamics of the independent patches is constantly updated by a random noise when quantum cosmological fluctuations cross out the radius of coarse-graining. Stochastic inflation was proved to be in agreement with quantum-field-theoretic calculations in the context of single-field, slow-roll inflation \cite{Starobinsky_1994,Tsamis_2005,Finelli_2010,Garbrecht_2014,Onemli_2015,Burgess_2016,Kamenshchik_2022}. It can further be combined with the $\delta\mathcal{N}$-formalism in order to account for the quantum backreaction in the calculation of cosmic inhomogeneities \cite{Fujita:2013cna,Vennin:2015hra}. There, stochastic effects play an important role in shaping the tail of the statistics of cosmological fluctuations with significant consequences on the abundances of primordial black holes \cite{Pattison_2017,Biagetti_2018,Ezquiaga_2020} and on the statistics of large-scale structures \cite{Ezquiaga:2022qpw,Coulton:2024vot}. This is because producing substantial amounts of primordial black holes requires a strong boost in the power spectrum of curvature perturbations which inevitably leads to a phase with strong stochastic diffusion. 

The radius of coarse-graining is set by the radius above which the separate-universe picture holds. Its identification is important not only to establish the range of validity of stochastic inflation but also because it partly fixes the amplitude of the stochastic noise, hence of the strength of the stochastic effects. Given that boosts in the power spectrum can be obtained in multifield inflation, it becomes important to assess the impact of stochastic effects in these models, hence to establish its scale of coarse-graining as a prerequisite. (Note that the knowledge of this scale is also needed in order to compare stochastic inflation with computations from quantum field theory.)

In practice, many calculations based on the separate-universe picture are done in a specific gauge. For instance, the stochastic $\delta\mathcal{N}$-formalism needs to be used in the uniform-expansion gauge \cite{Pattison:2019hef}. Because of that, the conditions for which the separate-universe picture properly describes cosmological fluctuations need to be established in a gauge-independent manner and our goal is also to establish the separate-universe picture for multiple fields considering several gauges of cosmological interests in addition to gauge-invariant variables.

A final goal of our study is to extend the phase-space construction of the separate-universe approximation to models with multiple scalar fields, allowing us to identify the conditions of validity for both the fields configurations and their associated momenta. In many situations the Hamiltonian and the Lagrangian formalisms are equivalent. Yet, a phase-space approach proves to be useful for several reasons. First, it is crucial for dealing with cosmological fluctuations at the quantum level since \eg the choice of the initial vacuum is tightly related to the choice of the parametrization of the phase space \cite{Grain:2019vnq}. Though our analysis is classical, it lays the ground for a quantum treatment. Second, the background dynamics in multifield inflation may not enjoy the presence of an attractor. In this case, it is important to keep track of the velocities and the comparison of cosmological perturbation theory with the separate-universe picture is preferentially done in the phase space. Third, a first-order formalism is required when it comes to stochastic inflation \cite{Grain:2017dqa,Miranda:2019ara,Pattison_2017}. This is because stochastic dynamical processes are formalized with first-order equations. In addition, the stochastic noise originates from quantum fluctuations and its properties are naturally defined in the phase space.

To do so, we consider non-linear sigma models as a proxy for multifield inflation with non-linear, derivative couplings \cite{Pinol_2021}. It also presents the nice feature of being covariant in the field-space. We derive the separate-universe picture in a manifestly covariant manner and in the basis of adiabatic (\ie curvature) and entropic (\ie isocurvature) perturbations.

The paper is organised as follows. In Sec. \ref{sec:cpt}, we review the Hamiltonian formulation of general relativity in the context of a homogeneous and isotropic cosmology. We then compute the fully covariant perturbations of the fields and their momentas, before computing the covariant Hamiltonian giving the dynamics of these variables. In Sec. \ref{sec:dynCPT}, we study said dynamics and explore the remaining gauge degrees of freedom. On the one hand we study the dynamics of gauge invariant variables, and on the other hand we study several gauge fixing procedures and their consequences. In Sec. \ref{sec:separate} we follow the same procedure, this time incorporating the separate-universe approach, before comparing results from both sections. Finally, we summarize our results in Sec. \ref{sec:conclusion}. We end our paper with various appendices in which we cover the more technical aspects of our work.

\section{Covariant cosmological perturbations}
\label{sec:cpt}
\subsection{Hamiltonian description}
\label{ssec:hamiltonian description}
   We start by introducing the Hamiltonian formulation of general relativity (see \eg \cite{thieman_book}). In this work we aim to study multifield models of inflation in the most generic way possible. We call $\phi^I$ the fields with $I$ running from $1$ to $n$ the number of fields, $G\Tsr$ is the field dependent coupling metric, and $g\tsr$ is the four dimensional metric of the curved space time. The scalar fields are all minimally coupled to gravity. The action reads
    \beq \label{einstein frame action}
         S = \int \dd^4x \sqrt{-g} \left[\frac{\Mp^2}{2}  R - \frac 1 2 g\utsr \,G\Tsr(\phi^K)\, \partial_\mu \phi^I \partial_\nu \phi^J - V(\phi^I) \right],
    \eeq
    where $R$ is the four dimensional Ricci scalar, $g$ the determinant of the metric $g\tsr$, $\Mp=1/\sqrt{8\pi G}$ is the reduced Planck mass and $V(\phi^I)$ is the scalar field potential that can depend on all fields.  Indices in the four dimensional space time are raised and lowered with the metric $g\tsr$ and its inverse $g\utsr$. In the field space, one raises and lowers indices with the coupling metric $G\Tsr$ and its inverse $G\uTsr$. 
    
    To get the Hamiltonian formulation of general relativity, we foliate our four dimensional space time into sheets of three dimensional space-like hypersurfaces. The foliation is defined by a pair of Lagrange multipliers: the lapse function $N(\tau,\vec x)$ and the shift vector $N_i(\tau, \vec x)$, with $\tau$ the time variable and $\vec x$ the spatial coordinates on the hypersurfaces. One can then express the metric with these Lagrange multipliers leading to its ADM form, \ie
    \begin{equation}
        \dd s^2 = -N^2(\tau, \vec x) \dd\tau^2 + \gamma\stsr(\tau,\vec x)\left[\dd x^i + N^i(\tau,\vec x)\dd\tau\right]\left[\dd x^j + N^j(\tau,\vec x)\dd\tau\right]
    \end{equation}
    with $\gamma\stsr$ the three dimensional metric induced on the hypersurfaces, which is used (along with its inverse $\gamma^{ij}$) to raise and lower indices on space-like sheets.  In the Hamiltonian formalism, one further needs to define the canonical variables for each field by defining the conjugate momentas $\pi_I := \delta S/ \delta \dot{\phi}^I$, with $\dot f := \dd f/\dd\tau$ the time derivative. Note here that $\phi^I$ is covariant and $\pi_I$ contravariant in the field space. The gravitational sector works similarly, and one can define $\pi\ustsr := \delta S / \delta \dot\gamma\stsr$. Using the phase-space variables, the action can be expressed using its Hamiltonian form
    \beq
    	S=\int \dd \tau\int\dd^3x\left(\pi_I\dot{\phi}^I+\pi^{ij}\dot{\gamma}_{ij}-N\mathcal{C}-N^i\mathcal{D}_i\right), \label{eq:SHam}
	\eeq
where $\mathcal{C}$ and $\mathcal{D}_i$ are the so-called scalar and diffeomorphism constraints. Each of them receives a contribution from the gravitational sector (labelled by a superscript $G$) and by the matter sector (labelled by a superscript $\phi$) and we write them as $\mathcal{C}=\mathcal{C}^G+\mathcal{C}^\phi$ (and similarly for the diffeomorphism constraints) where
    \begin{align}
            \mathcal C^G =& \frac 2 {\Mp^2 \sqrt \gamma} [\pi\stsr\pi\ustsr - \frac 1 2 \pi^2] - \frac {\Mp^2\sqrt \gamma} 2 R^{(3)}, \label{scalar constraint grav}\\
            \mathcal C^\phi =& \frac{1}{2 \sqrt \gamma} G\uTsr \pi_I\pi_J + \frac { \sqrt \gamma }{2}\gamma^{ij}G\Tsr  \partial_i\phi^I\partial_j\phi^J + \sqrt{\gamma}V, \label{scalar constraint field}\\
            \mathcal D_i^G =& -2\partial_m(\gamma\stsr\pi^{jm}) + \pi^{mn}\partial_i\gamma_{mn}, \label{diff constraint grav}\\
            \mathcal D_i^\phi =& \pi_I\partial_i\phi^I. \label{diff constraint field}
    \end{align}
In the above, $\pi:=\pi^{ij}\gamma_{ij}$ is the trace of the gravitational momentum and $R^{(3)}$ is the Ricci scalar of the space-like hypersurfaces.

	The dynamics of the scalar fields and of the gravitational fields is generated by the following Hamiltonian
    \begin{equation} \label{Hamiltoninan}
            H[N, N^i] = \int \dd^3 \vec x \left( N\mathcal C\ + N^i \mathcal D_i \right).
    \end{equation}
    The pairs $(\phi^I,\pi_J)$ and $(\gamma_{ij},\pi^{mn})$ form canonical pairs with Poisson brackets reading $\{\phi^I(\tau,\vec x);\pi_J(\tau,\vec y)\} = \delta^I_J \delta^3(\vec x - \vec y) $ and $\{\gamma\stsr(\tau,\vec x),\pi^{mn}(\tau,\vec y )\} = \frac 1 2 (\delta^m_i\delta^n_j + \delta^n_i\delta^m_j)\delta(\vec x - \vec y)$.  For any given functions $f$ and $g$ in this context, the time evolution is obtained from
    \begin{equation}
        \dot f(\phi^I,\pi_I;\gamma\stsr,\pi\ustsr) = \left\{ f(\phi^I,\pi_I;\gamma\stsr,\pi\ustsr),H[N, N^i]\right\},
    \end{equation}
    with the full Poisson bracket defined as
    \begin{equation}
        \{f,g\} = \int \dd^3x\left[\left(\frac{\delta f}{\delta\gamma\stsr}\frac{\delta g}{\delta\pi\ustsr}-\frac{\delta g}{\delta\gamma\stsr}\frac{\delta f}{\delta\pi\ustsr}\right)+\left(\frac{\delta f}{\delta\phi^I}\frac{\delta g}{\delta\pi_I}-\frac{\delta g}{\delta\phi^I}\frac{\delta f}{\delta\pi_I}\right) \right].
    \end{equation}
 As mentioned precedently, the lapse function and the shift vector are Lagrange multipliers associated with the freedom of coordinates system and they do not take part in the dynamics since they have no associated momenta. Because of that, varying the action with respect to the lapse and the shift enforces the constraints to vanish \ie
    \beq
    	\mathcal{C}=0=\mathcal{D}_i.
\eeq
 This imposes to search for solutions of the dynamical equations such that the above constraints hold and physical solutions are said to lie on the surface of constraints. It is worth stressing that the constraints are first-class\footnote{First class contraints commute i.e. their Poisson bracket is equal to zero. In our case this indicates that not all degrees of freedom are physical ones and some gauge degrees of freedom remain, as discussed in Sec.\ref{sec:dynCPT}} which guarantees them to be preserved through evolution of the fields. As a consequence, any solutions which is initially selected to lie on the surface of constraints will remain so through its evolution generated by $H[N,N^i]$.

Finally, let us provide the Hamilton equations which are the dynamical equations for the phase-space variables. First, we consider the scalar fields and their momentas and we get
    \begin{align}
        \dot \phi^I =& \frac N{\sqrt\gamma}G\uTsr\pi_J + N^i\partial_i\phi^I \label{eq:dotphigen} \\
        D_\tau  \pi_I =& - \sqrt \gamma N \frac{\partial V}{\partial\phi^I} + D_i (\sqrt \gamma N G\Tsr \gamma\ustsr \partial_j \phi^I) + D_i(N^i\pi_I),
    \end{align}
     where $D_\mu$ is the covariant derivatives in the field space of vectors $U^I$ and covectors $W_I$ defined as
    \begin{equation}
        D_\mu U^I = \partial_\mu U^I + \Gamma^I_{LK}\partial_\mu\phi^LU^K \quad \mathrm{and} \quad D_\mu W_I = \partial_\mu - \Gamma^K_{IL}\partial_\mu\phi^LW_K,
    \end{equation}
    with $\Gamma^K\Tsr$ the Christoffel coefficients associated to the coupling metric $G\Tsr$ [note that $D_\mu$ is easily expressed as a function of the momentas $\pi_I$ rather than the velocities $\dot{\phi}^I$ using \Eq{eq:dotphigen}]. From now on, partial derivatives in the space of scalar fields $\phi^I$ will be denoted by a comma \ie $U^J_{\,\,,I}$, and covariant derivatives  by a semicolon \ie $U^J_{\,\,;I}$. Second, the gravitational degrees of freedom evolves according to
    \bea
    	\dot{\gamma}_{ij}&=&\frac{2N}{\Mp^2\sqrt{\gamma}}\left[2\pi_{ij}-\gamma_{ij}\pi\right]+2\gamma_{mj}\partial_iN^m+N^m\partial_m\gamma_{ij}, \\
        \dot{\pi}^{ij}&=&\frac{2N\gamma^{ij}}{\Mp^2\sqrt{\gamma}}\left(\pi_{mn}\pi^{mn}-\frac{1}{2}\pi^2\right)+\frac{2N}{\Mp^2\sqrt{\gamma}}\left(\pi\pi^{ij}-2\gamma_{mn}\pi^{im}\pi^{jn}\right)+\frac{N\Mp^2\sqrt{\gamma}}{2}\frac{\delta R^{(3)}}{\delta\gamma_{ij}} \nonumber \\
	&&+\frac{N}{2 \sqrt \gamma} \gamma^{ij}G\uTsr \pi_I\pi_J +\frac{N}{2}\gamma^{im}\gamma^{jn}G\Tsr  \partial_m\phi^I\partial_n\phi^J-2\pi^{jm}\partial_mN^i-\partial_m\left(N^m\pi^{ij}\right),
\eea
     where the complexity of Einstein's field equations is mostly reflected in the dynamical equation of $\pi^{ij}$. The latter equation has been simplified by making use of the constraints $\mathcal{C}=0$. Yet, there is no simple manner to write down the equation of motion of the gravitational momenta despite such potential simplifications.
    % \bea
   %  	\dot{\pi}^{ij}&=&\frac{N\gamma^{ij}}{2}\left[\frac{2}{\Mp^2\sqrt{\gamma}}\left(\pi_{mn}\pi^{mn}-\frac{1}{2}\pi^2\right)+\frac{\Mp^2\sqrt{\gamma}}{2}R^{(3)}\right]+\frac{2N}{\Mp^2\sqrt{\gamma}}\left(\pi\pi^{ij}-2\gamma_{mn}\pi^{im}\pi^{jn}\right) \nonumber \\ 
%	&&+\frac{N\Mp^2\sqrt{\gamma}}{2}\frac{\delta R^{(3)}}{\delta\gamma_{ij}} +\frac{N}{4 \sqrt \gamma} \gamma^{ij}G\uTsr \pi_I\pi_J +\frac{N}{2}\gamma^{im}\gamma^{jn}G\Tsr  \partial_m\phi^I\partial_n\phi^J \nonumber \\ 
	%&&-\frac{N}{2}\gamma^{ij}\left(\frac { \sqrt \gamma }{2}\gamma^{mn}G\Tsr  \partial_m\phi^I\partial_n\phi^J + \sqrt{\gamma}V\right)-2\pi^{jm}\partial_mN^i-\partial_m\left(N^m\pi^{ij}\right),
%\eea

\subsection{Degrees of freedom}
\label{ssec:dof}
We now wish to apply to above formalism in the context of inflationary space times. To this end, the fields and the Lagrange multipliers are split into background degrees of freedom which are homogeneous and isotropic, and cosmological perturbations which are small fluctuations encoding deviations from homogeneity and isotropy. 

\subsubsection*{Homogeneous and isotropic degrees of freedom}
The background space time is described by the FLRW metric which is further assumed to be spatially flat. Hence, the metric reduces to
\beq
	\dd s^2=-N^2(\tau)\dd\tau^2+v^{2/3}(\tau)\tilde{\gamma}_{ij}\dd x^i\dd x^j,
\eeq 
where $N$ is the lapse function and $v$ is the volume variable defined as $v:=a^3$, $a$ being the usual scale factor. Both are functions of time only. We also introduce the time-independent flat metric $\tilde{\gamma}_{ij}$ which raises and lowers indices for comoving quantities. Different choices of the lapse function correspond to different choices for the time variable. For instance, $N=1$ stands for the cosmic time while $N=a$ stands for conformal time. In the following, we will keep $N$ arbitrary and one should bear in mind that $\dot{f}$ does not necessarily means differentiation with respect to cosmic time. Finally, we stress that isotropy imposes the shift vector to be zero at the background level, $N^i(\tau)=0$. 

The momenta $\theta(\tau)$ conjugated to $v$ is built as follows. Homogeneity and isotropy impose $\pi^{ij}(\tau,\vec{x})=f(\theta,v)\,\tilde{\gamma}^{ij}$ where $f$ is a function constrained such that $(v,\theta)$ form a canonical pair, \ie $\{v,\theta\}=1$. Plugging the homogeneous and isotropic expression of $\gamma_{ij}$ and $\pi^{ij}$ into their Poisson bracket and imposing $(v,\theta)$ to be a canonical pair, boils down to $\partial f/\partial\theta=v^{1/3}/2$. Hence it leads to $f=v^{1/3}\theta/2$ where the constant of integration is set equal to zero for simplicity. 

Finally, the background scalar fields are described by the pairs $\phi^I(\tau)$ and $\pi_I(\tau)$ which are time-dependent only and whose Poisson bracket reduces to $\{\phi^I,\pi_J\}=\delta^I_J$.

\subsubsection*{Cosmological perturbations}
Deviations from homogeneity and isotropy are formally introduced by subtracting the homogeneous and isotropic background solution to the phase-space variables, \ie
\bea
	\delta\phi^I(\tau,\vec{x})&=&\phi^I(\tau,\vec{x})-\phi^I(\tau), \label{eq:deltaphi} \\
	\delta\pi_J(\tau,\vec{x})&=&\pi_J(\tau,\vec{x})-\pi_J(\tau), \label{eq:deltapi} \\
	\delta\gamma_{ij}(\tau,\vec{x})&=&\gamma_{ij}(\tau,\vec{x})-v^{1/3}(\tau)\,\tilde{\gamma}_{ij}, \\
	\delta\pi^{ij}(\tau,\vec{x})&=&\pi^{ij}(\tau,\vec{x})-\frac{1}{2}v^{1/3}(\tau)\theta(\tau)\,\tilde{\gamma}^{ij}. 
\eea
As such, the above definition consists in performing a translation in the phase space given by fixed and time-dependent functions. Hence, it is a canonical transformation and $(\delta\phi^I,\delta\gamma_{ij},\delta\pi_I,\delta\pi^{ij})$ have the same Poisson brackets as $(\phi^I,\gamma_{ij},\pi_I,\pi^{ij})$. 

In the following, perturbations of any field $\Phi(\tau,\vec{x})$ will systematically be denoted as $\delta\Phi(\tau,\vec{x})$. The notation $\Phi$ will be kept for its homogeneous and isotropic, background value, \ie $\Phi(\tau)$.

Cosmological perturbations are preferentially treated in Fourier space since invariance of the background under spatial translation enforce different Fourier modes to decouple at the leading order in perturbation. In practice, any tensor field is Fourier transformed on the spatially-flat hypersurfaces according to
\bea
	T_{i\cdots j}(\tau,\vec{k})=\ds\int\frac{\dd^3\vec{x}}{(2\pi)^{3/2}}T_{i\cdots j}(\tau,\vec{x})e^{-i\vec{k}\cdot\vec{x}}.
\eea
Here, $\vec{k}$ stands for the comoving wavevector whose indices are raised and lowered by the flat metric $\tilde{\gamma}_{ij}$ and whose norm reads $k^2=\tilde\gamma_{ij}k^ik^j$. Let us also note that fields being real-valued, their Fourier modes are constrained to satisfy $T^\star_{i\cdots j}(\tau,\vec{k})=T_{i\cdots j}(\tau,-\vec{k})$, where a star denotes the complex conjugate. Starting from two real-valued tensors which have the following Poisson bracket in real space 
\beq
	\left\{T^1_{i\cdots j}(\tau,\vec{x}),T^{m\cdots n}_2(\tau,\vec{y})\right\}=\Delta^{m\cdots n}_{i\cdots j}\delta\left(\vec{x}-\vec{y}\right)
\eeq
with $\Delta^{m\cdots n}_{i\cdots j}$ independent of space, it can be shown that
\beq
	\left\{T^1_{i\cdots j}(\tau,\vec{k}),T^{m\cdots n\,\star}_2(\tau,\vec{q})\right\}=\Delta^{m\cdots n}_{i\cdots j}\delta^3\left(\vec{k}-\vec{q}\right).
\eeq

In full generality, cosmological perturbations are decomposed into scalar, vector and tensor degrees of freedom which decouples at leading order in perturbation. In the following, we will focus on scalar perturbations which are the main purposes of the separate-universe picture considered here (see \eg \cite{Tanaka_2021,Tanaka:2023gul,Tanaka:2024mzw} for tentative extensions). The scalar fields variables are scalar quantities and so are their fluctuations. In Fourier space, their Poisson bracket reads
\beq
	\left\{\delta\phi^I(\tau,\vec{k}),\delta\pi^\star_J(\tau,\vec{q})\right\}=\delta^I_J\delta^3(\vec{k}-\vec{q}).
\eeq
Fluctuations of the induced metric and its momenta are tensors in the hypersurfaces and according to the scalar-vector-tensor decomposition, they are expanded into a pure trace part which is a scalar quantity $(C\gamma_{ij})$, the double gradient of a second scalar function $(\partial_i\partial_j E)$, the gradient of a divergence-free vector $(\partial_iF_j+\partial_jF_i)$,  and a divergence-free and traceless tensor $(h_{ij})$. Setting the vector and tensor components to zero $F_i=0=h_{ij}$, it is convenient to express the fluctuations of the scalar gravitational degrees of freedom in Fourier space as
\bea
	\delta\gamma_{ij}(\tau,\vec{k})&=&\delta\gamma_1(\tau,\vec{k})M^1_{ij}(\vec{k})+\delta\gamma_2(\tau,\vec{k})M^2_{ij}(\vec{k}), \\
	\delta\pi_{ij}(\tau,\vec{k})&=&\delta\pi_1(\tau,\vec{k})M_1^{ij}(\vec{k})+\delta\pi_2(\tau,\vec{k})M_2^{ij}(\vec{k}),
\eea
where
\bea
	M^1_{ij}(\vec{k})=\frac{\tilde{\gamma}_{ij}}{\sqrt{3}}&\,\,\,\mathrm{and}\,\,\,& M^2_{ij}(\vec{k})=\sqrt{\frac{3}{2}}\left(\frac{k_ik_j}{k^2}-\frac{\tilde{\gamma}_{ij}}{3}\right).
\eea
The matrices $M^A_{ij}$ form an orthonormal basis of the scalar components of the metric since $M^A_{ij}M_B^{ij}=\delta^A_B$, and their indices are raised and lowered by the flat metric $\tilde{\gamma}_{ij}$. The matrix $M^1_{ij}$ is pure trace and one has $\delta\gamma_1(\tau,\vec{k})=v^{2/3}\sqrt{3}C(\tau,\vec{k})-k^2E(\tau,\vec{k})/\sqrt{3}$. It is an isotropic degree of freedom which can be interpreted as local variations of the scale factor.  The matrix $M^2_{ij}$ being traceless, the second degree of freedom is purely anisotropic. It can be interpreted as local deformations of a fiducial volume keeping its value unchanged, and one has $\delta\gamma_2(\tau,\vec{k})=-\sqrt{2/3}k^2E(\tau,\vec{k})$. Their Poisson bracket is obtained from $\delta\gamma_A=M_A^{ij}\delta\gamma_{ij}$ and $\delta\pi_A=M^A_{ij}\delta\pi^{ij}$ which leads to $\{\delta\gamma_A(\tau,\vec{k}),\delta\pi^\star_B(\tau,\vec{q}\}=M_A^{ij}M^{mn}_B\{\delta\gamma_{ij}(\tau,\vec{k}),\delta\pi^{mn\,\star}(\tau,\vec{q})\}$. Using the expression of the Poisson brackets of the induced metric and its conjugate momenta, it is shown that
\beq
	\left\{\delta\gamma_A(\tau,\vec{k}),\delta\pi^\star_B(\tau,\vec{q})\right\}=\delta_{A,B}\,\delta^3\left(\vec{k}-\vec{q}\right).
\eeq 
All other Poisson brackets vanish. Note that we can define the scalar gravitational degrees of freedom in real space by inverse Fourier transform, \ie
\beq
	\delta\gamma_A(\tau,\vec{x})=\ds\int\frac{\dd^3\vec{k}}{(2\pi)^{3/2}}\delta\gamma_A(\tau,\vec{k})e^{i\vec{k}\cdot\vec{x}},
\eeq
and similarly for the momentas. It yields 
\bea
    \delta\gamma_{ij}=\delta\gamma_1(\tau,\vec{x})\,\frac{\tilde{\gamma}_{ij}}{\sqrt{3}}+\sqrt{\frac{3}{2}}\left(\frac{\partial_i\partial_j}{\partial^2}-\frac{\tilde{\gamma}_{ij}}{{3}}\right)\delta\gamma_2(\tau,\vec{x}),
\eea
where $\partial^2\equiv\tilde{\gamma}^{ij}\partial_i\partial_j$  stands for the Laplace operator in flat space, and $\partial^{-2}$ for its inverse.

Finally, one should  introduce inhomogeneous and anisotropic deviations for the Lagrange multipliers
\bea
	\delta N(\tau,\vec{x})&=&N(\tau,\vec{x})-N(\tau), \\
	\delta N^i(\tau,\vec{x})&=&N^i(\tau,\vec{x})-N^i(\tau).
\eea
The lapse function is by construction a scalar quantity. The shift vector can be expanded as a curl-free part which derives from a scalar quantity $(\partial_i\delta N_1)$, and a divergence-free vector $(\delta N_2^i)$. Setting the divergence-free vector to zero, fluctuations in the shift vector reduces to $\delta N_i=\partial_i\delta N_1$.  In Fourier space, it gives
\beq
	\delta N^i(\tau,\vec{k})=i(k^i/k)\delta N_1(\tau,\vec{k}),
\eeq
which satisfies $\delta N^\star_1(\tau,\vec{k})=\delta N_1(\tau,-\vec{k})$.

\subsection{Second-order action}
The dynamics of the perturbed FLRW space time at leading order in perturbation is obtained by expanding the action \Eq{eq:SHam} around the homogeneous and isotropic degrees of freedom and up to quadratic order in cosmological perturbations, \ie $S=S^{(0)}+S^{(1)}+S^{(2)}$. The background evolution is given by $S^{(0)}$ and its extremum leads to the Friedman and Klein-Gordon equations. Assuming the homogeneous and isotropic background is solved, $S^{(1)}$ is vanishing owing to the least action principle. Hence, the dynamics of cosmological perturbations is derived from $S^{(2)}$. In the Hamiltonian framework, this means that we need to compute the constraints for the homogeneous and isotropic background, $H^{(0)}=N(\tau)\mathcal{C}^{(0)}$ (note that the background diffeomorphism constraint is zero), and at quadratic order $ H^{(2)}=\int\dd^3x [\delta N \,\mathcal{C}^{(1)}+\delta N^i \,\mathcal{D}^{(1)}_i+N\,\mathcal{C}^{(2)}]$. A direct expansion of the constraints up to quadratic order has been applied to the case of single-field inflation \cite{Artigas_2022} and here, we present its extension to the multifiled context. 

\subsubsection{Homogeneous and isotropic background}
At the background level, the action is greatly simplified thanks to isotropy which leads to $N^i=0$, and to homogeneity which leads to $\mathcal{D}_i=0$ and $R^{(3)}=0$. Plugging $\gamma_{ij}=v^{2/3}\tilde{\gamma}_{ij}$ and $\pi^{ij}=v^{1/3}\theta\tilde{\gamma}^{ij}/2$ into \Eq{eq:SHam}, one easily obtains
\beq
	S^{(0)}=\ds\int\dd \tau\left[\pi_I\dot{\phi}^I+\theta\dot{v}-N\mathcal{C}^{(0)}\right].
\eeq
where the scalar constraint is given by
\beq
	\mathcal{C}^{(0)}=\frac{1}{2v}G^{IJ}\pi_I\pi_J+vV-\frac{3}{4\Mp^2}v\theta^2. \label{eq:C0}
\eeq
As a result, the equations of motion for the scalar fields and the gravitational field are
\bea
	\dot\phi^I&=&\frac{N}{v}G^{IJ}\pi_J, \label{eq:dotphi} \\
	D_\tau \pi_I&=&-NvV_{;I}, \label{eq:dotpi}
\eea
and 
\bea
	\dot{v}&=&-\frac{3N}{2\Mp^2}v\theta, \label{eq:dotgamma} \\
	%\dot{\theta}&=&N\left(\frac{3}{4\Mp^2}\theta^2+\frac{1}{2v^2}G^{IJ}\pi_I\pi_J-V\right), \label{eq:dotpig}
	\dot{\theta}&=&N\frac{G^{IJ}\pi_I\pi_J}{v^2}. \label{eq:dotpig}
\eea
 These equations of motions have to be solved under the constraint $\mathcal{C}^{(0)}= 0$.\footnote{By injecting $\pi_I=(v/N)G_{IJ}\dot\phi^J$ in the equation of motion for $\pi_I$, one recovers the Klein-Gordon equation (assuming any arbitrary time variable here) 
\beq
	D_\tau \dot\phi^I+\left(\frac{\dot{v}}{v}-\frac{\dot{N}}{N}\right)\dot\phi^I+N^2G^{IJ}V_{,J}=0.
\eeq}
%Note also that using the constraint allows us to get $\dot{\theta}=NG^{IJ}\pi_I\pi_J/v^2$.

In the following, it will be useful to introduce the energy density and the pressure of the matter sector as
\bea
	\rho&:=&\frac{1}{2v^2}G^{IJ}\pi_I\pi_J+V, \label{eq:rho} \\
	p&:=&\frac{1}{2v^2}G^{IJ}\pi_I\pi_J-V, \label{eq:p}
\eea
which here should be interpreted as functions of the phase space, \ie of both $\phi^I$ and $\pi_I$.%\footnote{The usual expressions $\rho=G_{IJ}\dot\phi^I\dot\phi^J/2+V$ and $p=G_{IJ}\dot\phi^I\dot\phi^J/2-V$ are recovered by setting $N=1$ where here a dot stands for deriving with respect to the cosmic time.} In single field inflation, only the potential energy is a function of the field configuration. In this context however, the kinetic energy is also a function of the fields configurations through the metric $G_{IJ}$.
%Let us also introduce $H_N:=\dot{a}/a=\dot{v}/(3v)$ (note that $H$ will be used when working with cosmic time). With such variables, the dynamical equation for the volume reads $H_N=-N\theta/(2\Mp^2)$ and the constraint is $H_N^2=N^2\rho/(3\Mp^2)$ which is no more than the Friedman equation assuming an arbitrary choice for the time variable. Similarly, the equation of motion for $\theta$ (which is related to the expansion rate) reads $\dot\theta=N(\rho+p)$. In the context of inflation, it is common to introduce the first slow-roll parameter $\epsilon_1=-H^{-2}\dd H/\dd t$. Using our phase-space variables, it reads $N\epsilon_1=2\Mp^2\dot\theta/\theta^2$ and the Raychaudhuri equation now reads $2\Mp^2(\rho+p)=\theta^2\epsilon_1$.

\subsubsection{Cosmological perturbations}
\label{sssec:cosmopert}
Let us now consider cosmological perturbations.  The linear action is vanishing on the background solution (see App. \ref{app:linactionnaive}) and their dynamics is given by the quadratic action. It has the following form
\beq
	S^{(2)}=\ds\int\dd \tau\int\dd^3x\left(\delta\pi^{ij}\dot{\delta\gamma}_{ij}+\delta\pi_I\dot{\delta\phi^I}-\delta N\mathcal{C}^{(1)}-\delta N^i\mathcal{D}^{(1)}_i-N\mathcal{C}^{(2)}\right). \label{eq:2ndactgen}
\eeq
Varying the action with respect to $\delta N$ and $\delta N^i$ enforces $\mathcal{C}^{(1)}$ and $\mathcal{D}^{(1)}_i$  to vanish, hence constraining the region of phase-space trajectories generated by $N\mathcal{C}^{(2)}$. Hereafter, we will call them the linear constraints. Strictly speaking, $\mathcal{C}^{(2)}$ is not a constraint anymore even though it emanates from a totally constrained system. Nonetheless, we will call it the quadratic constraint for simplicity.

As explained previously, the quadratic action is most preferentially dealt with in Fourier space. However, one should avoid double counting of the degrees of freedom by taking into account that in Fourier space all fields are constrained by $\delta\Phi^\star(\tau,\vec{k})=\delta\Phi(\tau,-\vec{k})$, and similarly for the Lagrange multipliers. This is done by splitting the integral over $\mathbb{R}^3$ into an integral over $\mathbb{R}^{3+}:=\mathbb{R}^2\times\mathbb{R}^+$ and an integral over $\mathbb{R}^{3-}:=\mathbb{R}^2\times\mathbb{R}^-$. The second integral can be expressed as an integral over $\mathbb{R}^{3+}$ by performing the change of integration variable from $\vec{k}$ to $\vec{k}'=-\vec{k}$, and then by making use of $\delta\Phi^\star(\tau,\vec{k})=\delta\Phi(\tau,-\vec{k})$. As a result, the quadratic action in Fourier space boils down to
\bea
\label{second order action}	S^{(2)}&=&\ds\int\dd \tau\int_{\mathbb{R}^{3+}}\dd^3k\left[\left(\delta\pi^\star_I\dot{\delta\phi^I}+\mathrm{c.c.}\right)+\sum_{A=1}^2\left(\delta\pi^\star_A\dot{\delta\gamma}_A+\mathrm{c.c.}\right)\right. \\
	&&\left.-\left(\delta N^\star\,\mathcal{C}^{(1)}+\mathrm{c.c.}\right)+\frac{k^i}{k}\left(i\delta N^\star_1\,\mathcal{D}^{(1)}_i+\mathrm{c.c.}\right)-2N\mathcal{C}^{(2)}\right], \nonumber
\eea
where c.c. means the complex conjugate and which comes from the integral over $\mathbb{R}^{3-}$, and where all the quantities entering are now expressed in Fourier space. 

The constraints entering the above are derived in App. \ref{app:constraints} and we only report their expressions below. The scalar constraint at first order in Fourier space is given by
\bea
	\mathcal{C}^{(1)}&=&v\rho_{,I}\,\delta\phi^I+\frac{1}{v}G^{KI}\pi_K\delta\pi_I-\frac{\sqrt{3}}{2}{v^{1/3}}\left(\rho-\frac{\theta^2}{4\Mp^2}\right)\delta\gamma_1-\frac{\sqrt{3}}{\Mp^2}v^{2/3}\theta\,\delta\pi_1 \nonumber \\
	&&-\Mp^2\frac{k^2}{v^{2/3}}\left[\delta\gamma_1-\frac{1}{\sqrt{2}}\delta\gamma_2\right].
\eea
The diffeomorphism constraint at first order reads $\mathcal{D}^{(1)}_i=ik_i\mathcal{D}^{(1)}$ with
\beq
	\mathcal{D}^{(1)}=\pi_I\delta\phi^I+\frac{1}{\sqrt{3}}v^{1/3}\theta\left(\frac{1}{2}\delta\gamma_1-\sqrt{2}\delta\gamma_2\right)-\frac{2}{\sqrt{3}}v^{2/3}\left(\delta\pi_1+\sqrt{2}\delta\pi_2\right).
\eeq
Finally, the scalar constraint at quadratic order can be written as $\mathcal{C}^{(2)}=\mathcal{C}^{(2)\,G}+\mathcal{C}^{(2)\,\phi}+\mathcal{C}^{(2)\,V}$ where $\mathcal{C}^{(2)\,G}$ and $\mathcal{C}^{(2)\,\phi}$ are the pure gravitational and pure matter sector, and where $\mathcal{C}^{(2)\,V}$ encodes the coupling between matter and gravity. They are given by
\bea
	C^{(2)\,G}&=& \frac{v^{1/3}}{\Mp^2}\left(-|\delta\pi_1|^2 + 2|\delta\pi_2|^2\right)+\frac{1}{12v^{1/3}}\left(5\rho+3p-\frac{\Mp^2k^2}{v^{2/3}}\right)|\delta\gamma_1|^2 \nonumber \\
	&&+ \frac{1}{12v^{1/3}}\left(5\rho+3p-\frac{\Mp^2k^2}{2v^{2/3}}\right)|\delta\gamma_2|^2+ \frac{\sqrt{2}\Mp^2k^2}{24v}\left(\delta\gamma_1\delta\gamma_2^\star+\mathrm{c.c.}\right) \nonumber \\ 
	&&-\frac{\theta}{4\Mp^2}\left(\delta\pi_1\delta\gamma_1^\star + \mathrm{c.c.}\right) + \frac{\theta}{2\Mp^2}\left(\delta\pi_2\delta\gamma_2^\star + \mathrm{c.c.}\right) \label{eq:C2G} \\
	\mathcal{C}^{(2)\,\phi}&=&\frac{1}{4v}G\uTsr\left(\delta\pi_I\delta\pi_J^\star+ \mathrm{c.c.}\right) + \frac{v}{4}\left(\frac{k^2}{v^{2/3}}G_{IJ} + \rho,_{IJ}\right)\left(\delta\phi^I\delta\phi^{J\,\star}+ \mathrm{c.c.}\right) \nonumber \\
	&&+ \frac{1}{2v}G\uTsr,_K\pi_J\left(\delta\phi^K\delta\pi_I^\star + \mathrm{c.c.}\right) \label{eq:C2phi} \\
	\mathcal{C}^{(2)\,V}&=&- \frac{\sqrt{3}v^{1/3}}{4}\left[\left(p_{,I}\delta\phi^I +\frac{1}{v^2}G\uTsr\pi_J\delta\pi_I\right)\delta\gamma_1^\star+ \mathrm{c.c.}\right] \label{eq:C2V}
\eea
where we made use of $\mathcal{C}^{(0)}=0$ to simplify some of the background functions entering $\mathcal{C}^{(2)\,G}$. We remind that the expression of $\rho$ and $p$ are given in Eqs. (\ref{eq:rho}) and (\ref{eq:p}). 

These constraints are very similar to the case of single-field inflation which can be retrieved from the above setting $G_{IJ}=\delta_{IJ}$ and replacing $\rho$ and $p$ by their expression for one single field, \ie $\rho=\pi^2_\phi/(2v^2)+V(\phi)$ and $p=\pi^2_\phi/(2v^2)-V(\phi)$ (see Sec. 3.2 of \cite{Artigas_2022}). An important point to stress is that in the multifield context, not only the potential energy in $\rho$ and $p$ depends on the fields configurations, but also the kinetic energy through $G_{IJ}$. For instance, the term $\rho_{,IJ}$ which stands for the effective masses of the scalar fields, has contributions from both the kinetic energy and the potential in the multifield context while it only has a contribution from $V$ for single-field inflation. This similarly applies to the term $\rho_{,I}+3p_{,I}$ which quantifies the strength of the coupling between the scalar field configurations and the gravitational field, and this also explains the very last term in $\mathcal{C}^{(2)\,\phi}$ which is otherwise absent in the single-field framework. A part from that, we recover that the scalar fields couple to the isotropic degrees of freedom only, as it should since scalar fields are isotropic by definition, and that the two gravitational degrees of freedom are coupled through a gradient interaction, \ie the term $k^2(\delta\gamma_1)(\delta\gamma_2)$ which in real-space is of the form $\tilde\gamma^{ij}(\partial_i\delta\gamma_1)(\partial_j\delta\gamma_2)$. 

\subsection{Covariant Hamiltonian}\label{manifestly covariant variables}
The second-order action should be written  in a manifestly covariant way in the field-space since it results from the expansion of a full Lagrangian which is manifestly covariant. However, this is not the case of Eq. (\ref{eq:2ndactgen}). Indeed, the perturbative degrees of freedom $(\delta\phi^I,\delta\pi_I)$ introduced in Sec. \ref{ssec:dof} do not transform covariantly under fields redefinition \cite{Gong_2011,Pinol:2020cdp}. Moreover, $\delta\pi_I\dot{\delta\phi^I}$ is made of partial time-derivative rather than of covariant one. Because of that, the Hamiltonian of cosmological perturbations $\mathcal{H}^{(2)}\equiv\delta N \,\mathcal{C}^{(1)}+\delta N^i \,\mathcal{D}^{(1)}_i+N\,\mathcal{C}^{(2)}$ is not manifestly covariant neither. 

\subsubsection*{Covariant variables}
The naive variables $(\delta\phi^I,\delta\pi_I)$ are defined as finite differences between two neighbouring points in the field space and they can be related to covariant derivatives of the fields along the geodesic in the field-space connecting such points. The latter being covariant by construction, they provide us with a set of covariant variables which describe deviations from homogeneous and isotropic configurations of the fields, which is exactly what we are aiming for. Details of such a construction can be found in App. \ref{app:covQP} and here we only reported the most important features. (Note that the gravitational degrees of freedom are scalars in the field-space, hence they are already covariant quantities, and the field redefinition we apply to the naive variables does not affect the metric nor its perturbation.)

Because the covariant variables --that we name $(Q^I,P_I)$ hereafter-- are differential by nature, the finite differences defining $\delta\phi^I$ and $\delta\pi_I$ are related to them through a Taylor expansion. As a result, the transformation going from the naive variables to the covariant ones is non-linear. Such a Taylor expansion is performed up to quadratic order in App. \ref{app:covQP} and it is shown that at the linear order, covariance is enough to get pairs of canonically conjugate variables. However, requiring covariance is not sufficient anymore to uniquely define momentum-like variables at the quadratic order as it only leads to a one-parameter family of $P_I$'s instead \cite{Pinol:2020cdp}. Interestingly, this parameter can be fixed to a unique value otherwise the Poisson bracket $\{P_I,P_J\}$ is nonzero for $I\neq J$, and the covariant variables would not form a set of canonical pairs. Hence, the naive variables are given as functions of the covariant ones by
\bea
	\delta\phi^I &=& Q^I -\frac 1 2 \Gamma^I_{LK} Q^L Q^K+\mathcal{O}(\lambda^3) \label{mapping phi main}, \\
            \delta\pi_I &=& P_I + \Gamma^K\Tsr \pi_K  Q^J + \Gamma^K\Tsr P_K Q^J \label{mapping pi main} \\ \nonumber
            &&+ \frac 1 2 \left(\Gamma^S_{IJ,K} - \Gamma^S_{IR}\Gamma^R_{JK} + \Gamma^R\Tsr\Gamma^S_{RK}+\frac{1}{3}{R_{IJK}}^S\right)\pi_S Q^J Q^K+\mathcal{O}(\lambda^3),
\eea
where $\Gamma^K_{IJ}$ and $R_{IJKL}$ are the Christoffel symbols and the Riemann curvature tensor associated to the metric in the field space, $G_{IJ}$. 

\subsubsection*{Generating function}
The above relation, Eqs. (\ref{mapping phi main}) \& (\ref{mapping pi main}), defines a canonical transformation since $\{Q^I,P_J\}=\delta^I_J$ while all other Poisson brackets vanish. A standard way to implement it in the Hamiltonian action is through the use of generating functions. Here we consider such a function of the form $G=-Q^IP_I+F(\delta\phi^I,P_J,\tau)$ where $F$ can explicitly depend on time and is further built requiring \bea
 	\frac{\partial F}{\partial \delta\phi^I}=\delta\pi_I & \,\,\,\mathrm{and}\,\,\,& \frac{\partial F}{\partial P_I}=Q^I. \label{eq:dFdP main}
\eea
In order to generate the quadratic relation between the old and the new phase-space coordinates given in Eqs. (\ref{mapping phi main}) \& (\ref{mapping pi main}) and requiring Eq. (\ref{eq:dFdP main}), the function $F$ should be cubic in $\delta\phi^I$ and $P_J$, and its lowest order should be quadratic. Its construction is detailed in App. \ref{app:genfunc} and we arrived at 
\bea
	F(\delta\phi^I,P_J,t)&=&P_I\delta\phi^I+\frac{1}{2}\Gamma^K_{IJ}\pi_K\,\delta\phi^I\delta\phi^J+\frac{1}{2}\Gamma^K_{IJ}\,P_K\delta\phi^I\delta\phi^J \nonumber \\
	&&+\frac{1}{6}\left(\Gamma^S_{IJ,K}  + \Gamma^R_{IJ}\Gamma^S_{RK}\right)\pi_S\,\delta\phi^I\delta\phi^J\delta\phi^K, \label{eq:Fcubic}
\eea
where the two first terms generate the linear order in Eqs. (\ref{mapping phi main}) \& (\ref{mapping pi main}), and the two remaining ones the quadratic order.

\subsubsection*{Hamiltonian for covariant variables}
A priori, the Hamiltonian for the covariant variables should be derived from the Hamiltonian action of the naive variables including both the linear and the quadratic order in $(\delta\phi^I,\delta\pi_I)$. Indeed, the action linear in the naive variables bears contributions which are quadratic in $Q^I$ and $P_I$ owing to their non-linear relation to $\delta\phi^I$ and $\delta\pi_I$. The analysis keeping track of both the linear action and the quadratic action in the naive variable is done in App. \ref{app:actionQP}. There, it is shown that contributions originating from the action linear in the naive variables are vanishing at any order in the covariant variables once the isotropic and homogeneous degrees of freedom are taken to be solutions of the background evolution. This has to be so since the linear action is zero solely because of imposing the equations of motion for the background variables to hold and irrespectively of whether $\delta\phi^I$ and $\delta\pi_I$ are linearly or non-linearly related to any other kind of variables. Hence, it is sufficient to implement the canonical transformation considering the second-order action only. Thanks to that, the covariant variables can be restricted to their linear relation to the naive one by discarding any quadratic term in Eqs. (\ref{mapping phi main}) \& (\ref{mapping pi main}). Similarly, the canonical transformation is generated by restricting $F$ to be quadratic, thus given by Eq. (\ref{eq:Fcubic}) where cubic terms are removed.

The action in the covariant variable is related to the action in the naive variable through a total time-derivative which ensures that the two actions have the same extremum, \ie
\bea
	\ds\int\dd^3x\left[\dot{\delta\phi^I}\delta\pi_I-\mathcal{H}^{(2)}\left(\delta\phi^I,\delta\pi_I\right)\right]&=&\ds\int\dd^3x\left[\dot{Q^I}P_I-\mathcal{K}^{(2)}\left(Q^I,P_I\right)+\frac{\dd G}{\dd\tau}\right], \label{eq:actionFsec}
\eea
where $\mathcal{K}^{(2)}$ is the new Hamiltonian in $(Q^I,P_I)$. We stress that the background is not affected by the canonical transformation, \ie $\mathcal{K}^{(0)}=\mathcal{H}^{(0)}$, and similarly for the gravitational perturbations which leads to $\mathcal{K}^{G}=\mathcal{H}^{G}$, hence the pure gravitational part of the Hamiltonian can be safely ignored in the above equality. Note however that one should keep the coupling between scalar fields perturbations and gravitational perturbations.  Considering the generating function of type 2 introduced earlier, we are left with
\bea
	\ds\int\dd^3x\,\mathcal{K}^{(2)}&=&\ds\int\dd^3x\left[\mathcal{H}^{(2)}(\delta_1\phi^I,\delta_1\pi_I)+\frac{\partial F}{\partial\tau}(\delta_1\phi^I,\delta_1\pi_I)\right],
\eea
where $\delta_1\phi^I$ and $\delta_1\pi_I$ means that the naive variables should be replaced by their expressions as functions of the covariant variables truncated at first order. Since $\mathcal{H}^{(2)}$ has been derived in Fourier space, the above relation has to be Fourier transformed paying attention not to double-count the degrees of freedom. 

The derivation of $\mathcal{K}^{(2)}$ is detailed in App. \ref{app:actionQP}. First, one simply replaces all of the $(\delta\phi^I,\delta\pi_J)$ terms in the Hamiltonian given in Eq. (\ref{second order action}) with the linear part of Eqs. (\ref{mapping phi main}) \& (\ref{mapping pi main}). %To this end we remind that 
%\bea
%	\ds\int\dd^3x\mathcal{H}^{(2)}=\ds\int_{\mathbb{R}^{3+}}\dd^3k\left[\left(\delta N^\star\mathcal{C}^{(1)}+\mathrm{c.c.}\right)+k\left(\delta N_1^\star\mathcal{D}^{(1)}+\mathrm{c.c.}\right)+2N\mathcal{C}^{(2)}\right].
%\eea
Second we will have quadratic terms coming from the partial time-derivative of the generating function $F$, \ie 
\bea
    \frac{\partial F}{\partial \tau} &=& \frac 1 2 \left(\partial_t \Gamma^K_{IJ}\pi_K + \Gamma^K_{IJ}\dot \pi_K\right)Q^IQ^J, \label{replacement last}
\eea
in which $\delta\phi^I$ has been replaced by its expression as a function of $Q^I$, restricting to quadratic orders in $\partial F/\partial\tau$, hence to the linear order in Eq. (\ref{mapping phi main}).
Collecting all the different contributions, the new Hamiltonian in Fourier space is expressed as
\bea
	\mathcal{K}^{(2)}=\left(\delta N^\star\mathcal{C}^{(1)}+\mathrm{c.c.}\right)+k\left(\delta N^\star_1\mathcal{D}^{(1)}+\mathrm{c.c.}\right)+2N\left(\mathcal{K}^{(2)\,\phi}+\mathcal{K}^{(2)\,G}+\mathcal{K}^{(2)\,V}\right),
\eea
where the linear constraints are given in Eqs. (\ref{eq:diffeocov}) \& (\ref{eq:scalarcov}), and where the gravitational part of the quadratic constraints remains unmodified, \ie $\mathcal{K}^{(2)\,G}=\mathcal{C}^{(2)\,G}$ [see Eq. (\ref{eq:C2G})]. The coupling between the scalar fields perturbations and the gravitational perturbations comes from $\mathcal{C}^{(2)\,V}$ now expressed as a function of the new variables, and with no contributions from the generating function. The details of these computations are given in appendix \ref{app:actionQP}. The end result reads
\bea
	\mathcal{K}^{(2)\,V}&=&\frac{\sqrt{3}v^{1/3}}{4}\left[\left(V;_IQ^I-\frac{1}{v^2}G\uTsr\pi_JP_I\right)\delta\gamma_1^\star+ \mathrm{c.c.}\right], \label{eq:KVcov}
\eea
which is manifestly covariant. Again, it is easy to show that $-V_{;I}=p_{;I}$ and the covariant version of $\mathcal{K}^{(2)\,V}$ could have been inferred by replacing standard derivatives with covariant ones. The last term entering the new Hamiltonian, $\mathcal{K}^{(2)\,\phi}$, is the most intricate one since it collects terms from the two contributions, \ie 
\bea
	\mathcal{K}^{(2)\,\phi}&=&\frac{1}{4v}G\uTsr\left(P_IP_J^\star+ \mathrm{c.c.}\right)+\frac{1}{2v}H^{LI}_M\pi_L\left(P_IQ^{M\star}+ \mathrm{c.c.}\right) \\
	&&+\left\{\frac{v}{4}\left(\frac{k^2}{v^{2/3}}G_{IJ}+V_{;IJ}-\frac{1}{v^2}{{R_I}^{MN}}_J\,\pi_M\pi_N\right)\right\}\left(Q^IQ^{J\star}+\mathrm{c.c.}\right), \nonumber 
\eea
where\footnote{\label{ftnt} This is obtained combining $G^{KL}_{\,\,\,\,,I}=-G^{KM}G^{LN}G_{MN,I}$ and $G_{MN;I}=0$ which yields $G_{MN,I}=G_{MS}\Gamma^{S}_{NI}+G_{SN}\Gamma^{S}_{IM}$. Hence, one arrives at ${G^{KL}},_I=-G^{KJ}\Gamma^L_{JI}-G^{JL}\Gamma^K_{IJ}$ and at ${G^{KL}},_I\pi_K\pi_L=-2G^{KJ}\Gamma^L_{JI}\pi_K\pi_L$ using the symmetry of $\pi_K\pi_L$.}
\bea
	H^{LI}_M&=&G^{IL},_M+G\uTsr\Gamma^{L}_{JM}=-G^{JL}\Gamma^I_{JM}.
\eea
Let us note that the term multiplying $Q^IQ^J$ is manifestly covariant, and shows that the curvature in the field space now contribute to the effective mass.

\subsubsection*{Manifestly covariant Hamiltonian}
At this stage, the Hamiltonian is still not manifestly covariant because of the term $-G^{JL}\Gamma^I_{JM}\pi_L\left(P_IQ^{M\star}+ \mathrm{c.c.}\right)/(2v)$ in $\mathcal{K}^{(2)\,\phi}$. However, this term can be combined with $(P^\star_I\dot{Q}^{I}+\mathrm{c.c.})$ in the second-order action to get $(P^\star_ID_\tau {Q}^{I}+\mathrm{c.c.})$ which allows us to identify the covariant Hamiltonian in the second-order action. Indeed, the Hamiltonian action at second order now reads
\beq
	S^{(2)}=\ds\int\dd \tau\int_{\mathbb{R}^{3+}}\dd^3k\left[\left(P^\star_ID_\tau {Q}^{I}+\mathrm{c.c.}\right)-\mathcal{K}^{(2)}_{\mathrm{cov}}\right],
\eeq
where $\mathcal{K}^{(2)}_{\mathrm{cov}}$ is a manifestly covariant Hamiltonian. We write it as
\bea
	\mathcal{K}^{(2)}_{\mathrm{cov}}=\left(\delta N^\star\mathcal{C}^{(1)}_{\mathrm{cov}}+\mathrm{c.c.}\right)+k\left(\delta N^\star_1\mathcal{D}^{(1)}_{\mathrm{cov}}+\mathrm{c.c.}\right)+2N\left(\mathcal{K}^{(2)\,\phi}_{\mathrm{cov}}+\mathcal{K}^{(2)\,G}_{\mathrm{cov}}+\mathcal{K}^{(2)\,V}_{\mathrm{cov}}\right),
\eea
where $\mathcal{C}^{(1)}_{\mathrm{cov}}=\mathcal{C}^{(1)}$, $\mathcal{D}^{(1)}_{\mathrm{cov}}=\mathcal{D}^{(1)}$ and $\mathcal{K}^{(2)\,V}_{\mathrm{cov}}=\mathcal{K}^{(2)\,V}$ are given by Eqs. (\ref{eq:diffeocov}), (\ref{eq:scalarcov}) and (\ref{eq:KVcov}), respectively, and which are manifestly covariant. As already mentioned, the gravitational Hamiltonian is unchanged, \ie $\mathcal{K}^{(2)\,G}_{\mathrm{cov}}=\mathcal{C}^{(2)\,G}$ which can be found in Eq. (\ref{eq:C2G}), and it is by construction covariant in the field space. Finally, the Hamiltonian of the scalar-fields fluctuations is
\beq
	\mathcal{K}^{(2)\,\phi}_{\mathrm{cov}}=\frac{1}{4v}G\uTsr\left(P_IP_J^\star+ \mathrm{c.c.}\right) + \frac{v}{4}\left(\frac{k^2}{v^{2/3}}\delta_{IJ} + V;_{IJ} -\frac{1}{v^2}R_I{}^{KL}{}_J\bar\pi_K\bar\pi_L \right)\left(Q^IQ^{J\,\star}+ \mathrm{c.c.}\right), \label{eq:Kphicov}
\eeq
which is now built from covariant quantities only.

\section{Dynamics of scalar perturbations}
\label{sec:dynCPT}
\subsection{Equations of motion}
\subsubsection*{Equations of motion in the field basis}
The equations of motion for the fluctuations are given by the Hamilton equations, \ie $\dot{\delta z}=\left\{\delta z,\mathcal{K}^{(2)}\right\}$, where $\delta z\in\{Q^I,\delta\gamma_1,\delta\gamma_2,P_I,\delta\pi_1,\delta\pi_2\}$ and where the Hamiltonian that should be used here is the one which is not manifestly covariant. However, and since one starts from an action which is covariant, the Hamilton equations can be casted in a manifestly covariant manner as $D_\tau {\delta z}=\left\{\delta z,\mathcal{K}^{(2)}_\mathrm{cov}\right\},$ where the covariant time-derivative is given by the usual time-derivative for the gravitational degrees of freedom since these are scalars with respect to the field-space metric.

Explicitly, the dynamical equations read 
\bea
	&&\left\{\begin{array}{rcl}
	D_\tau \delta\gamma_1&=&\ds\frac{-2}{\sqrt{3}}v^{2/3}k\delta N_1-\frac{\sqrt{3}}{\Mp^2}v^{2/3}\theta\delta N-\frac{N}{\Mp^2}\left(2v^{1/3}\delta\pi_1+\frac{\theta}{2}\delta\gamma_1\right) \\
	D_\tau \delta\pi_1&=&\ds\frac{-v^{1/3}\theta}{2\sqrt{3}}k\delta N_1+\frac{v^{1/3}}{\sqrt{3}}\left[\frac{1}{2}\left(\rho+3p\right)+\Mp^2\frac{k^2}{v^{2/3}}\right]\delta N \\
	&&\ds-N\left[\frac{1}{6v^{1/3}}\left(5\rho+3p-\Mp^2\frac{k^2}{v^{2/3}}\right)\delta\gamma_1-\frac{\theta}{2\Mp^2}\delta\pi_1+\frac{\Mp^2\sqrt{2}}{12v}k^2\delta\gamma_2\right]  \\
	&&\ds-\frac{\sqrt{3}}{2}v^{1/3}N\left(V_{;I}Q^I-\frac{G^{IJ}\pi_J}{v^2}P_I\right),
	\end{array}\right. \label{eq:CovDiff1} \\
	&&\left\{\begin{array}{rcl}
	D_\tau \delta\gamma_2&=&\ds-2\sqrt{\frac{2}{3}}v^{2/3}k\delta N_1+\frac{N}{\Mp^2}\left(4v^{1/3}\delta\pi_2+{\theta}\delta\gamma_2\right) \\
	D_\tau \delta\pi_2&=&\ds\sqrt{\frac{2}{3}}v^{1/3}k\delta N_1-\frac{\Mp^2}{v^{1/3}\sqrt{6}}k^2\delta N \\
	&&\ds-N\left[\frac{1}{6v^{1/3}}\left(5\rho+3p-\Mp^2\frac{k^2}{2v^{2/3}}\right)\delta\gamma_2+\frac{\theta}{\Mp^2}\delta\pi_2+\frac{\Mp^2\sqrt{2}}{12v}k^2\delta\gamma_1\right], 
	\end{array}\right. \label{eq:CovDiff2} \\
	&&\left\{\begin{array}{rcl}
	D_\tau Q^I&=&\ds\frac{1}{v}G^{IJ}\pi_J \delta N + N\left(\frac{1}{v}G^{IJ}P_J-\frac{\sqrt{3}}{2v^{5/3}}G^{IJ}\pi_J\delta\gamma_1\right) \\
	D_\tau P_I&=&\ds-\pi_I k\delta N_1-vV_{;I}\delta N \\
	&&\ds-N\left[v\left(\frac{k^2}{v^{2/3}}+V_{;IJ}-\frac{1}{v^2}R_I{}^{KL}{}_J\pi_K\pi_L\right)Q^J+\frac{\sqrt{3}}{2}v^{1/3}V_{;I}\delta\gamma_1\right].
	 \end{array}\right. \label{eq:CovDiffQP}
\eea
The full dynamics of cosmological fluctuations has to be solved on surface of constraints and the above equations are complemented with the the linear constrains reading
\beq
	\left\{\begin{array}{rcl}
	0&=&\ds vV_{;I}\,Q^I+\frac{1}{v}G^{KI}\pi_KP_I-\frac{v^{1/3}}{\sqrt{3}}\left[\frac{1}{2}\left(\rho+3p\right)+\Mp^2\frac{k^2}{v^{2/3}}\right]\delta\gamma_1+\frac{\Mp^2}{\sqrt{6}}\frac{k^2}{v^{1/3}}\delta\gamma_2 \\
	&&\ds-\frac{\sqrt{3}}{\Mp^2}v^{2/3}\theta\,\delta\pi_1 \\
	0&=&\ds\pi_IQ^I+\frac{1}{\sqrt{3}}v^{1/3}\theta\left(\frac{1}{2}\delta\gamma_1-\sqrt{2}\delta\gamma_2\right)-\frac{2}{\sqrt{3}}v^{2/3}\left(\delta\pi_1+\sqrt{2}\delta\pi_2\right).
	\end{array}\right. \label{eq:constcov}
\eeq
These generalize the Hamiltonian dynamics of cosmological perturbations derived in \cite{Artigas_2022} to the multifield context in a covariant manner. Compared to the dynamics of the naive variables, the key difference lies in the effective mass of the scalar-field perturbations which is now given by a combination of the covariant derivatives of the potential and the Riemann curvature tensor. 

\subsubsection*{Adiabatic and entropic modes}
In the multifield context, cosmological perturbations are usually decomposed over the adiabatic and entropic directions. To this end, let us first introduce the background adiabatic momenta $\pi_\sigma=\sqrt{G^{IJ}\pi_I\pi_J}$ from which the adiabatic direction is defined by
\bea
	e_I^\sigma=\frac{\pi_I}{\pi_\sigma},
\eea
which is equivalent to the standard definition, $e_I^\sigma\equiv\dot\phi^I/\sqrt{G_{MN}\dot{\phi}^M\dot{\phi}^N}$, owing to the background equations of motion. The remaining $n -1$ directions are the entropic ones that we build using as a Frenet basis (see \eg \cite{Kaiser_2013,Achucarro:2018ngj,Pinol:2020kvw}). They are labelled by $s$ which runs from 1 to $n-1$. The first entropic direction is defined as the covariant rate turn of the adiabatic one. Coordinate wise, it gives $N\omega_1e^{1}_I=D_\tau e^\sigma_I$ where $\omega_1$ is the turning rate of the adiabatic direction and  $e^{1}_I$ have a unit-norm. (Note that we add an extra background lapse in order to account for any choice of the time variable.) From a physical standpoint, it measures the coupling between the adiabatic direction and the first entropic direction. The rest of the entropic directions are defined as the wedge product of the two previous vectors. This construction leads to the following expression of the covariant rate turn
\bea
	D_\tau e^a_I=N{\Omega^a}_b\,e^b_I, \label{eq:dotAEbasis}
\eea
where the matrix ${\Omega^a}_b$ is antisymmetric and made of the covariant rate turns, \ie
\beq
	{\Omega^a}_b=\begin{pmatrix}
                0 & \omega_1 & 0 & \cdots & 0 \\
                -\omega_1 & 0 & \omega_2 & \cdots & 0 \\
                0 & -\omega_2 & 0 & \cdots & 0 \\
                \vdots & \ddots & \ddots & \ddots & \vdots \\
                  0 & 0 & \cdots & 0& \omega_{n-1} \\
              0 & 0 & \cdots & -\omega_{n-1} & 0 \\
            \end{pmatrix},
            \label{eq:omegamatrix}
\eeq
and where $a$ runs over $(\sigma,s)$. The adiabatic and entropic directions form a vielbein in the field space hence $G_{IJ}=\delta_{ab}e^a_Ie^b_J$ with $\delta_{ab}$ denoting the flat metric. This yields $e^I_a=\delta_{ab}G^{IJ}e^b_J$ and $ D_\tau e_a^I={\Omega_a}^b\,e_b^I$ where we used that $D_\tau G_{IJ}=0$ for the second equality. We note that the entropic directions are defined up to a rotation. However, the peculiar choice adopted here is convenient since the couplings between the adiabatic mode and the entropic sector are all carried by the first entropic mode. 

Reprojecting the Klein-Gordon equations in this basis gives the following background equations 
\bea
	\ddot{\sigma}+3H\dot{\sigma}+V_{;\sigma}&=&0,
\eea
where we work in cosmic time, \ie $N=1$, and where we introduce $v=a^3$, $H=\dot{a}/a$ as well as $\dot{\sigma}=\pi_\sigma/a^3$ in order to match with the variables commonly-used in the second-order formalism. In the above, covariant derivatives of the potential are projected on the adiabatic/entropic basis via $V_{;a}\equiv e^I_aV_{;I}$ (note that $V_{;s\geq2}=0$). The Friedman and Raychaudhuri equations read
\bea
	H^2&=&\frac{1}{3\Mp^2}\left[\frac{\dot\sigma^2}{2}+V\right], \\
	\dot{H}&=&\frac{-1}{2\Mp^2}\dot\sigma^2,
\eea
and  the first slow-roll parameter is given by $\epsilon_1=\dot\sigma^2/(2\Mp^2H^2)$. From the definition of the turning rate, one also arrives at
\beq
	\sqrt{2\epsilon_1}\Mp \omega_1 H\delta_{s,1}+V_{;s}=0,
\eeq
where we further made use of the Friedman equation (see Ref. \cite{Kaiser_2013} and App. \ref{app:AEbasis} for details). This shows that the first turning rate  is tightly related to the first derivative of the potential. Similarly, we showed in App. \ref{app:AEbasis} that $V_{;\sigma 2}=-\omega_1\omega_2$ and that $V_{;\sigma s}=0$ for $s\geq3$, where $V_{;ab}\equiv e^{I}_ae^{J}_bV_{;IJ}$. As explained in that appendix, the turning rate $\omega_s$ is related to the $s$-th order derivative of the potential, as a result of the fact that $e^{s}_I$ is built from the $s$-derivative of the adiabatic vector. For the sake of completeness, let us introduce the second slow-roll parameter as $\epsilon_2\equiv\dd\ln\epsilon_1/\dd\ln v^{2/3}$. Using the equations of motion and working with cosmic time, one has $\epsilon_2=2\epsilon_1+2\ddot{\sigma}/(H\dot\sigma)$.

\subsubsection*{Equations of motion in the adiabatic/entropic basis}
Adiabatic and entropic perturbations are obtained by projecting on the adiabatic and entropic directions, \ie $Q^a=e^a_IQ^I$ and $P_a=e^I_aP_I$ and the inverse relation is $Q^I=e^I_aQ^a$ and $P_I=e^a_IP_a$. (Note that curvature perturbations are solely sourced by fields fluctuations along the adiabatic direction while field fluctuations in the entropic subspace are isocurvature.) It is easily checked that it corresponds to a canonical transformations, \ie $\{Q^a,P_b\}=e^a_Ie_b^J\delta^I_J=\delta^a_b$, which is further linear. Since the equations of motion in the field basis are expressed using covariant derivatives, this canonical transformation is most easily introduced directly at that level. Indeed, one has $D_\tau Q^a=N{\Omega^a}_bQ^b+e^a_ID_\tau Q^I$ and $D_\tau P_a=N{\Omega_a}^bP_b+e^I_aD_\tau P_I$. Replacing the covariant derivatives by the right-hand-side of Eq. (\ref{eq:CovDiffQP}) and further reintroducing the variables $(Q^a,P_b)$, we arrived at 
\bea
	&&\left\{\begin{array}{rcl}
		\dot{Q}^\sigma&=&\ds\frac{\pi_\sigma}{v} \delta N + N\left(\frac{1}{v}P^\sigma-\frac{\sqrt{3}}{2v^{5/3}}\pi_\sigma\delta\gamma_1-\omega_1Q_1\right), \\
		\dot{P}_\sigma&=&\ds-vV_{;\sigma}\delta N-\pi_\sigma k\delta N_1 -N\left[v\left(\frac{k^2}{v^{2/3}}\delta_{\sigma\sigma}+V_{;\sigma\sigma}\right)Q^\sigma+\frac{\sqrt{3}}{2}v^{1/3}V_{;\sigma}\delta\gamma_1\right]\\
	&&\ds-N\left(\omega_1P_1+vV_{;1\sigma}Q^1+vV_{;2\sigma}Q^{2}\right),
	 \end{array}\right. \label{eq:dynA} \\
	 &&\left\{\begin{array}{rcl}
		\dot{Q}^s&=&\ds\frac{N}{v}P^s+N\left({\Omega^s}_{s'}Q^{s'}+\delta_{1,s}\omega_1Q_\sigma\right) \\
		\dot{P}_s&=&\ds-vV_{;1}\delta_{1,s}\delta N-Nv\left(\frac{k^2}{v^{2/3}}\delta_{s,s'}+V_{;ss'}-\frac{1}{v^2}R_s{}^{\sigma\sigma}{}_{s'}\pi_\sigma\pi_\sigma\right)Q^{s'}\\
		&&\ds-N{\Omega_{s}}^{s'}P_{s'}+\delta_{1,s}N\left(\omega_1P_\sigma-vV_{;1\sigma}Q^\sigma+\frac{\sqrt{3}}{2}v^{1/3}V_{;1}\delta\gamma_1\right)-\delta_{2,s}NvV_{;2\sigma}Q^\sigma
	 \end{array}\right. \label{eq:dynE}
\eea
 where we separate the adiabatic direction and the entropic ones for clarity. Note that in the adiabatic/entropic basis, indices are raised and lowered with the flat metric $\delta_{ab}$ and there is no ambiguity in using variables with either upper or lower indices.

A few comments are in order. First, covariant time-derivative have been replaced by standard ones. Indeed, the adiabatic and entropic directions define a vielbeins hence the Christoffel coefficients vanish in that basis. Second, the curvature in the field space impact the entropic perturbations only. This comes from the algebraic properties of the Riemann tensor and it is so irrespectively of the choice of the entropic basis. Third, our specific choice of the entropic directions yields $V_{;s}=0$ for $s\geq2$ and $V_{;\sigma s}=0$ for $s\geq3$. As a result, adiabatic fluctuations directly couple to the first and the second entropic directions only, through $\omega_1$, $V_{;1\sigma}$ and $V_{;\sigma 2}$. Their couplings to entropic modes for $s\geq3$ are mediated by the second entropic directions. For the same reason, only the adiabatic direction and the first entropic ones directly couple to the isotropic gravitational perturbations. 

The case of gravitational perturbations is easily obtained by replacing $(Q^I,P_I)$ in the right-hand-side of Eqs. (\ref{eq:CovDiff1}) \&(\ref{eq:CovDiff2}) by their expressions as functions of $(Q^a,P_a)$. It boils down to
\beq
	\left\{\begin{array}{rcl}
	\dot{\delta\gamma}_1&=&\ds\frac{-2}{\sqrt{3}}v^{2/3}k\delta N_1-\frac{\sqrt{3}}{\Mp^2}v^{2/3}\theta\delta N-\frac{N}{\Mp^2}\left(2v^{1/3}\delta\pi_1+\frac{\theta}{2}\delta\gamma_1\right) \\
	\dot{\delta\pi}_1&=&\ds\frac{-v^{1/3}\theta}{2\sqrt{3}}k\delta N_1+\frac{v^{1/3}}{\sqrt{3}}\left(\frac{\pi^2_\sigma}{v^2}-V+\Mp^2\frac{k^2}{v^{2/3}}\right)\delta N \\
	&&\ds-N\left[\frac{2}{3v^{1/3}}\left(\frac{\pi^2_\sigma}{v^2}+\frac{V}{2}-\frac{\Mp^2}{4}\frac{k^2}{v^{2/3}}\right)\delta\gamma_1-\frac{\theta}{2\Mp^2}\delta\pi_1+\frac{\Mp^2\sqrt{2}}{12v}k^2\delta\gamma_2\right]  \\
	&&\ds-\frac{\sqrt{3}}{2}v^{1/3}N\left(V_{;\sigma}Q^\sigma+V_{;1}Q^1-\frac{\pi_\sigma}{v^2}P_\sigma\right),
	\end{array}\right. \label{eq:dyn1}
\eeq
and
\beq
	\left\{\begin{array}{rcl}
	\dot{\delta\gamma}_2&=&\ds-2\sqrt{\frac{2}{3}}v^{2/3}k\delta N_1+\frac{N}{\Mp^2}\left(4v^{1/3}\delta\pi_2+{\theta}\delta\gamma_2\right) \\
	\dot{\delta\pi}_2&=&\ds\sqrt{\frac{2}{3}}v^{1/3}k\delta N_1-\frac{\Mp^2}{v^{1/3}\sqrt{6}}k^2\delta N \\
	&&\ds-N\left[\frac{2}{3v^{1/3}}\left(\frac{\pi^2_\sigma}{v^2}+\frac{V}{2}-\frac{\Mp^2}{8}\frac{k^2}{v^{2/3}}\right)\delta\gamma_2+\frac{\theta}{\Mp^2}\delta\pi_2+\frac{\Mp^2\sqrt{2}}{12v}k^2\delta\gamma_1\right], 
	\end{array}\right. \label{eq:dyn2}
\eeq
where it is recovered that the isotropic fluctuations are solely coupled to the adiabatic perturbation and the first of the entropic ones. Finally, the two linear constrains are
\bea
	\mathcal{C}^{(1)}&=&vV_{;\sigma}\,Q^\sigma+vV_{;1}Q^1+\frac{\pi_\sigma}{v}P_\sigma-\frac{v^{1/3}}{\sqrt{3}}\left[\frac{\pi^2_\sigma}{v^2}-V+\Mp^2\frac{k^2}{v^{2/3}}\right]\delta\gamma_1 \label{eq:C1AE} \\
	&&+\frac{\Mp^2}{\sqrt{6}}\frac{k^2}{v^{1/3}}\delta\gamma_2-\frac{\sqrt{3}}{\Mp^2}v^{2/3}\theta\,\delta\pi_1 \nonumber \\
	\mathcal{D}^{(1)}&=&\pi_\sigma Q^\sigma+\frac{1}{\sqrt{3}}v^{1/3}\theta\left(\frac{1}{2}\delta\gamma_1-\sqrt{2}\delta\gamma_2\right)-\frac{2}{\sqrt{3}}v^{2/3}\left(\delta\pi_1+\sqrt{2}\delta\pi_2\right), \label{eq:D1AE}
\eea
where it is worth noticing that only the adiabatic mode contributes to the diffeomorphism constraint. Let us finally stress that under the flow generated by the above equations of motion, one arrives at $\dot{\mathcal{C}}^{(1)}=-N(k^2/v^{2/3})\mathcal{D}^{(1)}$ and $\dot{\mathcal{D}}^{(1)}=0$. As a result, if the constraints are initially vanishing, they remain so over evolution.

\subsection{Gauge-invariant variables}
\label{ssec:dynCPTGI}
Among the $2n+4$ degrees of freedom in the phase space, only $2n$ truly corresponds to physical degrees of freedom. This is because the theory should be independent of the choice of space-time coordinates. In the Hamiltonian language, this arbitrary choice is encoded by our  freedom to chose the perturbed lapse function and the perturbed shift vector. The four extra degrees of freedom are given by the two constraints, $Q_\mathcal{C}\equiv\mathcal{C}^{(1)}$ and $Q_\mathcal{D}\equiv\mathcal{D}^{(1)}$, and their canonically conjugate momenta. They are given by 
\bea
	P_\mathcal{C}&=&\lambda_\mathcal{C}\left[vV_{;\sigma}\,P_\sigma+vV_{;1}P_1-\frac{\pi_\sigma}{v}Q^\sigma-\frac{v^{1/3}}{\sqrt{3}}\left(\frac{\pi^2_\sigma}{v^2}-V+\Mp^2\frac{k^2}{v^{2/3}}\right)\delta\pi_1\right. \\
	&&\left.+\frac{\Mp^2}{\sqrt{6}}\frac{k^2}{v^{1/3}}\delta\pi_2+\frac{\sqrt{3}}{\Mp^2}v^{2/3}\theta\,\delta\gamma_1\right] \nonumber \\
	P_\mathcal{D}&=&\lambda_\mathcal{D}\left[\pi_\sigma P_\sigma+\frac{1}{\sqrt{3}}v^{1/3}\theta\left(\frac{1}{2}\delta\pi_1-\sqrt{2}\delta\pi_2\right)+\frac{2}{\sqrt{3}}v^{2/3}\left(\delta\gamma_1+\sqrt{2}\delta\gamma_2\right)\right],
\eea
where $\lambda_{\mathcal{C}/\mathcal{D}}$ are set to ensure that $\left\{Q_{\mathcal{C}/\mathcal{D}}, P_{\mathcal{C}/\mathcal{D}}\right\}=1$, and all other brackets vanish. The variables $Q_{\mathcal{C}/\mathcal{D}}$ do not bear any physical information since they are the constraints imposed to equal zero. The variables $P_{\mathcal{C}/\mathcal{D}}$ are the gauge degrees of freedom since they are the only ones whose values depend on the choice of $\delta N$ and $\delta N^i$, hence they should not carry any physical information either (see \cite{Artigas:2023kyo} for details). In fact, it is easy to check that $P_{\mathcal{C}}$ depends on the choice of $\delta N$ only and $P_{\mathcal{D}}$ on the choice of $\delta N_1$.

The remaining (and physical) degrees of freedom can be obtained either resorting to gauge-fixing or to gauge-invariant variables. In the latter case, a convenient and common choice is the Mukhanov-Sasaki variables, \ie
\bea
	\mathcal{Q}^\sigma=Q^\sigma+\frac{\Mp^2\pi_\sigma}{\sqrt{6}v^{5/3}\theta}\left(\sqrt{2}\delta\gamma_1-\delta\gamma_2\right),
\eea
and $\mathcal{Q}^s=Q^s$. This can be easily checked by verifying that $\{\mathcal{Q}^a,N\xi^0\mathcal{C}^{(1)}+k\xi\mathcal{D}^{(1)}\}=0$ where $(\xi^0,\xi)$ are the gauge parameters. The configuration variables of the entropic modes, $Q^s$, are known to be automatically gauge-invariant\footnote{This can be readily seen from their Hamilton equation which is free of any Lagrange multipliers; see Eq. (\ref{eq:dynE}).} and this holds true irrespectively of the choice for the entropic basis. However, the momentum for the first entropic mode, \ie $P_1$, is not gauge-invariant while the remaining entropic momenta are. This is tied to our specific choice for the entropic basis and we note that part or all entropic momenta would not be gauge-invariant assuming another choice for the entropic basis.

In the spatially-flat gauge where $\delta\gamma_1=0=\delta\gamma_2$, the gauge-invariant variables coincide with the $Q^a$'s. Hence, it is sufficient to derive autonomous equations for $Q^a$ in that gauge to obtain the equations of motion for the Mukhanov-Sasaki variables. In practice, it is done as follows. First one implement the gauge conditions in the constraints and in the equations of motion for both the scalar-field fluctuations and the gravitational fluctuations. Second, one eliminates $\delta\pi_1$ and $\delta\pi_2$ using the constraints, \ie $\mathcal{C}^{(1)}=0=\mathcal{D}^{(1)}$. Third, the gauge-conditions further impose that $\dot{\delta\gamma}_1=0=\dot{\delta\gamma}_2$ which can be used to eliminate $\delta N$ and $\delta N_1$. One is left with a closed set of $2n$ equations of first order for $(Q^a,P_a)$ which can be combined to give a closed set of $n$ equations of second order for $Q^a$. Following these steps, we arrived at 
\bea
	&\ddot{\mathcal{Q}}_\sigma+3H\dot{\mathcal{Q}}_\sigma+\left(k^2+M_{\sigma\sigma}\right)\mathcal{Q}_\sigma=2\left[\frac{\dd}{\dd t}\left(\frac{\omega_1\mathcal{Q}_1}{H}\right)-\frac{V_{;\sigma}}{\dot\sigma}\omega_1\mathcal{Q}_1\right],& \label{eq:MSad}\\
	&\ddot{\mathcal{Q}}_s+\ds\sum_{s'}\left(3H\delta_{s,s'}+2\Omega_{ss'}\right)\dot{\mathcal{Q}}_{s'}+\ds\sum_{s'}\left(k^2\delta_{s,s'}+M_{ss'}\right)\mathcal{Q}_{s'}=2\delta_{1,s}\left[\frac{\dd}{\dd t}\left(\frac{\omega_1\mathcal{Q}_1}{H}\right)-\frac{V_{;\sigma}}{\dot\sigma}\omega_1\mathcal{Q}_1\right].& \nonumber \\ \label{eq:MSent}
\eea
Note that we write the above using lower indices only and making the summations explicit. The mass matrices read 
\bea
	M_{\sigma\sigma}&=&V_{;\sigma\sigma}-\omega_1^2-\frac{1}{\Mp^2a^3}\frac{\dd}{\dd t}\left(\frac{a^3\dot\sigma}{H}\right), \\
	M_{ss'}&=&V_{;ss'}-R_{s\sigma\sigma s'}+3H\Omega_{ss'}+\dot{\Omega}_{ss'}+\ds\sum_{s''}\Omega_{ss''}\Omega_{s''s'},
\eea
which have mass-dimension two.\footnote{Note that Mukhanov-Sasaki variables can equally be defined in the field basis, \ie
\beq
	\mathcal{Q}^I = Q^I + \frac{\Mp^2 G\uTsr \pi_J}{\sqrt{6} v^{5/3} \theta}\left(\sqrt{2} \delta\gamma_1 - \delta \gamma_2\right),
\eeq
whose Poisson bracket with the linear constraints is easily shown to equal zero.  Following the same strategy than in the adiabatic/entropic basis, the equations of motion are
\beq
	D_t^2 \mathcal{Q}^I + 3HD_t  \mathcal{Q}^I + \left[\frac{k^2}{v^{1/3}}\delta^I_J +  M^I_J - \frac{1}{\MM^2v^2} D_\tau  \left(\frac{v^2}{H}G_{KL}\dot\phi^K\dot\phi^L\right)\right ]\mathcal{Q}^J = 0,
\eeq
where
\beq
	 M^I_J=G^{IK}V_{;KJ}-{R^I}_{KLJ}\dot\phi^K\dot\phi^L
\eeq	
which has mass-dimension two.} 

As compared to the single-field setup, the Lagrangian effective mass of the adiabatic mode gets corrected by its coupling with the first entropic mode through the term $\omega_1^2$. It is also worth stressing that the matrix $\Omega$ being antisymmetric, the terms proportional to $\Omega$ and $\dot{\Omega}$ in the equations of motion for the entropic modes bear the couplings between the different entropic modes, and as such they do not contribute either to the friction induced by the expansion or to the effective mass of a given entropic mode. However, the symmetric matrix $\Omega^2$ has both diagonal elements and non-diagonal ones; see Eq. (2.17) in Ref. \cite{Pinol:2020kvw}. Its diagonal elements thus contribute to the effective mass of a given entropic mode while its off-diagonal elements contribute to their couplings.

\subsection{Gauge-fixed dynamics}
Another way of studying the $2n$ physical degrees of freedom is to go through gauge fixing procedures. Gauge fixing our theory corresponds to imposing two linear combinations of our $2n+4$ initial degrees of freedom to zero and asking their equations of motions to also be vanishing.Let us start by studying the unitary gauge since the curvature perturbation is easily expressed in this case. We will also study the spatially flat gauge since it is the one used to compute scalar field correlators. Finally, since we are conducting this analysis in the context of stochastic inflation, we will study the uniform expansion gauge where the stochastic noise should be expressed.

\subsubsection*{Unitary gauge}
The unitary gauge consists in imposing $Q^\sigma=0$ and $\delta\gamma_2=0$. Hence, the adiabatic mode is given by $\delta\gamma_1$ which in this gauge is related to the curvature perturbation $\delta\gamma_1=2\sqrt{3}v^{2/3}\zeta$. As a result of such a gauge choice, the dynamics is given by an autonomous set of Hamilton equations for $(\delta\gamma_1,\delta\pi_1)$ for the adiabatic mode and $(Q^s,P_s)$ for the entropic mode, as we now show following the procedure described in Ref. \cite{Artigas:2023kyo}.

Combining the gauge conditions with the linear constraints lead to
\bea
	P_\sigma&=&\frac{v}{\pi_\sigma}\left[-vV_{;1}Q^1+\frac{v^{1/3}}{\sqrt{3}}\left(\frac{\pi_\sigma^2}{v^2}-V+\Mp^2\frac{k^2}{v^{2/3}}\right)\delta\gamma_1+\frac{\sqrt{3}}{\Mp^2}v^{2/3}\theta\delta\pi_1\right], \label{eq:Punitary} \\
	\delta\pi_2&=&\frac{1}{\sqrt{2}}\left(\frac{\theta}{4v^{1/3}}\delta\gamma_1-\delta\pi_1\right). \label{eq:piunitary}
\eea
Further imposing $\dot{Q}^\sigma=0$ and $\dot{\delta\gamma}_2=0$ allows us to express the Lagrange multipliers as functions of $(\delta\gamma_1,\delta\pi_1)$ and $(Q^s,P_s)$, \ie
\bea
	\frac{\delta N}{N}&=&\frac{1}{\sqrt{3}}\frac{v^{4/3}}{\pi^2_\sigma}\left[\left(\frac{\pi^2_\sigma}{2v^2}+V-\Mp^2\frac{k^2}{v^{2/3}}\right)\delta\gamma_1-\frac{3}{\Mp^2}v^{1/3}\theta\delta\pi_1\right], \label{eq:Nunitary} \\
	\frac{k\delta N_1}{N}&=&\frac{\sqrt{3}}{\Mp^2v^{1/3}}\left(\frac{\theta}{4v^{1/3}}\delta\gamma_1-\delta\pi_1\right), \label{eq:N1unitary}
\eea
where the expressions of $P_\sigma$ and $\delta\pi_2$ have been injected in and where we use Eq. (\ref{eq:eomV1}). Finally, the above expressions as well as the gauge conditions can be plugged in Eqs. (\ref{eq:dynE}) \& (\ref{eq:dyn1}) to obtain a set of autonomous Hamilton equations for $(\delta\gamma_1,\delta\pi_1)$ and $(Q^s,P_s)$.

\subsubsection*{Spatially-flat gauge}
The spatially-flat gauge amounts to impose that the metric does not fluctuate, hence $\delta\gamma_1=0=\delta\gamma_2$. This is sufficient to eliminate all redundant degrees of freedom. Here, the adiabatic mode is described by $(Q^\sigma,P_\sigma)$ and the entropic modes by $(Q^s,P_s)$, and for which one can derive an autonomous set of Hamilton equations. 

First, the linear constraints given in Eqs. (\ref{eq:C1AE}) \& (\ref{eq:D1AE}), are used to express the gravitational momenta since both vanish on physical solutions. Indeed, imposing $\mathcal{C}^{(1)}=0=\mathcal{D}^{(1)}$ complemented with the gauge conditions yields
\bea
	\delta\pi_1&=&\frac{\Mp^2}{\sqrt{3}v^{5/3}\theta}\pi_\sigma P_\sigma+\frac{\Mp^2v^{1/3}}{\sqrt{3}\theta}V_{;a}Q^a, \label{eq:pi1SF} \\
	\delta\pi_2&=&-\frac{\Mp^2}{\sqrt{6}v^{5/3}\theta}\pi_\sigma P_\sigma-\frac{\Mp^2v^{1/3}}{\sqrt{6}\theta}V_{;a}Q^a+\frac{1}{2v^{2/3}}\sqrt{\frac{3}{2}}\pi_\sigma Q^\sigma. \label{eq:pi2SF}
\eea
Second, the gauge conditions are further constrained to hold across evolution leading to $\dot{\delta\gamma}_1=0=\dot{\delta\gamma}_2$. These derived conditions can be imposed thanks to our freedom to choose the lapse function and the shift vector. Indeed this means that the dynamical equations for the gravitational fluctuations are now algebraic ones which fixes the values of the Lagrange multipliers for the $\delta\gamma_A$'s to be conserved quantities. The conditions $\dot{\delta\gamma}_2=0$ allows us to express $\delta N_1$ as a function of $\delta\pi_2$ and the conditions $\dot{\delta\gamma}_1=0$ to express $\delta N$ as a function of $\delta N_1$ and $\delta\pi_1$. Combining them with the expressions of the gravitational momenta as functions of the scalar-field fluctuations boils down to
\bea
	\frac{\delta N}{N}&=&\frac{-1}{v\theta}\pi_\sigma Q^\sigma, \label{eq:NSF} \\
	\frac{k\delta N_1}{N}&=&\frac{3}{2\Mp^2 v}\pi_\sigma Q^\sigma-\frac{1}{\theta}V_{;a}Q^a-\frac{1}{v^2\theta}\pi_\sigma P_\sigma. \label{eq:N1SF}
\eea
This finalizes to prove that all degrees of freedom are expressed as functions of the scalar-field fluctuations $(Q^a,P_a)$. We note that because $V_{;a}Q^a=V_{;\sigma}Q^\sigma+V_{;1}Q^1$, the redundant degrees of freedom are in fact all fixed by the adiabatic mode and the first entropic mode only, in agreement with the fact that all the entropic modes but the first one are automatically gauge-invariant. Finally, a set of closed Hamilton equations for $(Q^a,P_a)$ is derived by plugging the gauge conditions as well as the expressions of $\delta\pi_{1(2)}$, $\delta N$ and $k\delta N_1$ in Eqs (\ref{eq:dynA}) \& (\ref{eq:dynE}).

\subsubsection*{Uniform-expansion gauge}
The uniform-expansion gauge is defined as the gauge in which the fluctuations of the integrated expansion rate are imposed to be zero, \ie
\bea
	\mathcal{N}_{\mathrm{int}}=\frac{1}{3}\ds\int\nabla_\mu n^\mu n_\nu\dd x^\nu,
\eea
where $\nabla_\mu$ is the spacetime covariant derivative and $n^\mu=(-1/N,N^i/N)$ the unit vector orthogonal to the hypersurfaces. At the background level, this is no more than the number of e-folds. This gauge choice is particularly convenient in the context of the stochastic $\delta N$-formalism \cite{Fujita:2013cna,Vennin:2015hra}. There exists several ways to implement this gauge \cite{Artigas_2022,Artigas:2023kyo}. For instance, it can be shown that
\bea
	\delta\mathcal{N}_{\mathrm{int}}=\frac{1}{2\sqrt{3}v^{2/3}}\delta\gamma_1+\frac{k}{3}\ds\int\delta N_1\dd\tau,
\eea
and the uniform-expansion gauge can be realized provided $\delta\gamma_1=0$ and $\delta N_1=0$. 

In practice, one uses $\mathcal{C}^{(1)}=0=\mathcal{D}^{(1)}$ in which we plug $\delta\gamma_1=0$ in order to express $\delta\pi_1$ and $\delta\pi_2$ as functions of $(Q^a,P_a)$ and $\delta\gamma_2$, \ie
\bea
	\delta\pi_1&=&\frac{\Mp^2}{\sqrt{3}}\left(\frac{v^{1/3}}{\theta}\right)V_{;a}Q^a+\frac{\Mp^2}{\sqrt{3}}\left(\frac{\pi_\sigma}{v^{5/3}\theta}\right)P_\sigma+\frac{\Mp^4}{3\sqrt{2}}\left(\frac{k^2}{v\theta}\right)\delta\gamma_2, \label{eq:pi1UEG} \\
	\delta\pi_2&=&-\frac{\Mp^2}{\sqrt{6}}\left(\frac{v^{1/3}}{\theta}\right)V_{;a}Q^a-\frac{1}{2}\sqrt{\frac{3}{2}}\left(\frac{\pi_\sigma}{v^{2/3}}\right)Q^\sigma-\frac{\Mp^2}{\sqrt{6}}\left(\frac{\pi_\sigma}{v^{5/3}\theta}\right)P_\sigma  \nonumber \\
	&&-\left[\frac{\Mp^4}{6}\left(\frac{k^2}{v\theta}\right)+\frac{1}{2}\left(\frac{\theta}{v^{1/3}}\right)\right]\delta\gamma_2. \label{eq:pi2UEG}
\eea
This can be combined with $\dot{\delta\gamma}_1=0$ and $\delta N_1=0$ to write $\delta N$ as a function of $(Q^a,P_a)$ and $\delta\gamma_2$. It reads \beq
	\delta N=-\frac{2}{\sqrt{3}}\left(\frac{N}{v^{1/3}\theta}\right)\delta\pi_1, \label{eq:NUEG}
\eeq
which can further be expressed as a function of $(Q^a,P_a)$ and $\delta\gamma_2$ using Eq. (\ref{eq:pi1UEG}). However, one is now running out of constraints to eliminate one variable among the $2n+1$ remaining ones, and we are left with the following set of dynamical equations
\bea
	&&\left\{\begin{array}{rcl}
		\dot{Q}^\sigma&=&\ds-N\left[\frac{2\Mp^2}{3}\frac{\pi_\sigma V_{;\sigma}}{v\theta^2}Q^\sigma+\frac{1}{v}\left(\frac{2\Mp^2}{3}\frac{\pi^2_\sigma}{v^2\theta^2}-1\right)P_\sigma+\left(\omega_1+\frac{2\Mp^2}{3}\frac{\pi_\sigma V_{;1}}{v\theta^2}\right)Q^1\right] \\
		&&\ds-N\frac{\Mp^4}{3}\sqrt{\frac{2}{3}}\left(\frac{\pi_\sigma}{v^{5/3}\theta^2}\right)\left(\frac{k^2}{v^{2/3}}\right)\delta\gamma_2, \\
		\dot{P}_\sigma&=&\ds N\left[v\left(\frac{2\Mp^2}{3}\frac{V^2_{;\sigma}}{\theta^2}-\frac{k^2}{v^{2/3}}-V_{;\sigma\sigma}\right)Q^\sigma+\frac{2\Mp^2}{3}\frac{\pi_\sigma V_{;\sigma}}{v\theta^2}P_\sigma\right] \\ 
		&&\ds -N\left[v\left(V_{;1\sigma}-\frac{3\Mp^2}{2}\frac{V_{;\sigma}V_{;1}}{\theta^2}\right)Q^1+\omega_1P_1+vV_{;2\sigma}Q^2\right] \\
		&&\ds+N\frac{\Mp^4}{3}\sqrt{\frac{2}{3}}\left(\frac{v^{1/3}V_{;\sigma}}{\theta^2}\right)\left(\frac{k^2}{v^{2/3}}\right)\delta\gamma_2, \\
	 \end{array}\right. \label{eq:dynAueg} \\
	 &&\left\{\begin{array}{rcl}
		\dot{Q}^s&=&\ds\frac{N}{v}P^s+N\left({\Omega^s}_{s'}Q^{s'}+\delta_{1,s}\omega_1Q_\sigma\right), \\
		\dot{P}_s&=&\ds N\left[v\left(\frac{2\Mp^2}{3}\frac{V^2_{;1}}{\theta^2}\delta_{1,s}-\frac{k^2}{v^{2/3}}\delta_{s,s'}-V_{;ss'}+\frac{1}{v^2}R_s{}^{\sigma\sigma}{}_{s'}\pi_\sigma\pi_\sigma\right)Q^{s'}+\delta_{1,s}\frac{2\Mp^2}{3}\frac{\pi_\sigma V_{;1}}{v\theta^2}P_\sigma\right] \\ 
		&&\ds -N{\Omega_{s}}^{s'}P_{s'}-\delta_{1,s}N\left[v\left(V_{;1\sigma}-\frac{3\Mp^2}{2}\frac{V_{;\sigma}V_{;1}}{\theta^2}\right)Q^\sigma-\omega_1P_\sigma\right]-\delta_{2,s}NvV_{;2\sigma}Q^2 \\
		&&\ds+N\frac{\Mp^4}{3}\sqrt{\frac{2}{3}}\left(\frac{v^{1/3}V_{;1}}{\theta^2}\right)\left(\frac{k^2}{v^{2/3}}\right)\delta\gamma_2, \\
	 \end{array}\right. \label{eq:dynEueg}
\eea
and
\bea
	\dot{\delta\gamma}_2&=&-\frac{2\Mp^2}{3}\left(\frac{N}{\theta}\right)\left(\frac{3\theta^2}{2\Mp^4}+\frac{k^2}{v^{2/3}}\right)\delta\gamma_2 \nonumber \\
	&&-\frac{4N}{\Mp^2}v^{1/3}\left[\frac{\Mp^2}{\sqrt{6}}\left(\frac{v^{1/3}}{\theta}\right)V_{;a}Q^a+\frac{1}{2}\sqrt{\frac{3}{2}}\left(\frac{\pi_\sigma}{v^{2/3}}\right)Q^\sigma+\frac{\Mp^2}{\sqrt{6}}\left(\frac{\pi_\sigma}{v^{5/3}\theta}\right)P_\sigma\right]. \label{eq:dyn2ueg}
\eea
It clearly shows that the dynamics of $(Q^a,P_a)$ cannot be decoupled from the one of $\delta\gamma_2$ and that their couplings in the dynamical equations of $\delta\gamma_2$ are not $k$-suppressed; see the second line of Eq. (\ref{eq:dyn2ueg}). Because of that, not all the gauge degrees are fixed, extending this known result \cite{Artigas_2022,Artigas:2023kyo} to the case of multifield models.

\section{Separate universe} \label{sec:separate}
\subsection{The separate-universe expansion}
\subsubsection*{Degrees of freedom}
When cosmological perturbation theory is developed around a homogeneous and isotropic background, one can introduce an important simplification at large scales, \textit{i.e.} scales larger than the length scale associated with the expanding \textemdash or contracting universe. We can describe the universe at these scales as an ensemble of homogeneous and isotropic patches that evolve independently from one another as a first approximation. This is called the separate universe approach, or quasi-isotropic picture. With this simplification we thus only need to solve the homogeneous and isotropic problem with different initial conditions depending on the patch we are considering. At the level of the perturbations themselves, the separate universe picture consists in introducing small deviations assumed to be homogeneous and isotropic as a proxy to describe cosmological perturbations at large scales. Hence, the gravitational degrees of freedom read 
\bea
	\gamma_{ij}&=&\left(v+\delta v\right)\widetilde{\gamma}_{ij}, \label{eq:deltav}\\
	\pi^{ij}&=&\frac{1}{2}\left[v^{1/3}\theta+\left(v^{1/3}\delta\theta+\frac{1}{3v^{2/3}}\theta\delta v\right)\right]\widetilde{\gamma}^{ij} \label{eq:deltatheta}
\eea
 where $\delta v$ and $\delta\theta$ solely depend on time.  Similarly, the field variables are $\phi^I+\delta\phi^I$ and $\pi_I+\delta\pi_I$ where $(\delta\phi^I,\delta\pi_I)$ are now independent of space. Covariant variables describing homogeneous and isotropic deviations are defined in the very same way as in full perturbation theory. It gives at linear order $Q^I=\delta\phi^I$ and $P_I=\delta\pi_I-\Gamma^K_{IJ}\pi_K\delta\phi^J$. Finally, the Lagrange multiplier are $N+\delta N$ and $\delta N_1=0$ which owes to the fact that isotropy is now imposed at the perturbative level too. 

A priori, the perturbations introduced above have no reason to match the ones introduced in the full perturbative problem and the aim is precisely to identify the conditions needed for matching the two. Hence, one should first establish a dictionary between the two sets of perturbative variables and Lagrange multipliers \cite{Artigas_2022}. Because isotropy is imposed in the separate-universe picture, $\delta\gamma_{ij}$ and $\delta\pi^{ij}$ are pure traces and the gravitational perturbations are only composed of isotropic degrees of freedom. This leads to the dictionary given in Tab. \ref{tab:dict} which straightforwardly extended the one for single-field inflation to the multifield context (note that the gravitational degrees of freedom have been expanded in $\delta v$ and $\delta\theta$). From now on, perturbations in the separate-universe approach will be denoted with a bar over them in order to avoid confusion with the inhomogeneous and (possibly) anisotropic deviations introduced in Sec. \ref{sec:cpt}. 

\begin{table}
\begin{center}
\begin{tabular}{l | l}
Perturbation theory & Separate universe \\ \hline\hline
 & \\
$\delta N(\tau,\vec{x})$ & $\overline{\delta N}(\tau)$ \\
$\delta N_1(\tau,\vec{x})$ & $\overline{\delta N}_1(\tau)=0$ \\ \hline \\
$\delta\gamma_1(\tau,\vec{x})$ & $\overline{\delta\gamma}_1(\tau)=\sqrt{3}\left[\left(v+\delta v\right)^{2/3}-v^{2/3}\right]\simeq\frac{2}{\sqrt{3}}\frac{\delta v}{v^{1/3}}$ \\
$\delta\pi_1(\tau,\vec{x})$ & $\overline{\delta\pi}_1(\tau)=\frac{\sqrt{3}}{2}\left[\left(v+\delta v\right)^{1/3}\left(\theta+\delta\theta\right)-v^{1/3}\theta\right]\simeq\frac{\sqrt{3}}{2}v^{1/3}\theta\left(\frac{\delta\theta}{\theta}+\frac{\delta v}{3v}\right)$ \\
$\delta\gamma_2(\tau,\vec{x})$ & $\overline{\delta\gamma}_2(\tau)=0$ \\
$\delta\pi_2(\tau,\vec{x})$ & $\overline{\delta\pi}_2(\tau)=0$ \\ \hline
 & \\
$\delta\phi^I(\tau,\vec{x})$ & $\overline{\delta\phi}^I(\tau)$ \\
$\delta\pi_I(\tau,\vec{x})$ & $\overline{\delta\pi}_I(\tau)$  \\
$Q^I(\tau,\vec{x})=\delta\phi^I$ & $\overline{Q}^I(\tau)=\overline{\delta\phi}^I$ \\
$P^I(\tau,\vec{x})=\delta\pi_I-\Gamma^K_{IJ}\delta\phi^J$ & $\overline{P}_I(\tau)=\overline{\delta\pi}_I-\Gamma^K_{IJ}\overline{\delta\phi}^J$
\end{tabular}
\caption{Dictionary between cosmological perturbations and homogeneous and isotropic deviations as described by the separate-universe picture. For the gravitational degrees of freedom, we provide their non-linear definition and their first-order expansion}
\label{tab:dict}
\end{center}
\end{table}

\subsubsection*{Dynamical equations}
The equations of motion in the separate-universe picture can be obtained following the very same strategy as the one employed for the full perturbative problem, with the huge simplification of having no gradient terms nor anisotropic degrees of freedom. Following this path, the dynamical equations for $(\overline{\delta\gamma}_1,\overline{\delta\pi}_1)$ and $(\overline{Q}^I,\overline{P}_I)$ are given by Eqs. (\ref{eq:CovDiff1}) \& (\ref{eq:CovDiffQP}) where the dynamical variables and the Lagrange multipliers are replaced by their analog in the separate universe using the dictionary in Tab. \ref{tab:dict}, and where all gradient terms are set to zero. This applies similarly to the constraints, \ie Eq. (\ref{eq:constcov}). Note however that in the separate-universe picture, there is no dynamical equations for anisotropic degrees of freedom nor any diffeomorphism constraint. 

Alternatively, these equations can be directly obtained from the background equations by replacing the background variables by their expression including isotropic and homogeneous deviations, and subsequently expand at first order in these small deviations. This is possible since the small deviations in the separate-universe picture enjoy the same symmetries as the background. Explicitly, we perform the following replacements in Eqs. (\ref{eq:dotphi}), (\ref{eq:dotpi}), (\ref{eq:dotgamma}), (\ref{eq:dotpig}), and (\ref{eq:C0}):
\bea
	v\to v+\frac{\sqrt{3}}{2}v^{1/3}\overline{\delta\gamma}_1 & \quad\mathrm{and}\quad& \theta\to\theta-\frac{\sqrt{3}}{2}v^{2/3}\theta\overline{\delta\gamma}_1+\frac{2}{\sqrt{3}v^{1/3}}\overline{\delta\pi}_1, \\
	\phi^I\to\phi^I+\overline{Q}^I & \quad\mathrm{and}\quad& \pi_I\to\pi_I+\overline{P}_I+\Gamma^K_{IJ}\pi_K\overline{Q}^J
\eea
as well as 
\beq
	N\to N+\overline{\delta N}.
\eeq
In the above, $\overline{\delta v}$ and $\overline{\delta\theta}$ are expressed using $\overline{\delta\gamma}_1$ and $\overline{\delta\pi}_1$ in order to formally match with the variables used in cosmological perturbation theory. This simply amounts to perform a canonical transformation \cite{Artigas_2022}. Linearizing the homogeneous and isotropic equations of motion for the gravitational and scalar fluctuations, \ie Eqs. (\ref{eq:dotphi}), (\ref{eq:dotpi}), (\ref{eq:dotgamma}) \& (\ref{eq:dotpig}) is done in App. \ref{app:separate universe}. This leads to the following set of equations in the adiabatic/entropic basis (note that switching from one basis to another employs the same strategy as in the full perturbative problem), \ie 
\bea
	&&\left\{\begin{array}{rcl}
		\dot{\overline{Q}}^\sigma&=&\ds\frac{\pi_\sigma}{v} \overline{\delta N} + N\left(\frac{1}{v}\overline{P}^\sigma-\frac{\sqrt{3}}{2v^{5/3}}\pi_\sigma\overline{\delta\gamma}_1+\omega_1\overline{Q}_1\right), \\
		\dot{\overline{P}}_\sigma&=&\ds-vV_{;\sigma}\overline{\delta N} -N\left(vV_{;\sigma\sigma}\overline{Q}^\sigma+\frac{\sqrt{3}}{2}v^{1/3}V_{;\sigma}\overline{\delta\gamma}_1\right) \\
		&&\ds+N\left(\omega_1\overline{P}_1-vV_{;1\sigma}\overline{Q}^1-vV_{;2\sigma}\overline{Q}^2\right),
	 \end{array}\right. \label{eq:dynAsu} \\
	 &&\left\{\begin{array}{rcl}
		\dot{\overline{Q}}^s&=&\ds\frac{N}{v}\overline{P}^s+{\Omega^s}_{s'}\overline{Q}^{s'}-N\omega_1\delta_{1,s}\overline{Q}_\sigma, \\
		\dot{\overline{P}}_s&=&\ds-vV_{;1}\delta_{1,s}\overline{\delta N}-Nv\left(V_{;ss'}-\frac{1}{v^2}R_s{}^{\sigma\sigma}{}_{s'}\pi_\sigma\pi_\sigma\right)\overline{Q}^{s'}\\
		&&\ds+{\Omega_{s}}^{s'}\overline{P}_{s'}-\delta_{1,s}N\left(\omega_1\overline{P}_\sigma+vV_{;1\sigma}\overline{Q}^\sigma-\frac{\sqrt{3}}{2}v^{1/3}V_{;1}\overline{\delta\gamma}_1\right)-\delta_{s,2}NvV_{;2\sigma}\overline{Q}^\sigma,
	 \end{array}\right. \label{eq:dynEsu} \\
	 &&\left\{\begin{array}{rcl}
	\dot{\overline{\delta\gamma}}_1&=&\ds-\frac{\sqrt{3}}{\Mp^2}v^{2/3}\theta\overline{\delta N}-\frac{N}{\Mp^2}\left(2v^{1/3}\overline{\delta\pi}_1+\frac{\theta}{2}\overline{\delta\gamma}_1\right) \\
	\dot{\overline{\delta\pi}}_1&=&\ds\frac{v^{1/3}}{\sqrt{3}}\left(\frac{\pi^2_\sigma}{v^2}-V\right)\overline{\delta N} -N\left[\frac{2}{3v^{1/3}}\left(\frac{\pi^2_\sigma}{v^2}+\frac{V}{2}\right)\overline{\delta\gamma}_1-\frac{\theta}{2\Mp^2}\overline{\delta\pi}_1\right]  \\
	&&\ds-\frac{\sqrt{3}}{2}v^{1/3}N\left(V_{;\sigma}\overline{Q}^\sigma+V_{;1}\overline{Q}^1-\frac{\pi_\sigma}{v^2}\overline{P}_\sigma\right),
	\end{array}\right. \label{eq:dyn1su}
\eea
where we separate the adiabatic mode and the entropic modes for clarity. This is complemented with the scalar constraint reading
\beq
	\overline{\mathcal{C}}^{(1)}=vV_{;\sigma}\,\overline{Q}^\sigma+vV_{;1}\overline{Q}^1+\frac{\pi_\sigma}{v}\overline{P}_\sigma-\frac{v^{1/3}}{\sqrt{3}}\left(\frac{\pi^2_\sigma}{v^2}-V\right)\overline{\delta\gamma}_1-\frac{\sqrt{3}}{\Mp^2}v^{2/3}\theta\,\overline{\delta\pi}_1,  \label{eq:C1AEsu} 
\eeq
where we express $\rho+3p$ as functions of the kinetic energy and the potential energy of the scalar fields.
 
Finally, let us mention that the above dynamical equations result from the following second-order action
\bea
	\overline{S}^{(2)}=\ds\int\dd\tau\left[\overline{P}_a\dot{\overline{Q}}^a+\overline{\delta\pi}_1\dot{\overline{\delta\gamma}}_1-\overline{\delta N}\,\overline{\mathcal{C}}^{(1)}-2N\overline{\mathcal{K}}^{(2)}_\mathrm{cov}\right],
\eea
where the linear constraint is given in Eq. (\ref{eq:C1AEsu}) and where
\bea
\overline{\mathcal{K}}^{(2)}_\mathrm{cov}&=&-\frac{v^{1/3}}{\Mp^2}\overline{\delta\pi_1}^{\,2}+\frac{1}{3v^{1/3}}\left(\frac{\pi^2_\sigma}{v^2}+\frac{V}{2}\right)\overline{\delta\gamma_1}^{\,2}-\frac{\theta}{2\Mp^2}\overline{\delta\gamma_1}\,\overline{\delta\pi_1}+\frac{1}{2v}\delta^{ab}\overline{P}_a\,\overline{P}_b \label{eq:Kcovsu}\\
	&& +\frac{v}{2}\left(V_{;ab}-\frac{1}{v^2}R_a{}^{\sigma\sigma}{}_{b}\pi_\sigma\pi_\sigma\right)\overline{Q}^a\,\overline{Q}^b+{\Omega_a}^b\,\overline{Q}^a\,\overline{P}_b -\frac{\sqrt{3}v^{1/3}}{2}\left(V_{;a}\overline{Q}^a-\frac{\pi_\sigma}{v^2}\overline{P}_\sigma\right)\overline{\delta\gamma}_1. \nonumber
\eea
Just as cosmological perturbation theory is, the separate-universe picture is a constrained theory with one constraint, $\overline{\mathcal{C}}^{(1)}$, which is first class. Indeed, it is easily checked that $\dot{\overline{\mathcal{C}}}^{(1)}=0$ under the flow generated by the separate-universe dynamics, ensuring the solutions to lie on the surface of constraints (which is 1-dimensional here) defined as $\mathcal{C}^{(1)}=0$.

\subsection{Matching to cosmological perturbations}
\label{ssec:SUvsCPT}

\subsubsection{Arbitrary gauge}
\label{sssec:arbgauge}
Having established the dynamics of cosmological fluctuations both at the level of the full perturbative problem and in the separate-universe approximation, one can identify the conditions needed to match the two approaches. To this end, we implement the requirements that gradients should be neglected in full perturbation theory.

Let us start with the case of the diffeomorphism constraints. This is the more subtle one since this constraint is totally lost in the separate-universe approximation.  On the one hand, we note that the diffeomorphism constraint is a pure gradient since $\delta N^i\mathcal{D}_i=k\delta N_1\mathcal{D}$ and one could be tempted to argue that it is necessary $k$-suppressed. However this statement is gauge-dependent and depending on such a choice, $k\delta N_1$ might not be $k$-suppressed (two concrete examples will be given hereafter). On the other hand, the dictionary established in Tab. \ref{tab:dict} imposes to set the fluctuations of the shift vector to zero in the separate-universe picture which is possible in the full perturbative problem providing that one selects an appropriate gauge. Moreover, a fictitious diffeomorphism constraint can be defined in the separate-universe picture using Tab. \ref{tab:dict}. It reads
\beq
		\overline{\mathcal{D}}^{(1)}=\pi_\sigma \overline{Q}^\sigma+\frac{1}{2\sqrt{3}}v^{1/3}\theta\,\overline{\delta\gamma}_1-\frac{2}{\sqrt{3}}v^{2/3}\,\overline{\delta\pi}_1. \label{eq:D1AEsu}
\eeq
Interestingly, $\dot{\overline{\mathcal{D}}}^{(1)}=0$ providing that $\overline{\mathcal{C}}^{(1)}$ is vanishing. As a result, if there is a mismatch between $\mathcal{D}^{(1)}$ and $\overline{\mathcal{D}}^{(1)}$ which is small and of order $(k^2/v^{2/3})$, it should remain small once cosmological perturbations are evolved through the separate-universe dynamics. Finally, it is also possible to work with variables which are diffeomorphism-invariant, hence insensitive to $k\delta N_1\mathcal{D}^{(1)}$ (a strategy that will be employed later on). As a result, the contribution of the diffeomorphism constraint can be dropped out at large scales being either gradient-suppressed, or absent but one should bear in mind that this is possible providing a suitable choice of gauge and/or of phase-space variables.

Inspecting the scalar constraints $\mathcal{C}^{(1)}$ given in Eq. (\ref{eq:C1AE}), a first condition is that
\beq
	\frac{k^2}{v^{2/3}}\ll\frac{1}{\Mp^2}\left|\frac{\pi_\sigma}{v^2}-V\right|. \label{eq:gradC1}
\eeq
In this case indeed, the scalar constraints reads as $\mathcal{C}^{(1)}=\mathcal{C}^{(\mathrm{iso})}+\mathcal{C}^{(\mathrm{aniso})}$ where
\bea
	\mathcal{C}^{(\mathrm{iso})}&=& vV_{;\sigma}\,Q^\sigma+vV_{;1}Q^1+\frac{\pi_\sigma}{v}P_\sigma-\frac{v^{1/3}}{\sqrt{3}}\left(\frac{\pi^2_\sigma}{v^2}-V\right)\delta\gamma_1-\frac{\sqrt{3}}{\Mp^2}v^{2/3}\theta\,\delta\pi_1\nonumber \\
    &&+\mathcal{O}\left(\frac{k^2}{v^{1/3}}\delta\gamma_1\right), \\
	\mathcal{C}^{(\mathrm{aniso})}&=&\frac{\Mp^2}{\sqrt{6}}\frac{k^2}{v^{1/3}}\delta\gamma_2.
\eea
This is easily matched to the scalar constraint in the separate-universe picture, Eq. (\ref{eq:C1AEsu}), using the dictionary established in Tab. \ref{tab:dict} (note that it is important to set $\delta\gamma_2=0$ in the above for performing the matching, as established by our dictionary). As is the case of the diffeomorphism constraint, the matching here is gauge-dependent since the fluctuations of the lapse function needs to match at large scales. This first condition is identical to the one derived in the case of single-field inflation where $\pi_\phi$ is now replaced by $\pi_\sigma$. Working in cosmic time, the above easily translates into a condition on $(k/aH)$, \ie
\bea
	\left(\frac{k}{aH}\right)^2\ll\left|1-\epsilon_1\right|,
\eea
where we remind that $\epsilon_1=-\dot{H}/H^2$ is the first slow-roll parameters and where we omit numerical factors of order one.

Neglecting gradients in $\mathcal{K}^{(2)}_\mathrm{cov}$ yields two sets of additional requirements. Indeed, one notes that gradients arise through the quadratic term $(Q^a,\delta\gamma_1,\delta\gamma_2) \mathcal{M} (Q^a,\delta\gamma_1,\delta\gamma_2)^\dag$ with
\beq
	\mathcal{M}=\left(\begin{array}{ccc}
		\frac{v}{4}\left(\frac{k^2}{v^{2/3}}\delta_{ab} + V;_{ab} -\mathcal{R}_{ab}\right) & \frac{\sqrt{3}v^{1/3}}{8}V_{;a} & 0 \\
		 \frac{\sqrt{3}v^{1/3}}{8}V^\mathrm{T}_{;a} & \frac{1}{v^{1/3}}\left(\frac{\pi^2_\sigma}{v^2}+\frac{V}{2}-\frac{\Mp^2k^2}{4v^{2/3}}\right) & \frac{\sqrt{2}\Mp^2k^2}{24v} \\
		 0 & \frac{\sqrt{2}\Mp^2k^2}{24v} & \frac{1}{v^{1/3}}\left(\frac{\pi^2_\sigma}{v^2}+\frac{V}{2}-\frac{\Mp^2k^2}{8v^{2/3}}\right)
	\end{array}\right),
\eeq
where $V_{;a}$ should be understood as an $n$-dimensional column vector and where we introduce $\mathcal{R}_{ab}\equiv R_a{}^{\sigma\sigma}{}_b\pi_\sigma\pi_\sigma/v^2$ to lighten the expressions. In principle, one should diagonalize the above matrix and highlight the set of conditions needed to neglect $(k^2/v^{2/3})$ in its eigenvalues. This task depends on the details of the model one is considering and in which the number of fields, their potential and their coupling metric are specified. Here, we rather follow a simpler procedure which however lead to a sufficient set of conditions to neglect gradients (see \cite{Artigas_2022}). 

Indeed, discarding couplings between fluctuations of the gravitational fields and fluctuations of the scalar fields, one first needs to diagonalize the matrix 
\beq
	\mathcal{M}_\mathrm{G}=\left(\begin{array}{cc}
		\frac{\pi^2_\sigma}{v^2}+\frac{V}{2}-\frac{\Mp^2k^2}{4v^{2/3}} & \frac{\sqrt{2}\Mp^2k^2}{24v^{2/3}} \\
		 \frac{\sqrt{2}\Mp^2k^2}{24v^{2/3}} & \frac{\pi^2_\sigma}{v^2}+\frac{V}{2}-\frac{\Mp^2k^2}{8v^{2/3}}
	\end{array}\right),
\eeq
whose eigenvalues are $\lambda_1=2(\pi_\sigma^2/v^2+V/2)/3$ and $\lambda_2=\lambda_1-[\Mp k/(2v^{1/3})]^2$. Hence, a second condition emerges for gradients to play a negligible role, \ie
\beq
	\frac{k^2}{v^{2/3}}\ll\frac{1}{\Mp^2}\left|\frac{\pi^2_\sigma}{v^2}+\frac{V}{2}\right|. \label{eq:gradC2G}
\eeq
This is similar to the single-field situation in which $\pi_\sigma$ plays the role of $\pi_\phi$. In cosmic time, this condition yields an additional constraint on $(k/aH)$ which is
\bea
	\left(\frac{k}{aH}\right)^2\ll\left|1+\epsilon_1\right|.
\eea
In the case of slow-roll inflation where $\epsilon_1\ll1$, this constraint has the same order of magnitude than the one derived by requiring to match $\mathcal{C}^{(1)}$ with $\overline{\mathcal{C}}^{(1)}$. This condition can be interpreted as the fact that $k/a$ must be much smaller than the effective mass of the gravitational degrees of freedom, $\delta\gamma_1$ and $\delta\gamma_2$, and which is given by $H$.

Second, one needs to diagonalize the matrix $\mathcal{M}_\phi=({k^2}/{v^{2/3}})\delta_{ab} + V_{;ab}- \mathcal{R}_{ab}$. The mass matrix $M^{(\phi)}_{ab}\equiv V_{;ab}-\mathcal{R}_{ab}$ is symmetric, hence diagonalizable by an orthogonal matrix that we note $\mathcal{O}^{(\phi)}$. The eigenvalues of $M^{(\phi)}$ are interpreted as the square of the effective masses of the adiabatic and entropic modes that we name $m_a$.\footnote{Note that effective masses here are obtained in the Hamiltonian framework and as such they can differ from the effective masses in the Lagrangian. This is because of the couplings between configuration variables and momentum variables which makes the mass matrix to be different in the Lagrangian language and in the Hamiltonian language. A key difference here is that the matrix $\Omega$ does not contribute to the mass matrix in the Hamiltonian framework but does in the Lagrangian formalism.} Since $M^{(\phi)}$  is added to a matrix which is proportional to the identity, the matrices $\mathcal{O}^{(\phi)}$ also diagonalizes $\frac{k^2}{v^{2/3}}\delta_{ab}+M^{(\phi)}_{ab}$. As a result, one arrives at
\beq
	\mathcal{M}_{\phi}=\mathcal{O}^{(\phi)}\left(\frac{k^2}{v^{2/3}}\mathcal{I}_n+\mathrm{diag}\left[m^2_\sigma,m^2_1,\cdots,m^2_{n-1}\right]\right)\mathcal{O}^{(\phi)\,\mathrm{T}},
\eeq
where $\mathcal{I}_n$ is the identity matrix with dimension $n$. Finally, this leads to a set of $n$ conditions for gradients to be neglected, \ie
\beq
	\frac{k^2}{v^{2/3}}\ll m^2_a, \label{eq:gradC2phi}
\eeq
where $a$ runs over $\{\sigma,1,\cdots,n-1\}$ and where the effective masses are to be interpreted in the sense of diagonalizing the matrices $M^{(\phi)}$. This straightforwardly generalizes the case of single-field inflation, in which situation the above set reduces to one condition $(k^2/v^{2/3})\ll V_{,\phi\phi}$ \cite{Artigas_2022}.

Implementing the conditions Eqs. (\ref{eq:gradC2G}) \& (\ref{eq:gradC2phi}), the quadratic constraint boils down to $\mathcal{K}^{(2)}_\mathrm{cov}= \mathcal{K}^{(\mathrm{iso})}+\mathcal{K}^{(\mathrm{aniso})}+\mathcal{O}[(k^2/v)\delta\gamma_1\delta\gamma_2]$ where
\bea
	\mathcal{K}^{(\mathrm{iso})}&=&-\frac{v^{1/3}}{\Mp^2}\left|\delta\pi_1\right|^2+\frac{1}{3v^{1/3}}\left(\frac{\pi^2_\sigma}{v^2}+\frac{V}{2}\right)\left|\delta\gamma_1\right|^2-\frac{\theta}{4\Mp^2}\left(\delta\gamma_1\delta\pi_1^\star+\mathrm{c.c.}\right)\\
	&&+\frac{1}{4v}\delta^{ab}\left(P_aP^\star_b+\mathrm{c.c.}\right) +\frac{v}{4}\left(V_{;ab}-\mathcal{R}_{ab}\right)\left(Q^aQ^{b\star}+\mathrm{c.c.}\right)+\frac{1}{2}{\Omega_a}^b\left(Q^aP^\star_b+\mathrm{c.c.}\right) \nonumber \\
	&&-\frac{\sqrt{3}v^{1/3}}{4}\left[\left(V_{;a}Q^a-\frac{\pi_\sigma}{v^2}_\sigma\right)\delta\gamma_1^\star+\mathrm{c.c.}\right]+\mathcal{O}\left[\frac{k^2}{v^{2/3}}\left(v\delta_{ab}Q^aQ^b+v^{-1/3}\delta\gamma_1^2\right)\right], \nonumber \\
	\mathcal{K}^{(\mathrm{aniso})}&=& \frac{2v^{1/3}}{\Mp^2}\left|\delta\pi_2\right|^2+\frac{1}{3v^{1/3}}\left(\frac{\pi^2_\sigma}{v^2}+\frac{V}{2}\right)\left|\delta\gamma_2\right|^2+\frac{\theta}{2\Mp^2}\left(\delta\gamma_2\delta\pi_2^\star+\mathrm{c.c.}\right)  \\
    &&+\mathcal{O}\left(\frac{k^2}{v}\delta\gamma_2^{\,2}\right). \nonumber
\eea
The above is easily matched to the separate-universe picture, Eq. (\ref{eq:Kcovsu}), using the dictionary of Tab. \ref{tab:dict}. It is worth stressing that the anisotropic degrees of freedom decouples from isotropic ones by neglecting gradient terms and it is not harmful to the separate-universe approximation to totally lose tracks of the anisotropic degrees of freedom. Let us finally note that the matching here is independent of the gauge choice in the sense that the Lagrange multipliers not need to be matched.

These considerations show that the separate-universe picture (in which case cosmological perturbations are further assumed to be homogeneous and isotropic) is equivalent to cosmological perturbation theory providing that (i) the conditions Eqs. (\ref{eq:gradC1}), (\ref{eq:gradC2G}) \&  (\ref{eq:gradC2phi}), (ii) anisotropic degrees of freedom are set to zero, and (iii) the Lagrange multipliers are appropriately matched. The first requirement is a direct consequence of expanding the action in small deviations (which are either inhomogeneous and anisotropic, or homogeneous and isotropic). However, the other two are gauge-dependent and not fixed by the perturbative expansion. This is why we now consider specific implementations of gauge-invariance and gauge fixing.

\subsubsection{Gauge-invariant variables}
\label{sssec:gi}
Because the separate-universe picture is a constrained theory, its $(2n+2)$-dimensional phase space contains only two physical degrees of freedom. One of them is algebraically fixed by imposing $\mathcal{C}^{(1)}=0$ and a second does not bear any physical information since its value is arbitrarily fixed by the choice of the lapse function $\overline{\delta N}$. The separate-universe is not a gauge theory with respect to spacetime diffeomorphism which is unlike full cosmological perturbation theory. Indeed spatial diffeormorphism are not permitted here. However, it remains invariant under local time reparametrization and in that respect, the physical degrees of freedom will be qualified as gauge-invariant variables. Since there are two of them in the phase space of the separate-universe, one could expect these to be matched to the gauge-invariant variables in full cosmological perturbation theory. 

To this end, let us introduce the Mukhanov-Sasaki-like variable
\beq
	\overline{\mathcal{Q}}^\sigma=\overline{Q}^\sigma+\frac{\Mp^2\pi_\sigma}{\sqrt{3}v^{5/3}\theta}\,\overline{\delta\gamma}_1,
\eeq
and $\overline{\mathcal{Q}}^s=\overline{Q}^s$. It is easily checked that $\left\{\overline{\mathcal{Q}}^a,N\xi^0\overline{\mathcal{C}}^{(1)}\right\}=0$ where $\xi^0$ parametrizes the local time change (note that by homogeneity, $\xi^0$ is here a function of $\tau$ only). As a result of the separate-universe equations of motion, their dynamics is governed by the following second-order differential equations
\bea
	\ddot{\overline{\mathcal{Q}}}_\sigma+3H\dot{\overline{\mathcal{Q}}}_\sigma+M_{\sigma\sigma}\overline{\mathcal{Q}}_\sigma&=&2\left[\frac{\dd}{\dd t}\left(\frac{\omega_1\mathcal{Q}_1}{H}\right)-\frac{V_{;\sigma}}{\dot\sigma}\omega_1\mathcal{Q}_1\right] \label{eq:dynMSSUad} \\
	&&+\sqrt{\frac{\epsilon_1}{2}}\epsilon_2\Mp H^2\overline{\mathcal{D}}^{(1)}, \nonumber\\
	\ddot{\mathcal{Q}}_s+\ds\sum_{s'}\left(3H\delta_{s,s'}+2\Omega_{ss'}\right)\dot{\mathcal{Q}}_{s'}+\ds\sum_{s'}M_{ss'}\mathcal{Q}_{s'}&=&2\delta_{1,s}\left[\frac{\dd}{\dd t}\left(\frac{\omega_1\mathcal{Q}_1}{H}\right)-\frac{V_{;\sigma}}{\dot\sigma}\omega_1\mathcal{Q}_1\right] \label{eq:dynMSSUen} \\
	&&+2\sqrt{2\epsilon_1}\Mp\omega_1 H\overline{\mathcal{D}}^{(1)}\delta_{1,s}. \nonumber
\eea
where $\epsilon_2\equiv\dd\ln\epsilon_1/\dd\ln v^{1/3}$ is the second slow-roll parameter and where the mass-matrices $M_{\sigma\sigma}$ and $M_{ss'}$ have been given in Sec. \ref{ssec:dynCPTGI}.  These have to be compared with the large-scale limit $(k\to0)$ of the equations of motion of the Mukhanov-Sasaki variables in perturbation theory, Eqs. (\ref{eq:MSad}) \& (\ref{eq:MSent}). Here, one easily notices the extra quasi-diffeomorphism constraint appearing in the equation of motion of the adiabatic direction, but also for the first entropic direction. This separate-universe diffeomorphism constraint can be dealt with in several ways. We can impose a specific constraint on the anisotropic sector by fixing the gauge to make it vanish, or we can see that it is either constant or decaying in slow-roll and ultra-slow-roll inflation \cite{caravano2024ultraslowrollinflationlatticebackreaction}. In any case, this term is expected to be much smaller than the left-hand-side which grows as $\sim v^{1/3}$. These arguments are easy to use on the adiabatic Mukhanov-Sasaki variable, notably because the prefactor is suppressed by both Hubble flow parameters. However let us take a closer look at the first entropic direction. The prefactor depends on the rate of turn $\omega_1$, which we cannot guaranty to be small, in particular in cases of shape turns \cite{Aragam_2021}. In this case, the diffeomorphism constraint will not only affect the sub dominant modes on the super-Hubble scales (which are in any case discarded by the gradient expansion), and the separate-universe approach could be jeopardised during this turn.

\subsubsection{Gauge-fixed dynamics}
\label{sssec:gf}
In the context of constrained Hamiltonian dynamics, the linear constraint $\overline{\mathcal{C}}^{(1)}$ can be considered as one of the configuration variables. As a consequence, its canonically conjugate momentum depends on the choice of the Lagrange mutliplier $\overline{\delta N}$. As such, it can be considered as the gauge degree of freedom and this ambiguity can be fixed by resorting to a gauge choice.  It remains to establish the map from the gauge fixed in the separate-universe picture to the gauge fixed in the full perturbative problem. (Note that such a mapping cannot be bijective because one has two constraints and two gauge degrees of freedom in the full perturbative problem and only one constraint and one gauge degree of freedom in the separate-universe picture.) Naively, one could expect that gauge choices at the level of perturbation theory in which one of the anisotropic degrees of freedom is set to zero to be particularly suited. However, the examples we consider hereafter show that it is a more subtle issue. 

\subsubsection*{Unitary gauge}
The most natural implementation of the unitary gauge in the separate-universe picture consists in imposing $\overline{Q}^\sigma=0$. Plugging this in $\overline{\mathcal{C}}^{(1)}=0$ leads to
\bea
	\overline{P}_\sigma&=&\frac{v}{\pi_\sigma}\left[-vV_{;1}\overline{Q}^1+\frac{v^{1/3}}{\sqrt{3}}\left(\frac{\pi_\sigma^2}{v^2}-V\right)\overline{\delta\gamma}_1+\frac{\sqrt{3}}{\Mp^2}v^{2/3}\theta\overline{\delta\pi}_1\right]. \label{eq:Punitarysu}
\eea
By further imposing $\dot{\overline{Q}^\sigma}=0$, one arrives at
\beq
	\frac{\overline{\delta N}}{N}=\frac{1}{\sqrt{3}}\frac{v^{4/3}}{\pi^2_\sigma}\left[\left(\frac{\pi^2_\sigma}{2v^2}+V\right)\overline{\delta\gamma}_1-\frac{3}{\Mp^2}v^{1/3}\theta\overline{\delta\pi}_1\right], \label{eq:Nunitarysu}
\eeq
where we plugged in the gauge condition and the expression of $\overline{P}_\sigma$. The above expressions and the gauge condition can be plugged into Eqs. (\ref{eq:dynEsu}) \& (\ref{eq:dyn1su}) to derive a set of closed Hamilton equations for $(\overline{\delta\gamma}_1,\overline{\delta\pi}_1)$ and $(\overline{Q}^s,\overline{P}_s)$. This proves that fixing $\overline{Q}^\sigma=0$ remove the redundant degree of freedom in the separate-universe picture. 

The large-scale limit of $P_\sigma$ and $\delta N$ in the unitary gauge of cosmological perturbations is easily matched to $\overline{P}_\sigma$ and $\overline{\delta N}$ [see Eqs. (\ref{eq:Punitary}) \& (\ref{eq:Nunitary}) to be compared with Eqs. (\ref{eq:Punitarysu}) \& (\ref{eq:Nunitarysu})]. However, $k\delta N_1$ is not $k$-suppressed in the full perturbative treatment [see Eq. (\ref{eq:N1unitary})]. As a result, its contribution to the dynamical equations of $(\delta\gamma_1,\delta\pi_1)$ remains at large scales but it is absent in the dynamical equations of $(\overline{\delta\gamma}_1,\overline{\delta\pi}_1)$. Because of that, the separate-universe approximation fails to be matched to cosmological perturbations in the unitary gauge. Note that the entropic modes $(Q^s,P_s)$ are invariant under spatial-diffeomorphism; hence their dynamical equations are free of $k\delta N_1$ and the separate-universe picture can be matched to cosmological perturbations. However, they are coupled to the adiabatic mode and they will inherit from the mismatch between full cosmological perturbation theory and the separate-universe picture along this direction.

Since the mismatch comes from the fact that $k\delta N_1$ is not $k$-suppressed in the unitary gauge, cosmological perturbation theory at large scales and the separate-universe picture can be reconciled by working with variables for the adiabatic mode which are diffeomorphism-invariant, \ie variables $\delta z$ such that $\{\delta z,\mathcal{D}\}=0$.\footnote{Here, by diffeomorphism-invariant we mean invariant under spatial diffeomorphism generated by $\mathcal{D}$ and not invariant under spacetime diffeomorphism which are gauge-invariant variables.} Indeed, the dynamical equations of such variables are free of any contribution from $k\delta N_1$. (Note that the entropic modes are readily diffeomorphism-invariant, hence not plagued by the contribution coming from $k\delta N_1\mathcal{D}$). To this end, let us introduce the following variables 
\bea
	\widetilde{Q}^\sigma&=&Q^\sigma+\frac{1}{\sqrt{3}}\frac{v^{1/3}}{\pi_\sigma}\left(\theta\delta\gamma_1-2v^{1/3}\delta\pi_1\right),\label{eq:unitaryDIQ}\\
	\widetilde{P}_\sigma&=&P_\sigma, \label{eq:unitaryDIP}\\
	\widetilde{\delta\gamma}_1&=&\delta\gamma_1-\frac{2}{\sqrt{3}}\frac{v^{2/3}}{\pi_\sigma}P_\sigma, \label{eq:unitaryDIGa}\\
	\widetilde{\delta\pi}_1&=&\delta\pi_1-\frac{1}{\sqrt{3}}\frac{v^{1/3}\theta}{\pi_\sigma}P_\sigma, \label{eq:unitaryDIPi}
\eea
as well as $(\widetilde{Q}^s,\widetilde{P}_s)=(Q^s,P_s)$ and $(\widetilde{\delta\gamma}_2,\widetilde{\delta\pi}_2)=(\delta\gamma_2,\delta\pi_2)$. One can readily check that these new variables are generated by a canonical transformation; \ie the only non-vanishing Poisson brackets are $\{\widetilde{Q}^\sigma,\widetilde{P}_\sigma\}$, $\{\widetilde{\delta\gamma}_1,\widetilde{\delta\pi}_1\}$, $\{\widetilde{Q}^s,\widetilde{P}_s\}$ and $\{\widetilde{\delta\gamma}_2,\widetilde{\delta\pi}_2\}$ which all equal one. This canonical transformation has several important properties. First, it is scale-independent and it does not mix the isotropic degrees of freedom with the anisotropic ones. As a consequence, it can be identically applied in the separate-universe picture using the dictionary of Tab. \ref{tab:dict}. Second, because it is scale-invariant, all the terms which are $k$-suppressed in the dynamics of the old variables $\delta \boldsymbol{z}\equiv(Q^a,\delta\gamma_1,\delta\gamma_2,P_a,\delta\pi_1,\delta\pi_2)^\mathrm{T}$ will be identically $k$-suppressed for the new variables $\delta\widetilde{ \boldsymbol{z}}\equiv(\widetilde{Q}^a,\widetilde{\delta\gamma}_1,\widetilde{\delta\gamma}_2,\widetilde{P}_a,\widetilde{\delta\pi}_1,\widetilde{\delta\pi}_2)^\mathrm{T}$. In particular, isotropic modes in the new variables will decouple from the anisotropic modes at large scales because these couplings are $k^2$-suppressed, and the analysis performed in Sec. \ref{sssec:arbgauge} for arbitrary gauges equally applies to this set of new variables. Consequently, the large-scale limit of the equations of motion for $\delta\widetilde{\boldsymbol{z}}$ in cosmological perturbation theory can be matched to the separate-universe equations for $\overline{\delta\widetilde{\boldsymbol{z}}}$ providing that $\delta N$ can be matched to $\overline{\delta N}$ and that $k\delta N_1$ is vanishing or $k$-suppressed. Finally, it is easy to check that $\{\widetilde{\delta\gamma}_1,\mathcal{D}\}=0=\{\widetilde{\delta\pi}_1,\mathcal{D}\}$. Hence, $(\widetilde{\delta\gamma}_1,\widetilde{\delta\pi}_1)$ forms a pair of canonical variables which are diffeomorphism-invariant and which can describe the adiabatic mode.\footnote{Note that $(\widetilde{Q}^\sigma,\widetilde{P}_\sigma)$ are not diffeomorphism-invariant.} Thanks to that, we have at our disposal variables for the adiabatic mode whose dynamics is free of contribution from $k\delta N_1$ and from the anisotropic modes, hence insensitive to the condition on the shift vector for matching cosmological perturbation theory at large scales to the separate-universe approximation.

Having defined this new set of variables, one can now implement the gauge conditions which select the unitary gauge. The properties identified before are gauge-independent and they hold when fixing the gauge. In the full perturbative treatment, the expressions of $\widetilde{P}_\sigma$, $\widetilde{\delta\pi}_2$, $\delta N$ and $k\delta N_1$ are given by Eqs. (\ref{eq:Punitary}), (\ref{eq:piunitary}), (\ref{eq:Nunitary}) \& (\ref{eq:N1unitary}) in which the old variables are replaced by their expressions as functions of the new variables, and in which we further plug $Q^\sigma=0$. Hence, all the redundant degrees of freedom are expressed as functions of $(\widetilde{\delta\gamma}_1,\widetilde{\delta\pi}_1)$ and $(\widetilde{Q}^s,\widetilde{P}_s)$ for which it is possible to derive a set of autonomous equations of motion. Now, because $(\widetilde{\delta\gamma}_1,\widetilde{\delta\pi}_1)$ and $(\widetilde{Q}^s,\widetilde{P}_s)$ are diffeomorphism-invariant, these dynamical equations are free of any contribution from $k\delta N_1\mathcal{D}$. 

The very same approach can be performed at the level of the separate universes. Here, the redundant degrees of freedom are given by Eqs. (\ref{eq:Punitarysu}) \& (\ref{eq:Nunitary}) by expressing the old variables as functions of the new ones. Then one obtains a closed set of equations of motion for the variables $(\overline{\widetilde{\delta\gamma}}_1,\overline{\widetilde{\delta\pi}}_1)$ and $(\overline{\widetilde{Q}^s},\overline{\widetilde{P}}_s)$. Because the canonical transformation is the same at the full perturbative level and at the separate-universe level, it is clear that the expressions of $\widetilde{P}_\sigma$ and of $\delta N$ are matched to the ones of $\overline{\widetilde{P}}_\sigma$ and of $\overline{\delta N}$ at large scales. As a result, the large-scale limit of the dynamical equations for $(\widetilde{\delta\gamma}_1,\widetilde{\delta\pi}_1)$ and $(\widetilde{Q}^s,\widetilde{P}_s)$ is matched to the ones of $(\overline{\widetilde{\delta\gamma}}_1,\overline{\widetilde{\delta\pi}}_1)$ and $(\overline{\widetilde{Q}}^s,\overline{\widetilde{P}}_s)$ in the separate-universe picture. This shows that one can use the separate-universe picture in the unitary gauge providing that the variables are diffeomorphism-invariant.

\subsubsection*{Spatially-flat and uniform-expansion gauges}
A natural implementation of the spatially-flat gauge in the separate universe picture would consist in enforcing $\overline{\delta\gamma}_1=0$. This is a priori sufficient since by construction $\overline{\delta\gamma}_2=0$ in this context. However, such a choice equally appears as a natural implementation of the uniform-expansion gauge since $\overline{\delta N}_1$ is also vanishing by construction of the separate-universe approximation. Hence, it is unclear which gauge in the full perturbative treatment is naturally related to $\overline{\delta\gamma}_1=0$ in the separate-universe approximation.

 In practice, fixing $\overline{\delta\gamma}_1=0$ indeed removes the gauge degree of freedom. To show this, we start by combining the gauge condition and $\overline{\mathcal{C}}^{(1)}=0$ to derive
\beq
	\overline{\delta\pi}_1=\frac{\Mp^2}{\sqrt{3}v^{2/3}\theta}\left(vV_{;a}\,\overline{Q}^a+\frac{\pi_\sigma}{v}\overline{P}_\sigma\right). \label{eq:pi1suUEG}
\eeq
The gauge condition is preserved through evolution; thus $\dot{\overline{\delta\gamma}}_1=0$. Making use of the separate-universe dynamics for $\overline{\delta\gamma}_1$ yields 
\bea
	\overline{\delta N}=-\frac{2\Mp^2}{3}\frac{N}{\theta^2}\left(V_{;a}\overline{Q}^a+\frac{\pi_\sigma}{v^2}\overline{P}_\sigma\right). \label{eq:NsuUEG}
\eea
All variables are now functions of the scalar-field fluctuations. Their dynamics is given by an autonomous set of equations obtained by plugging the above expressions into the dynamical equations of the fields fluctuations. They read
\bea
	&&\left\{\begin{array}{rcl}
		\dot{\overline{Q}}^\sigma&=&\ds-N\left[\frac{2\Mp^2}{3}\frac{\pi_\sigma V_{;\sigma}}{v\theta^2}\overline{Q}^\sigma+\frac{1}{v}\left(\frac{2\Mp^2}{3}\frac{\pi^2_\sigma}{v^2\theta^2}-1\right)\overline{P}_\sigma+\left(\omega_1+\frac{2\Mp^2}{3}\frac{\pi_\sigma V_{;1}}{v\theta^2}\right)\overline{Q}^1\right] \\
		\dot{\overline{P}}_\sigma&=&\ds N\left[v\left(\frac{2\Mp^2}{3}\frac{V^2_{;\sigma}}{\theta^2}-V_{;\sigma\sigma}\right)\overline{Q}^\sigma+\frac{2\Mp^2}{3}\frac{\pi_\sigma V_{;\sigma}}{v\theta^2}\overline{P}_\sigma\right] \\ 
		&&\ds -N\left[v\left(V_{;1\sigma}-\frac{3\Mp^2}{2}\frac{V_{;\sigma}V_{;1}}{\theta^2}\right)\overline{Q}^1+\omega_1\overline{P}_1+vV_{;2\sigma}\overline{Q}^2\right], \\
	 \end{array}\right. \label{eq:dynAsuUEG} \\
	 &&\left\{\begin{array}{rcl}
		\dot{\overline{Q}}^s&=&\ds\frac{N}{v}\overline{P}^s+N\left({\Omega^s}_{s'}\overline{Q}^{s'}+\delta_{1,s}\omega_1\overline{Q}_\sigma\right), \\
		\dot{\overline{P}}_s&=&\ds N\left[v\left(\delta_{1,s}\frac{2\Mp^2}{3}\frac{V^2_{;1}}{\theta^2}-V_{;ss'}+\mathcal{R}_{ss'}\right)\overline{Q}^{s'}+\delta_{1,s}\frac{2\Mp^2}{3}\frac{\pi_\sigma V_{;1}}{v\theta^2}\overline{P}_\sigma\right] \\ 
		&&\ds -N{\Omega_{s}}^{s'}\overline{P}_{s'}-\delta_{1,s}N\left[v\left(V_{;1\sigma}-\frac{3\Mp^2}{2}\frac{V_{;\sigma}V_{;1}}{\theta^2}\right)\overline{Q}^\sigma-\omega_1\overline{P}_\sigma\right]-\delta_{2,s}NvV_{;2\sigma}\overline{Q}^2,
	 \end{array}\right. \label{eq:dynEsuUEG}
\eea
and they provide us with the separate-universe dynamics with the gauge fixed requiring that $\overline{\delta\gamma}_1=0$.

Let us now compare the above with cosmological perturbation theory in the spatially-flat gauge and in the uniform-expansion gauge. The expression of $\delta\pi_1$ in the spatially-flat gauge, Eq. (\ref{eq:pi1SF}), is easily matched to its corresponding expression in the separate-universe picture,  Eq. (\ref{eq:pi1suUEG}). However, it is not the case of $\delta N$; see Eq. (\ref{eq:NSF}) in the spatially-flat gauge to be compared with Eq. (\ref{eq:NsuUEG}). As a result, imposing $\overline{\delta\gamma}_1=0$ is not the large-scale limit of the spatially-flat gauge and the mismatch between $\delta N$ and $\overline{\delta N}$ will reflect in the equations of motion of $(Q^a,P_a)$ whose large-scale limit is not the dynamical equations for $(\overline{Q}^a,\overline{P}_a)$.  In addition, one notes that $k\delta N_1$ is not $k$-suppressed in the spatially-flat gauge. Hence, this contribution will remain at large-scales in the Hamilton equation of $P_\sigma$ and this cannot be captured by the separate universe. Note that here, one cannot make use of the same technique as the one employed previously for the unitary gauge. This is because not only $k\delta N_1$ is not $k$-suppressed but also $\delta N$ cannot be matched to $\overline{\delta N}$.

On the contrary, Eqs. (\ref{eq:pi1suUEG}) \& (\ref{eq:NsuUEG}) are easily matched to the expressions of $\delta\pi_1$ and $\delta N$ in the uniform-expansion gauge, Eqs. (\ref{eq:pi1UEG}) \& (\ref{eq:NUEG}). This is indeed achieved by either employing the dictionary of Tab. \ref{tab:dict} or by taking the large-scale limit of Eqs. (\ref{eq:pi1UEG}) \& (\ref{eq:NUEG}). Moreover, $k\delta N_1$ is set to zero in the uniform-expansion gauge. Hence, the gauge condition $\overline{\delta\gamma}_1=0$ corresponds to the large-scale limit of the uniform-expansion gauge which is not pathological anymore in the separate-universe picture. As a consequence, the dynamical equation for $(Q^a,P_a)$ at large scales match the ones for $(\overline{Q}^a,\overline{P}_a)$ up to inhomogeneous terms which scale as $(k^2/v^{2/3})\delta\gamma_2$. These are the ones bearing the gauge degree of freedom unfixed by the uniform-expansion gauge but which decouples from the physical degrees of freedom at large scales [see Eqs. (\ref{eq:dynAueg}) \& (\ref{eq:dynEueg}) whose large-scale limit is to be compared to Eqs. (\ref{eq:dynAsuUEG}) \& (\ref{eq:dynEsuUEG})]. This finishes to prove that fixing $\overline{\delta\gamma}_1=0$ is the large-scale limit of the uniform-expansion gauge. 

\subsubsection*{Adequate variables}
The separate-universe picture may fail to trustfully reproduce the large-scale dynamics of cosmological perturbations when fixing the gauge because $k\delta N_1$ is not $k$-suppressed (or vanishing) and/or because $\overline{\delta N}$ cannot be matched to $\delta N$. However, a mismatch due to the shift vector can systematically be avoided by using diffeomorphism-invariant variables as the ones introduced in Eqs. (\ref{eq:unitaryDIGa}) \& (\ref{eq:unitaryDIPi}), and this is so whether the gauge leads to a $k$-suppressed shift vector or not. Indeed, the strategy we employed for the unitary gauge can be used in all the gauges but the ones for which the gauge conditions are $\widetilde{\delta\gamma}_1=0$ and/or $\widetilde{\delta\pi}_1=0$. In that cases, one can instead introduce
\bea
	\widetilde{Q}^\sigma_{(\alpha)}&=&Q^\sigma, \label{eq:altDIQ}\\
	\widetilde{P}_\sigma^{(\alpha)}&=&P_\sigma-\sqrt{3}\frac{\pi_\sigma}{v^{1/3}}\left[\alpha\frac{\delta\gamma_1}{2v^{1/3}}+\left(1-\alpha\right)\frac{\delta\pi_1}{\theta}\right], \label{eq:altDIP} \\
	\widetilde{\delta\gamma}_1^{(\alpha)}&=&\delta\gamma_1+\sqrt{3}\left(1-\alpha\right)\left(\frac{\pi_\sigma}{v^{1/3}\theta}\right)Q_\sigma, \\
	\widetilde{\delta\pi}_1^{(\alpha)}&=&\delta\pi_1-\alpha\frac{\sqrt{3}}{2}\left(\frac{\pi_\sigma}{v^{2/3}}\right)Q_\sigma,
\eea
as well as $(\widetilde{Q}^s,\widetilde{P}_s)=(Q^s,P_s)$ and $(\widetilde{\delta\gamma}_2,\widetilde{\delta\pi}_2)=(\delta\gamma_2,\delta\pi_2)$. In the above, $\alpha$ can be any function of time. These new variables are generated by a canonical transformation sharing the very same properties as the one given in Sec. \ref{sssec:gf}, except that it is now $(\widetilde{Q}^\sigma_{(\alpha)},\widetilde{P}_\sigma^{(\alpha)})$ which are diffeomorphism-invariant instead of $(\widetilde{\delta\gamma}_1^{(\alpha)},\widetilde{\delta\pi}_1^{(\alpha)})$. As a result, one can use $(\widetilde{Q}^\sigma_{(\alpha)},\widetilde{P}_\sigma^{(\alpha)})$ and $(\widetilde{Q}^s,\widetilde{P}_s)$ as a set of diffeomorphism-invariant variables to describe the adiabatic mode and the entropic modes, and these will never be plagued by mismatch in $k\delta N_1$.\footnote{Note that this choice is obviously not adapted to the unitary gauge where one imposes $\widetilde{Q}^\sigma=0$. This is why one preferentially use variables in Eqs. (\ref{eq:unitaryDIGa}) \& (\ref{eq:unitaryDIPi}) in that situation.} They are constructed from isotropic degrees of freedom, hence they can be defined in the separate-universe picture too.

This shows that by working with $(\widetilde{\delta\gamma}_1,\widetilde{\delta\pi}_1)$ as given by Eqs. (\ref{eq:unitaryDIGa}) \& (\ref{eq:unitaryDIPi}), or with $(\widetilde{Q}^\sigma_{(\alpha)},\widetilde{P}^{(\alpha)}_\sigma)$ as given by Eqs. (\ref{eq:altDIQ}) \& (\ref{eq:altDIP}), for the adiabatic mode, and with $(Q^s,P_s)$ for the entropic modes, one has at hand a diffeomorphism-invariant formulation of the separate-universe approximation, hence independent of the choice of $k\delta N_1$.\footnote{We note that other choices of such variables which are adequate to a diffeomorphism-invariant formulation of the separate-universe picture can be constructed. The building rules is that the canonical transformation generating them should be scale-independent and should not mix the isotropic sector with anisotropic one. These are necessary conditions for the diffeomorphism-invariant variables to have their equivalent version in the separate-universe approximation. At the end, one can pick up any set of isotropic, diffeomorphism-invariant variables paying attention that these are not part of the gauge conditions one is implmenting.}  Obviously, this cannot solve for a potential mismatch between $\overline{\delta N}$ and $\delta N$. However, it shows that the choice of $k\delta N_1$ is not a concern for matching cosmological perturbation theory at large scales with the separate-universe picture since there is a set of diffeomorphism-invariant variables which can systematically be used to do it. {\it As a result, we conclude that the separate-universe picture is equivalent to cosmological perturbation theory at large scales providing that the lapse function in both pictures can be matched, and that one properly picks up isotropic and diffeomorphism-invariant variables.}

In passing, we note that in the uniform-expansion gauge one has $\widetilde{Q}^\sigma_{(\alpha=1)}=Q_\sigma$ and $\widetilde{P}_\sigma^{(\alpha=1)}=P_\sigma$. This means that we are effectively dealing with diffeomorphism-invariant variables which in addition decouple from the anisotropic degrees of freedom at large scales, hence decouple from the gauge degree of freedom $P_\mathcal{D}$. This further explains why the separate-universe picture works in that gauge.

\subsection{Application to orbital inflation}
We apply the above results to the case of orbital inflation \cite{achucarro2019orbitalinflationinflatingangular} which is described by the following Lagrangian for two scalar fields $(\beta,\phi)$
\bea
	\mathcal{L}=-\frac{1}{2}g^{\mu\nu}\left[\partial_\mu\phi\partial_\nu\phi+f(\phi)\partial_\mu\beta\partial_\nu\beta\right]-V(\phi,\beta).
\eea
In the 2-dimensional field space, $\beta$ can be naively viewed as a polar coordinate and $\phi$ as the radial one. Since the metric in the field space is $G_{IJ}=\mathrm{diag}[1,f(\phi)]$, the kinetic term exhibits an isometry along $\beta$. In this context, it was shown that slow-roll inflation can proceed along the isometry direction, \ie $\dot{\phi}=0$ and $\dot{\beta}<0$ (where the sign of $\dot{\beta}$ is a matter of convention), and the inflationary trajectory now follows the Hubble gradient rather than the potential gradient since $\dot\beta=-2\partial_\beta H/f$ \cite{achucarro2019orbitalinflationinflatingangular}. As a result, the adiabatic direction is aligned with $\dot{\beta}$ and the entropic one with $\dot{\phi}$, \ie $e^I_\sigma=(-1/\sqrt{f},0)$ and $e^I_s=(0,1)$. Potentials admitting such solutions can be engineered extending the Hamilton-Jacobi approach \cite{PhysRevD.42.3936} to the case of multiple fields. In practice, one starts from an expression of $H$ as a function of the fields and then use $V=3\Mp^2H^2-2G^{IJ}(D_IH)(D_JH)$. As is done in \cite{achucarro2019orbitalinflationinflatingangular}, we choose $f(\phi)=e^{2\phi/R_0}$ where $R_0$ is the curvature radius of the field-space. As a result, the slow-roll trajectory has a turning radius which is constant $\kappa^2=R_0^2$, as is the case in \eg quasi-single field models \cite{Chen:2009we,Chen:2009zp,Chen:2012ge}. It is related to the turning rate by $\omega_1=-\sqrt{2\epsilon_1}\Mp H/\kappa$.

In this model, the mass-matrix of the adiabatic and entropic perturbations $M^{(\phi)}_{ab}=V_{ab}-\mathcal{R}_{ab}$ at leading order in slow-roll parameters reads
\bea
	M^{(\phi)}_{ab}=\left(\begin{array}{cc}
		H^2\left[6\epsilon_1-\frac{3}{2}\epsilon_2+2\epsilon_1\left(\frac{\Mp}{\kappa}\right)^2\right] & 3H^2\sqrt{2\epsilon_1}\left(1-\epsilon_1+\frac{\epsilon_2}{2}\right)\left(\frac{\Mp}{\kappa}\right) \\
		3H^2\sqrt{2\epsilon_1}\left(1-\epsilon_1+\frac{\epsilon_2}{2}\right)\left(\frac{\Mp}{\kappa}\right) & \mu^2-6H^2\epsilon_1\left(\frac{\Mp}{\kappa}\right)^2
	\end{array}\right),
\eea
where $\mu^2$ is the Lagrangian mass of the entropic mode. It can be approximately constant given a suitable choice of the potential (see \cite{achucarro2019orbitalinflationinflatingangular}). Here we consider slow-roll solutions hence $\epsilon_1\ll1$ and for simplicity we set $\epsilon_2=0$. We suppose that $\epsilon_1\ll\mu^2/H^2\leq9/4$ where the lower bound ensure the entropic mode to be heavier than the adiabatic one. Its diagonalization yields 
\bea
	m^2_\pm&=&\frac{1}{2}\left(6H^2\epsilon_1+\mu^2-4\epsilon_1H^2\left(\frac{\Mp}{\kappa}\right)^2\right)\pm\sqrt{\Delta}
\eea
where
\bea
	\Delta&=&\left(6H^2\epsilon_1-\mu^2\right)\left[6H^2\epsilon_1-\mu^2+16\epsilon_1H^2\left(\frac{\Mp}{\kappa}\right)^2\right]+36\epsilon_1(1-2\epsilon_1)H^4\left(\frac{\Mp}{\kappa}\right)^2 \nonumber \\
	&&+64\epsilon_1^2H^4\left(\frac{\Mp}{\kappa}\right)^4.
\eea
The single-field limit is obtained by taking $(\kappa/\Mp)^2\gg1/\epsilon_1$ in which case $M^{(\phi)}_{ab}$ is diagonal with $m^2_+=m^2_\sigma=6\epsilon_1H^2$ and $m^2_-=m^2_s=\mu^2$, and $\omega_1\to0$ as well. This guarantees adiabatic perturbations to decouple from entropic ones. 

Considering instead trajectories with a `small' radius and the hierarchy $\epsilon_1\ll(\kappa/\Mp)^2\ll1/\epsilon_1$, the predicted observables are dictated by the entropic mass and might significantly depart from single-field inflation (see Fig. 1 of \cite{achucarro2019orbitalinflationinflatingangular}). In this case, one arrives at 
\bea
	\left(\frac{m_\pm}{H}\right)^2=-2\epsilon_1\left(\frac{\Mp}{\kappa}\right)^2+\frac{1}{2}\left(\frac{\mu}{H}\right)^2\pm\frac{1}{2}\sqrt{\left(\frac{\mu}{H}\right)^4-16\epsilon_1\left(\frac{\Mp}{\kappa}\right)^2\left(\frac{\mu}{H}\right)^2+36\epsilon_1\left(\frac{\Mp}{\kappa}\right)^2}. \nonumber \\
\eea
If $\mu/H=\mathcal{O}(1)$, it is further simplifies to $m^2_+=\mu^2-10\epsilon_1H^2(\Mp/\kappa)^2$ and $m^2_-=6\epsilon_1H^2(\Mp/\kappa)^2$. Hence,  the separate-universe picture is valid providing that $k/aH\ll\sqrt{\epsilon_1}(\Mp/\kappa)$. Typical values of $(\Mp/\kappa)$ considered in \cite{achucarro2019orbitalinflationinflatingangular} ranges from 1 to $10^3$. Hence, the condition is less stringent than in single field where the separate-universe picture holds given that $k/aH\ll\sqrt{\epsilon_1}$. On the contrary, if $\epsilon_1(\Mp/\kappa)^2\geq\mu^2/H^2$, on arrives at $m^2_\pm=\pm3\sqrt{\epsilon_1}H^2(\Mp/\kappa)$.\footnote{Note that it implies that $\mu^2/H^2\ll1$ given that $\epsilon_1\ll(\kappa/\Mp)^2$.} The resulting condition for the separate-universe picture to hold is $k/aH\ll\epsilon_1^{1/4}\sqrt{\Mp/\kappa}$ which is again less stringent than in single-field. It is also worth noticing that in all these cases $\omega_1\ll1$ meaning that the mismatch in the dynamical equations of the Mukhanov-Sasaki variables, Eqs. (\ref{eq:dynMSSUad}) \& (\ref{eq:dynMSSUen}), can be safely neglected.

The case of strong turn is obtained in the regime where $(\kappa/\Mp)^2\ll\epsilon_1$. Here, the diagonalization leads to $m_+^2=2\epsilon_1H^2(\Mp/\kappa)^2$ and $m_-^2=-6\epsilon_1H^2(\Mp/\kappa)^2$. Conditions for the separate-universe picture to be valid is thus $(k/aH)^2\ll\epsilon_1(\Mp/\kappa)^2$ which is easily met since $\epsilon_1(\Mp/\kappa)^2\gg1$. Interestingly, it does not require to be deeply in the superhorizon regime. However, the strong turn leads to an important mismatch in the equations of motion of the Mukhanov-Sasaki variables and the separate-universe dynamics should be initiated with initial conditions such that $\overline{\mathcal{D}}^{(1)}=0$.

Observables can be obtained by solving for the separate-universe equations. Following \cite{achucarro2019orbitalinflationinflatingangular}, we work in the unitary gauge hence the adiabatic mode is given by the curvature perturbation $\zeta=\delta\gamma_1/(2\sqrt{3}a^2)$. Here, we introduce $P_\zeta=2\sqrt{3}a^2\delta\pi_1+\Mp^2\sqrt{3}(aH)\delta\gamma_1+4\Mp^2\epsilon_1a^3(H/\kappa)Q_1$. [Note that the canonical transformation relating $(\zeta,P_\zeta)$ to $(\delta\gamma_1,\delta\pi_1)$ only holds in the unitary gauge.] The Hamilton equations are
\bea
	\dot\zeta=\frac{P_\zeta}{2\Mp^2\epsilon_1a^3}+\frac{2H}{\kappa}Q_1 &\,\,\,\mathrm{and}\,\,\,& \dot{P}_\zeta=0
\eea
for curvature perturbations, and
\bea
	\dot{Q}_1=\frac{P_1}{a^3}  &\,\,\,\mathrm{and}\,\,\,& \dot{P}_1=-a^3{\mu^2}Q_1+\frac{2H}{\kappa}P_\zeta,
\eea
for the entropic mode. The momentum $P_\zeta$ is a constant of motion in the separate-universe picture that we note $P_\zeta^{(0)}$ (note that setting it to zero yields the Euler-Lagrange equation in which the decaying mode is neglected \cite{achucarro2019orbitalinflationinflatingangular}). For the curvature perturbations, one obtains
\bea
	\zeta&=&\zeta^{(0)}+\frac{2}{\kappa}\ds\int^t\dd s HQ_1+\frac{P^{(0)}_\zeta}{2\Mp^2}\ds\int^t\frac{\dd s}{\epsilon_1a^3},
\eea
with $\zeta^{(0)}$ an integration constant. The entropic mode is solved for to give
\bea
	Q_1&=&\frac{1}{a\sqrt{3}}\left\{\frac{C^{(0)}}{(aH)^{\nu_+}}\left[1-\left(\frac{2H^2P^{(0)}_\zeta}{\sqrt{3}\kappa}\right)\left(1+\nu_-\right)\ds\int^t\frac{\dd s}{(aH)^{1+\nu_-}}\right]\right. \nonumber \\
	&&+\left.\frac{D^{(0)}}{(aH)^{\nu_-}}\left[1+\left(\frac{2H^2P^{(0)}_\zeta}{\sqrt{3}\kappa}\right)\left(1+\nu_+\right)\ds\int^t\frac{\dd s}{(aH)^{1+\nu_+}}\right]\right\}, \\
	P_1&=&\frac{a}{\sqrt{3}}\left\{\left(1+\nu_+\right)\frac{C^{(0)}}{(aH)^{\nu_+-1}}\left[1-\left(\frac{2H^2P^{(0)}_\zeta}{\sqrt{3}\kappa}\right)\left(1+\nu_-\right)\ds\int^t\frac{\dd s}{(aH)^{1+\nu_-}}\right]\right. \nonumber \\
	&&\left.+\left(1+\nu_-\right)\frac{D^{(0)}}{(aH)^{\nu_--1}}\left[1+\left(\frac{2H^2P^{(0)}_\zeta}{\sqrt{3}\kappa}\right)\left(1+\nu_+\right)\ds\int^t\frac{\dd s}{(aH)^{1+\nu_+}}\right]\right\}
\eea
where $\nu_\pm=(1\pm3\sqrt{1-(2\mu/3H)^2+\mathcal{O}(\epsilon_1)})/2$ and $C^{(0)}$ and $D^{(0)}$ are two integration constants which are dimensionless. We recover that the contribution of the momentum $P_\zeta$ to the curvature perturbation is exponentially suppressed with the number of e-folds $\mathcal{N}$ counted from the onset of the separate-universe picture to the end of inflation. Similarly for the entropic mode, the contribution of $P_\zeta$ scales as $e^{-3\mathcal{N}}$, hence exponentially suppressed too. 

We first consider $\mu/H\ll1$. In this case $\nu_+\simeq2$ and $\nu_-\simeq1$ and $Q_1$ is dominated by the $D^{(0)}$ branch which is the growing mode. This leads to $Q_1\propto e^{-\mu^2\mathcal{N}/(3H^2)}$ and 
\bea
{\zeta}=\zeta_0+\left(Q^{(0)}_1-H^2P^{(0)}_1\right)\left(\frac{6}{\kappa}\right)\left(\frac{H}{\mu}\right)^2\left[1-e^{\left(\frac{\mu}{H}\right)^2\frac{\mathcal{N}}{3}}\right],
\eea
where $Q^{(0)}_1$ and $P^{(0)}_1$ are the values of the entropic mode at the onset of the separate-universe description, which is when modes cross the physical radius $H^{-1}\kappa/(\sqrt{\epsilon_1}\Mp)$ or $H^{-1}\sqrt{\kappa/\Mp}/\epsilon^{1/4}_1$ depending on the hierarchy between $(\mu/H)^2$ and $\epsilon_1\Mp/\kappa$. This matches the results in \cite{achucarro2019orbitalinflationinflatingangular} setting $P^{(0)}_1=0$ which corresponds to discard the decaying mode. The contribution of $Q_1$ to the curvature perturbation depends on the number of e-folds and this should modify the tilt $n_\mathrm{S}-1$. 

Second, we set $1-(2\mu/3H)^2\ll1$ which gives $\nu_\pm=1/2\pm\mathcal{O}[1-(2\mu/3H)^2]$. Here, one finds $Q_1\propto C^{(0)}e^{-(1+3\varepsilon)\mathcal{N}/2}+D^{(0)}e^{-(1-3\varepsilon)\mathcal{N}/2}$ where $\varepsilon\equiv\sqrt{1-(2\mu/3H)^2}$. Hence the entropic mode is rapidly decaying. Plugging this in the expression of the curvature perturbations boils down to 
 \bea
	 %\zeta=\zeta^{(0)}+\frac{4}{\kappa}(1-3\varepsilon)C^{(0)}+\frac{4}{\kappa}(1+3\varepsilon)D^{(0)}.
	 \zeta=\zeta^{(0)}+\frac{8\varepsilon}{\sqrt{3}\kappa}\left(Q^{(0)}_1-H^2P^{(0)}_1\right).
\eea
In this case, the corrections due to the coupling of $\zeta$ with the entropic mode are suppressed by $\varepsilon$ and fixed at the onset of the separate-universe picture, \ie when modes cross the physical radius $\kappa/(\sqrt{\epsilon_1}\Mp H)$. Because of that, it will only yield a small change in the amplitude of the power spectrum of curvature perturbations.

\section{Conclusion}
\label{sec:conclusion} 
In this paper we have explored the Hamiltonian description of cosmological perturbation theory and of the separate-universe approach for non-linear sigma models as a typical formulation of multifield inflation. We have restricted ourselves to scalar fields minimally coupled to gravity as described by general relativity. The separate universe approach that we have studied consists in discarding all anisotropies, even in the scalar perturbations. We have compared these two methods in hopes of matching them at large scales, and have done so in an arbitrary gauge and after gauge fixing in several manners. let us note that in this work no assumptions have been made on the background dynamics. These results are thus applicable outside of the slow-roll regime and even outside of inflating scenarios.

Our initial conclusion can be found in Sec. \ref{sssec:arbgauge}, and states that in an arbitrary gauge, the separate-universe picture matches cosmological perturbation theory if three conditions are met. The first condition is a  direct consequence of the expansion of the Hamiltonian which led to a set of equations on $k^2$, and indicate that for the separate-universe picture to work one needs to be working at large enough scales compared not only to the horizon scale but also compared to the effective masses of all fields. The second condition states that anisotropic degrees of freedom are set to zero, which is expected since they are entirely discarded in the separate universe approach. Finally, the Lagrange multipliers, so the perturbed lapse and the scalar degree of freedom of the perturbed shift, need to match in both frameworks. In reality, only the perturbed lapse needs to be matched since the scalar degree of freedom of the perturbed shift already matches once the second condition has been imposed. The conditions we uncover generalise to the multifield context those found for a single-field analysis \cite{Artigas_2022}.

Then, we have studied several specific gauges in order to find which ones are applicable in the separate-universe approach. In the unitary gauge we initially found it challenging to match both Lagrange multipliers, but making a field redefinition in order to handle diffeomorphism-invariant and isotropic variables only allowed us to bypass the problematic and not $k$-suppressed $k\delta N_1$ and verify all conditions given in the arbitrary gauge conclusions. These variables can be used to build a diffeomorphism-invariant version of the separate-universe approximation hence independent on the choice of perturbed shift vector. This means that by using these type of variables, one only needs to match the perturbed lapse function. We have tried following similar procedures for the spatially-flat gauge and the uniform-expansion gauge but to no avail. We initially noticed that any attempt to define the spatially-flat gauge in the separate-universe picture not only leads to unmatchable perturbed lapses in both frameworks, but actually leads to a definition of the uniform-expansion gauge instead. We thus decided to focus on the latter. In cosmological perturbation theory, the uniform-expansion gauge is not unequivocally defined, one spurious gauge mode remains, which explains the extra $\left(k^2/v^{2/3}\right)\delta\gamma_2$ that cannot be decoupled from the field equations of motions. However, the remaining gauge degree of freedom decouples at large scale making this gauge choice unambiguous in the separate-universe picture which is further shown to match cosmological perturbation theory.

However, the separate-universe picture fails to reproduce the dynamics of the gauge-invariant, Mukhanov-Sasaki variables at large scales with a mismatch proportional to the diffeomorphism constraint. This mismatch is slow-roll suppressed for the adiabatic mode (hence harmless in slow-roll and ultra slow-roll), and it has the very same expression as in single-field inflation \cite{Artigas_2022,Cruces_2023}. For the first entropic mode however, it scales with the turning-rate of the adiabatic direction, $\omega_1$, which can become sizable in models with sharp turns in the field space. The large-scale dynamics of the remaining entropic modes is nicely reproduced by the separate-universe approximation but this holds true in the specific choice of basis adopted in this work. Given a different basis, one would have a mismatch for possibly all the entropic modes. This also means that the mismatch will arise in all directions by working in the field basis and it reads $G^{IJ}[4\pi_J/(v\theta)-V_{,J}]\overline{\mathcal{D}}^{(1)}$. (Note however that the diffeomorphism constraint $\overline{\mathcal{D}}^{(1)}$ can be enforced to be vanishing in the separate-universe picture without affecting the dynamics of the perturbations.) 

Our work has opened up many questions, a few of which we mention here. We have looked at non-linear sigma models which encompass a variety of models but we believe it is important to take this work and apply the methods and results to several examples. In particular, models with non-minimal couplings are mapped from the Jordan frame where the non-minimal coupling is explicit, to the Einstein frame where minimal coupling is restored but introducing a coupling metric in the field space. In that respect, our study established the separate-universe picture for models with non-minimal coupling in the Einstein frame. Yet, it remains to establish it in the Jordan frame and to build the map of the separate-universe approximation from one frame to another. This is important from a formal perspective to assess the equivalence between the two frames after implementing approximation schemes (see for instance \cite{White_2012,White_2013}), and in particular regarding the conditions of validity of the separate-universe picture which might differ from one frame to another. This is interesting from a practical perspective too since one does not know a priori in which frame the separate-universe approximation is the easiest to handle.

Another important aspect is that in the current stochastic inflation models, short-wavelength modes are assumed to evolve linearly. This is only true at leading order in perturbation theory and in principle, one could include non-linear evolution of the short-wavelength modes. In order to do this we would need to take this work and extend it to the next order in perturbation theory, \ie to the third order. Such studies have been mostly done in the Lagrangian framework \cite{Sasaki_1996,KarimAMalik_2004,Gong_2011,Gao_2012,Tzavara:2013wxa} or using a geometrical approach \cite{Langlois:2005ii,Finelli_2006,Langlois:2006vv,Langlois:2010vx}, but not in the Hamiltonian framework (see however \cite{Dom_nech_2018,Braglia:2024zsl}). Conducting such an analysis would also give insight on the contribution of loop corrections to the power spectrum of the curvature perturbation \cite{Weinberg_2005,Senatore:2009cf,Assassi:2012et,Kristiano:2022zpn,Kristiano:2022maq,Kristiano:2023scm,Kristiano:2024vst,Fumagalli:2023hpa,Fumagalli:2024jzz,Ballesteros:2024zdp,Garcia-Saenz:2025jis} since such a computation would be doable in a gauge-invariant manner. Loops may also play an important role in the stochastic formalism because of their potential secular growth \cite{PhysRevD.99.086009}. This has been recently reexamined in the context of scalar field theories in de Sitter spaces \cite{gorbenko2019lambdaphi4ds,Cespedes:2023aal,Palma:2023uwo,Palma:2025oux} with interesting connections with stochastic inflation. These studies should be generalized to the case of cosmological perturbations (meaning derivative couplings and non-de Sitter background spacetimes), and extending the separate-universe approach to higher orders would provide some insights in that respect.

Finally one could question the coarse-graining scale chosen in our analysis. Usually, one chooses to compare modes to the comoving horizon in order to categorize them into short-wavelength and long-wavelength modes. However as we have concluded, in order to have a matching of the separate-universe approach to cosmological perturbation theory the typical scales studied have to be large compared to the inverse of the effective masses of the fields as well. One could thus ask why this is not the favoured scale at which to coarse grain our fields in the first place, and if so, what is the impact of such a choice on the phenomenological predictions of stochastic inflation. These questions will also be studied in future works. 

\acknowledgments
We would like to thank Lucas Pinol and Lucien Wasquez for very helpful discussions, as well as David Kaiser for several key insights.

\appendix
\addtocontents{toc}{\protect\setcounter{tocdepth}{1}}
\section{Linear action}
\label{app:linactionnaive}
By noticing that $\mathcal{D}^{(0)}_i=0$ and $N^i(\tau)=0$, the first order action reads
\bea
	S^{(1)}=\ds\int\dd \tau\int\dd^3x\left(\pi_I\dot{\delta\phi^I}+\delta\pi_I\dot\phi^I+\frac{1}{2}v^{1/3}\theta\tilde\gamma^{ij}\dot{\delta\gamma_{ij}}+\frac{2\dot{v}}{3v^{1/3}}\tilde\gamma_{ij}\delta\pi^{ij}-\delta N\,\mathcal{C}^{(0)}-N\mathcal{C}^{(1)}\right). \nonumber \\
\eea
Since contracting with $\tilde\gamma_{ij}$ projects out the anisotropic part of the gravitational degrees of freedom, we arrive at
\bea
	S^{(1)}=\ds\int\dd \tau\int\dd^3x\left(\pi_I\dot{\delta\phi^I}+\delta\pi_I\dot\phi^I+\frac{\sqrt{3}}{2}v^{1/3}\theta\,\dot{\delta\gamma_1}+\frac{2\dot{v}}{\sqrt{3}v^{1/3}}\delta\pi_1-\delta N\,\mathcal{C}^{(0)}-N\mathcal{C}^{(1)}\right). \nonumber \\
\eea
Performing an integration by parts on $\pi_I\dot{\delta\phi^I}+\frac{\sqrt{3}}{2}v^{1/3}\theta\dot{\delta\gamma_1}$, the above is further rewritten as 
\bea
	S^{(1)}&=&\ds\int\dd \tau\int\dd^3x\left[-\dot{\pi_I}\,{\delta\phi^I}+\dot\phi^I\,\delta\pi_I-\frac{\sqrt{3}}{2}v^{1/3}\left(\frac{\dot{v}}{3v}\theta+\dot{\theta}\right)\,{\delta\gamma_1}+\frac{2\dot{v}}{\sqrt{3}v^{1/3}}\,\delta\pi_1\right. \nonumber \\
	&&\left.-\delta N\,\mathcal{C}^{(0)}-N\mathcal{C}^{(1)}\right], \label{eq:S1}
\eea
where we suppose vanishing perturbations at the initial and final times. The constraint $\mathcal{C}^{(0)}$ is given in \Eq{eq:C0}. The scalar constraint at first order $\mathcal{C}^{(1)}$ receives a contribution from the gravitational sector, $\mathcal{C}^{(1)\,G}$, and the matter sector, $\mathcal{C}^{(1)\,\phi}$. They read
\bea
%	\mathcal{C}^{(1)\,G}&=&-\sqrt{3}{v^{1/3}}\left(\frac{1}{4v^2}G^{IJ}\pi_I\pi_J+\frac{1}{2}V-\frac{\theta^2}{8\Mp^2}\right)\delta\gamma_1(\tau,\vec{x})-\frac{\sqrt{3}}{\Mp^2}v^{2/3}\theta\,\delta\pi_1(\tau,\vec{x}) \nonumber \\
%	&&+\frac{\Mp^2}{v^{2/3}}\tilde\gamma^{ij}\partial_i\partial_j\left[\delta\gamma_1(\tau,\vec{x})-\frac{1}{\sqrt{2}}\delta\gamma_2(\tau,\vec{x})\right]. \label{eq:C1G} \\
	\mathcal{C}^{(1)\,G}&=&-\frac{\sqrt{3}}{2}{v^{1/3}}\left(\rho-\frac{\theta^2}{4\Mp^2}\right)\delta\gamma_1(\tau,\vec{x})-\frac{\sqrt{3}}{\Mp^2}v^{2/3}\theta\,\delta\pi_1(\tau,\vec{x}) \nonumber \\
	&&+\frac{\Mp^2}{v^{2/3}}\partial^2\left[\delta\gamma_1(\tau,\vec{x})-\frac{1}{\sqrt{2}}\delta\gamma_2(\tau,\vec{x})\right]. \label{eq:C1G}
\eea
and
\bea
%	\mathcal{C}^{(1)\,\phi}&=&\left(\frac{1}{2v}G^{KL}_{\,\,,I}\pi_K\pi_L+vV_{,I}\right)\delta\phi^I(\tau,\vec{x})+\frac{1}{v}G^{KI}\pi_K\delta\pi_I(\tau,\vec{x}), \label{eq:C1phi}
	\mathcal{C}^{(1)\,\phi}&=&v\rho_{,I}\,\delta\phi^I(\tau,\vec{x})+\frac{1}{v}G^{KI}\pi_K\delta\pi_I(\tau,\vec{x}), \label{eq:C1phi}
\eea
where we introduce the shorthand notation $\rho_{,I}=G^{KL}_{\,\,,I}\pi_I\pi_J/(2v^2)+V_{,I}$. Since $\mathcal{C}^{(1)}$ multiplies $N(\tau)$ which is homogeneous, the Laplacian term in the last line of $\mathcal{C}^{(1)\,G}$ will have a vanishing contribution to the linear action after integration over the whole space. This is easily shown using integration by parts and assuming vanishing perturbations at spatial infinity. Hence, we will drop this term in $S^{(1)}$ which receives no contribution from the anisotropic degree of freedom.

Let us explictly show that $S^{(1)}=0$ providing that the homogeneous and isotropic degrees of freedom are solutions of the background evolution. First, this imposes that $\mathcal{C}^{(0)}=0$ and the term $\delta N \mathcal{C}^{(0)}$ can be dropped from \Eq{eq:S1}. The constraint $\mathcal{C}^{(0)}=0$ can further be used to simplify the gravitational part of $\mathcal{C}^{(1)}$ which is expressed as
\bea
	\mathcal{C}^{(1)\,G}=-\frac{v^{1/3}}{\sqrt{3}}\left(\frac{1}{v^2}G^{IJ}\pi_I\pi_J-V\right)\delta\gamma_1(\tau,\vec{x})-\frac{\sqrt{3}}{\Mp^2}v^{2/3}\theta\,\delta\pi_1(\tau,\vec{x}).
\eea
It is instructive to note that $\frac{1}{v^2}G^{IJ}\pi_I\pi_J-V=(\rho+3p)/2$. As a result, the first order action is expressed as
\bea \label{first order action}
	S^{(1)}&=&\ds\int\dd \tau\int\dd^3x\left\{-\left(\dot{\pi}_I+Nv\rho_{,I}\right)\delta\phi^I+\left(\dot\phi^I-\frac{N}{v}G^{KI}\pi_K\right)\delta\pi_I\right. \\
	&&\left.-\left[\frac{\sqrt{3}}{2}v^{1/3}\left(\frac{\dot{v}}{3v}\theta+\dot{\theta}\right)-\frac{Nv^{1/3}}{2\sqrt{3}}\left(\rho+3p\right)\right]\delta\gamma_1+\left[\frac{2\dot{v}}{\sqrt{3}v^{1/3}}+\frac{\sqrt{3}N}{\Mp^2}v^{2/3}\theta\right]\delta\pi_1\right\}. \nonumber 
\eea
Finally, given that the equations of motion hold for the background variables, \ie Eqs. (\ref{eq:dotphi}), (\ref{eq:dotpi}), (\ref{eq:dotgamma}) \& (\ref{eq:dotpig}), it shows that $S^{(1)}$ is vanishing, a well-known fact indeed.

\section{Constraints}\label{app:constraints}

Let us split the scalar and diffeomorphism constraints into a gravitational contribution and a scalar field contribution, as expressed in Sec. \ref{ssec:hamiltonian description}. We will go one step further and separate the two scalar contributions into a kinetic and a potential part: $\mathcal C^G = \mathcal T + \mathcal W $ and  $\mathcal C^\phi = T + W $, with 
\bea
    \mathcal T &=& \frac 2 {\Mp^2 \sqrt \gamma} \left(\pi\stsr\pi\ustsr - \frac 1 2 \pi^2\right) \, ,\\
    \mathcal W &=& - \frac {\Mp^2\sqrt \gamma} 2 R^{(3)} \, , \\
    T &=& \frac{1}{2 \sqrt \gamma} G\uTsr \pi_I\pi_J \, , \label{eq:Tfull}\\
    W &=& \frac { \sqrt \gamma }{2}\gamma^{ij}G\Tsr  \partial_i\phi^I\partial_j\phi^J + \sqrt{\gamma}V \, , \label{eq:Wfull} 
\eea
and
\bea
    \mathcal D_i^G &=& -2\partial_m(\gamma\stsr\pi^{jm}) + \pi^{mn}\partial_i\gamma_{mn}\, ,\\
    \mathcal D_i^\phi &=& \pi_I\partial_i\phi^I \, . 
\eea

\subsubsection*{Gravitational sector}
All gravitational contributions have already been treated in this exact manner up to second order in perturbation theory in App. B of \cite{Artigas_2022}. We will only reproduce the final results here. At the background level we have:
\bea
    \mathcal{D}_i^{(0)\,G} &=& 0 ,\\
	\mathcal{T}^{(0)}&=&\frac{-3}{4\Mp^2}v\theta^2, \\
	\mathcal{W}^{(0)}&=&0.
\eea
At first order, one derives:
\bea
    \mathcal{D}_i^{(1)\,G}&=&ik_i\left(\pi_\phi\delta\phi+\pi^{mn}\delta\gamma_{mn}\right)-2ik_m\left(\pi^{jm}\delta\gamma_{ij}+\gamma_{ij}\delta\pi^{jm}\right) ,\\
    \mathcal{T}^{(1)}&=&\frac{-\sqrt{3}}{\Mp^2}v^{1/3}\theta\left(\frac{\theta}{8}\delta\gamma_1+v^{1/3}\delta\pi_1\right) ,\\
    \mathcal{W}^{(1)}&=&-\frac{\Mp^2}{\sqrt{3}}\left(\frac{k^2}{v^{1/3}}\right)\left(\delta\gamma_1-\frac{1}{\sqrt{2}}\delta\gamma_2\right) .
\eea
Finally at second order, discarding total derivative terms that appear along the way, we obtain:
\bea
    \mathcal{T}^{(2)}&=&\frac{v^{1/3}}{\Mp^2}\left(-\left|\delta\pi_1\right|^2+2\left|\delta\pi_2\right|^2\right)+\frac{\theta}{2\Mp^2}\left[-\left(\delta\gamma_1\right)\left(\delta\pi_1^\star\right)+2\left(\delta\gamma_2\right)\left(\delta\pi_2^\star\right)\right] \nonumber \\
	&&+\frac{1}{32\Mp^2}\frac{\theta^2}{v^{1/3}}\left(\left|\delta\gamma_1\right|^2+10\left|\delta\gamma_2\right|^2\right)\, , \\
    \mathcal{W}^{(2)}&=&\frac{\Mp^2}{24v}k^{2}\left[-2\left|\delta\gamma_1\right|^2-\left|\delta\gamma_2\right|^2+{10}{\sqrt{2}}\left(\delta\gamma_1\right)\left(\delta\gamma_2^\star\right)-8{\sqrt{2}}\left(\delta\gamma^\star_1\right)\left(\delta\gamma_2\right)\right].
\eea
The total derivative terms that have been discarded but need to be including if one wishes to look at the next order. We leave this for future works. We have no use for the second order diffeomorphism constraint. Let us note finally that this perturbative expansion has been done in Fourier space whereas equations for the first order constraint in Sec. \ref{sssec:cosmopert} are given in real space. The change between both is minimal and one only needs to be careful not to double count some contributions while performing the inverse Fourier transform. 

\subsubsection*{Scalar-fields sector}
Let us now treat the scalar field contributions. At the background level we have:
\bea
    D_i^{(0)\, \phi} &=& 0 \,,\\
    T^{(0)} &=& \frac{1}{2 v} G\uTsr \bar{\pi}_I\bar{\pi}_J\,, \\
    W^{(0)} &=& v^2 V .
\eea
Linearising at first order and dropping off the bars indicating background quantities, we get:
\bea
    \mathcal{D}_i^{(1)\, \phi} &=& k_i\pi_I \delta\phi^I,\\
    T^{(0)} &=& -\frac{\delta\gamma}{2\gamma}T^{(0)}+\frac{1}{v}G\uTsr \pi_I\,\delta\pi_J, + \frac{1}{v}G\uTsr_{,K}\pi_I\pi_J\,\delta\phi^K  \\
    W^{(1)}&=&\frac{\delta\gamma}{2\gamma}W^{(0)}+vV_{,K}\,\delta\phi^K\,,
\eea
where $\delta\gamma = \gamma\,\gamma^{ij}\,\delta\gamma_{ij}=\sqrt{3}v^{4/3}\,\delta\gamma_1$, leading to
\bea
	T^{(1)}&=&-\frac{\sqrt{3}}{4}\frac{\pi_\phi^2}{v^{5/3}}\,\delta\gamma_1+\frac{1}{v}G\uTsr \pi_I\,\delta\pi_J + \frac{1}{v}G\uTsr_{,K}\pi_I\pi_J\,\delta\phi^K  , \\
	W^{(1)}&=&\frac{\sqrt{3}}{2}v^{1/3}V\,\delta\gamma_1+vV_{,K}\,\delta\phi^K.
\eea	
Continuing the Taylor expansion of Eqs. (\ref{eq:Tfull}) \& (\ref{eq:Wfull}) at the second order in Fourier space, we get:
\bea
    T^{(2)}&=&\frac{1}{4v}G\uTsr\left(\delta\pi_I\delta\pi_J^\star+ \mathrm{c.c.}\right) + \frac{v}{4}\left(\frac{k^2}{v^{2/3}}G_{IJ} + \frac{1}{2v^2}G^{KL}{}_{,IJ}\pi_K\pi_K\right)\left(\delta\phi^I\delta\phi^{J\,\star}+ \mathrm{c.c.}\right) \label{eq:T2} \nonumber \\
	&&+ \frac{1}{2v}G\uTsr{}_{,K}\pi_J\left(\delta\phi^K\delta\pi_I^\star + \mathrm{c.c.}\right)  + \frac{\sqrt{3}}{8 v^{5/3}}G^{KL}{}
    _{,I}\pi_K\pi_L \left(\delta\phi^I\delta\gamma_1^* + \mathrm{c.c.}\right) \\
	W^{(2)}&=& \frac{\sqrt{3}v^{1/3}}{4}\left[\left(V_{,I}\delta\phi^I +\frac{1}{v^2}G\uTsr\pi_J\delta\pi_I\right)\delta\gamma_1^\star+ \mathrm{c.c.}\right] + \frac{v}{4}v_{,IJ} \left(\delta\phi^I\delta\phi^{J\,\star}+ \mathrm{c.c.}\right) \, . \label{eq:W2}
\eea
Let us note that the decomposition we have done here extends what was done in \cite{Artigas_2022} to non linear sigma models. However Eqs. (\ref{eq:T2}) \& (\ref{eq:W2}) need to be combined in order to identify Eqs. (\ref{eq:C2phi}) \& (\ref{eq:C2V}).

\section{Covariant variables}
\label{app:covQP}
Covariant variables in the field space are defined following \cite{Gong_2011,Pinol:2020cdp}. This is done considering that two neighboring points in the field space, say $\phi^I(\lambda=0)=\phi^I(\tau)$ and $\phi^I(\lambda=1)=\phi^I(\tau)+\delta\phi^I(\tau,\vec{x})$, are connected by a unique geodesic which is parametrized by the parameter $\lambda$. Similarly, the associated momentum is given by covectors along the geodesic \ie $\pi_I(\lambda=0)=\pi_I(\tau)$ and $\pi_I(\lambda=1)=\pi_I(\tau)+\delta\pi_I(\tau,\vec{x})$. The variables $Q^I\equiv\left.D_\lambda\phi^I(\lambda)\right|_{\lambda=0}$ and $P_I\equiv\left.D_\lambda\pi_I(\lambda)\right|_{\lambda=0}$, where covariant derivatives are evaluated at the origin of the geodesic, transform covariantly by construction and capture deviations from the purely homogeneous and isotropic background. Here $\lambda=1$ should be understood as a formal parameter controlling the expansion that we finally set equal to one. 

Since $\phi^I(\lambda=1)=\phi^I(\tau)+\delta\phi^I(\tau,\vec{x})$ and $\delta\phi^I$ is assumed to be small, one can expand around $\lambda=0$ to get
\bea
	\delta\phi^I(\tau,\vec{x})=\lambda\left.\frac{\dd \phi^I}{\dd\lambda}\right|_{\lambda=0}+\frac{\lambda^2}{2}\left.\frac{\dd^2 \phi^I}{\dd\lambda^2}\right|_{\lambda=0},
\eea
where the expansion is truncated at second order. The linear term is given by $Q^I$ and the quadratic term is obtained noticing that $\phi^I$ verifies the geodesic equation, \ie $D_\lambda^2 \phi^I(\lambda)=0$ where
\bea
    D_\lambda^2 \phi^I(\lambda) = \frac{\dd^2\phi^I}{\dd \lambda^2} + \Gamma^I_{JK}\frac{\dd \phi^J}{\dd \lambda}\frac{\dd \phi^K}{\dd \lambda}. \label{geodesic}
\eea
This leads to $\left.\frac{\dd^2 \phi^I}{\dd\lambda^2}\right|_{\lambda=0}=-\Gamma^I_{KL}Q^KQ^L$. A similar expansion is performed for the momentum variables
\bea
	\delta\pi_I(\tau,\vec{x})=\lambda\left.\frac{\dd \pi_I}{\dd\lambda}\right|_{\lambda=0}+\frac{\lambda^2}{2}\left.\frac{\dd^2 \pi_I}{\dd\lambda^2}\right|_{\lambda=0}.
\eea
One can further parallel transport $P_I$ along the geodesic to fix the coefficients of the expansion, \ie $D_\lambda P_I=0$, which boils down to
\bea
    \frac{\dd\pi_I}{\dd\lambda}\Big|_{\lambda = 0} &=& P_I + \Gamma^K_{IJ}\pi_KQ^J \\
    \frac{\dd^2\pi_I}{\dd\lambda^2}\Big|_{\lambda = 0} &=& 2\Gamma^K_{IJ}Q^JP_K + \left( \Gamma^S_{IJ,K}-\Gamma^S_{JR}\Gamma^R_{IK} + \Gamma^R_{IJ}\Gamma^S_{RK}\right)\pi_sQ^JQ^K.
\eea
As a result of the process described above, one arrives at
\bea
	\delta\phi^I &=& Q^I -\frac 1 2 \Gamma^I_{LK} Q^L Q^K+\mathcal{O}(\lambda^3) \label{mapping phi}, \\
            \delta\pi_I &=& P_I + \Gamma^K\Tsr \pi_K  Q^J + \Gamma^K\Tsr P_K Q^J \label{mapping pi} \\ \nonumber
            &&+ \frac 1 2 \left(\Gamma^S_{IJ,K} - \Gamma^S_{IR}\Gamma^R_{JK} + \Gamma^R\Tsr\Gamma^S_{RK}\right)\pi_S Q^J Q^K+\mathcal{O}(\lambda^3),
\eea
which clearly shows that the naive variables defined as finite differences are non-linearly related to manifestly covariant variables. It is worth stressing that there exists an entire family of covariant momentum perturbations which is
\bea
	P_I^{(\alpha)}=P_I-\frac{\alpha}{2} {R_{IJK}}^S\,\pi_SQ^JQ^K, \label{eq:PalphaP}
\eea
where $R_{IJKL}$ is the Riemann curvature tensor associated to the metric in the field space and $\alpha$ a real number parametrizing the family of momentum variables. This ambiguity in defining the covariant momentum owes to the fact that considering the parallel transport of $P_I$ is one choice among many others which guarantee covariance (see App. B of \cite{Pinol:2020cdp}) and we note that $P_I=P^{(\alpha=0)}_I$. As a result, the naive momentum reads
\bea
	 \delta\pi_I &=& P^{(\alpha)}_I + \Gamma^K\Tsr \pi_K  Q^J + \Gamma^K\Tsr P^{(\alpha)}_K Q^J \label{mapping pi alpha} \\ \nonumber
            &&+ \frac 1 2 \left(\Gamma^S_{IJ,K} - \Gamma^S_{IR}\Gamma^R_{JK} + \Gamma^R\Tsr\Gamma^S_{RK}+\alpha {R_{IJK}}^S\right)\pi_S Q^J Q^K+\mathcal{O}(\lambda^3), \label{eq:pialphaP}
\eea
where again, we truncate at the quadratic order.

The above can be inverted and truncating the inverted relation at second order, it gives 
\bea
	Q^I&=&\delta\phi^I+\frac{1}{2}\Gamma^I_{KL}\delta\phi^K\delta\phi^L+\mathcal{O}(\lambda^3), \label{eq:Qphipi}\\
	P^{(\alpha)}_I&=&\delta\pi_I-\Gamma^K_{IJ}\pi_K\delta\phi^J-\Gamma^K_{IJ}\delta\pi_K\delta\phi^J \nonumber \\
	&&-\frac{1}{2}\left(\Gamma^S_{IJ,K}- \Gamma^R\Tsr\Gamma^S_{RK}+\alpha {R_{IJK}}^S\right)\pi_S\delta\phi^J\delta\phi^K+\mathcal{O}(\lambda^3). \label{eq:Pphipi}
\eea
One should now check that it corresponds to a canonical transformation up to quadratic order in $\lambda$. In fact, one can safely compute the Poisson brackets up to order $\lambda^3$ starting from Eqs. (\ref{eq:Qphipi}) \& (\ref{eq:Pphipi}) since the lowest order in $\delta\phi^I$ and $\delta\pi_I$ is $\mathcal{O}(\lambda)$. Starting from the expressions of $Q^I$ and $P^{(\alpha)}_I$ and using the Poisson brackets of the naive variables, it is straightforward to show that 
\beq
\left\{Q^I,Q_J\right\}=\mathcal{O}(\lambda^4)
\eeq and that 
\beq
\left\{Q^I,P^{(\alpha)}_J\right\}=\lambda^2\delta^I_J+\mathcal{O}(\lambda^4),
\eeq
where here we explicitly mention the order in $\lambda$ for clarity. For the last bracket, it is useful to introduce ${\Lambda_{IJK}}^S=\Gamma^S_{IJ,K}- \Gamma^R\Tsr\Gamma^S_{RK}$ as well as $T_{(IJ)}=T_{IJ}+T_{JI}$ and $T_{[IJ]}=T_{IJ}-T_{JI}$ in order to lighten its expression. With such notations, we arrived at
\bea
	\left\{P^{(\alpha)}_I,P^{(\alpha)}_J\right\}&=&-\frac{\lambda^3}{2}\left\{{\Lambda_{I(JK)}}^S-{\Lambda_{J(IK)}}^S+2\Gamma^R_{[IK}\Gamma^S_{J]R}+\alpha\left[{R_{I(JK)}}^S-{R_{J(IK)}}^S\right]\right\}\pi_S\delta\phi^K \nonumber \\
	&&+\mathcal{O}(\lambda^4). 
\eea
An explicit calculation shows that ${\Lambda_{I(JK)}}^S-{\Lambda_{J(IK)}}^S+2\Gamma^R_{[IK}\Gamma^S_{J]R}=-{R_{IJK}}^L$ and the last Poisson bracket boils down to
\bea
    \left\{P^{(\alpha)}_I,P^{(\alpha)}_J\right\}&=&\frac{\lambda^3}{2}\left[\left(1-2\alpha\right){R_{IJK}}^S+\alpha {R_{KIJ}}^S+\alpha {R_{JKI}}^S\right]\pi_S\delta\phi^K+\mathcal{O}(\lambda^4). 
\eea
Note that from the antisymmetry of the Riemann curvature tensor, one obtains $\left\{P^{(\alpha)}_I,P^{(\alpha)}_I\right\}=0$ irrespectively of the value of $\alpha$, as it should since the Poisson bracket is antisymmetric. However, $\left\{P^{(\alpha)}_I,P^{(\alpha)}_{J\neq I}\right\}$ is not zero at order $\mathcal{O}(\lambda^3)$ unless $\alpha=1/3$. Indeed, fixing $\alpha$ to this value leads to
\bea
	\left\{P^{(\alpha)}_I,P^{(\alpha)}_J\right\}&=&\frac{\lambda^3}{6}\left[{R_{IJK}}^S+ {R_{KIJ}}^S+ {R_{JKI}}^S\right]\pi_S\delta\phi^K+\mathcal{O}(\lambda^4),
\eea
hence vanishing for all $I$ and $J$ as a result of the algebraic symmetries of the Riemann curvature tensor. This shows that for defining a proper canonical transformation up to order $\mathcal{O}(\lambda^3)$, one needs to select the momentum variables setting $\alpha=1/3$. 

Interestingly, it was shown in \cite{Pinol:2020cdp} that this specific value of $\alpha$ plays a peculiar role in the cubic action since it exactly cancels terms with covariant derivatives of $Q^I$ in the cubic 'Hamiltonian'.\footnote{The ambiguity is captured by a parameter called $\kappa$ in \cite{Pinol:2020cdp} and which relates to our ambiguity-parameter via $\alpha=2\kappa$. This can be obtained by comparing Eq. (\ref{eq:PalphaP}) and Eq. (B.4) of \cite{Pinol:2020cdp}. Note the different ordering of indices of the Riemann tensor which leads to an additional minus sign from the algebraic (anti-)symmetries of $R_{IJKL}$.} From the viewpoint of canonical transformation, this should not come as a surprise. Indeed, letting $\alpha$ unfixed does not yield a canonical transformation at $\mathcal{O}(\lambda^3)$ and this is why the cubic action cannot be expressed in an Hamiltonian form because of remaining terms with time-derivatives of the phase-space variables. On the contrary, fixing $\alpha=1/3$ ensures a canonical transformation, hence a proper Hamiltonian action without time-derivative of the phase-space variables at cubic order.

\section{Generating function}
\label{app:genfunc}
In order to implement the canonical transformation described in App. \ref{app:covQP} in the Hamiltonian action, we resort to generating functions which we present below. Since a proper canonical transformation is obtained for $\alpha=1/3$, one should expect that the generating function at cubic order can be found for such a value only. Here and as a way to double-check our previous finding, we will leave $\alpha$ unfixed and show that indeed, the generating function only exists for the peculiar value $\alpha=1/3$.

We introduce a generating function of type 2, \ie $G=-Q^IP^{(\alpha)}_I+F(\delta\phi^I,P^{(\alpha)}_J,t)$ where $F$ can explicitly depend on time and can be built such that 
 \bea
 	\frac{\partial F}{\partial \delta\phi^I}=\delta\pi_I & \,\,\,\mathrm{and}\,\,\,& \frac{\partial F}{\partial P^{(\alpha)}_I}=Q^I. \label{eq:dFdP}
\eea
In order to generate the quadratic relation between the old and the new phase-space coordinates given in Eqs. (\ref{mapping phi}) \& (\ref{mapping pi alpha}) and requiring Eq. (\ref{eq:dFdP}), the function $F$ should be a cubic function of $(\delta\phi^I,P^{(\alpha)}_J)$ and its lowest order should be quadratic. Hence, the function $F$ reads 
\beq
	F(\delta\phi^I,P^{(\alpha)}_J,t)=P^{(\alpha)}_I\delta\phi^I+\frac{1}{2}\Gamma^K_{IJ}\pi_K\delta\phi^I\delta\phi^J+\frac{1}{2}\Gamma^K_{IJ}P^{(\alpha)}_K\delta\phi^I\delta\phi^J+{T^{(\alpha)}_{IJK}}^S\pi_S\delta\phi^I\delta\phi^J\delta\phi^K,
\eeq
where ${T^{(\alpha)}_{IJK}}^S$ is constrained by $\partial F/\partial\delta\phi^I=\delta\pi_I$ which leads to
\bea
	\left({T^{(\alpha)}_{IKJ}}^S+{T^{(\alpha)}_{JKI}}^S+{T^{(\alpha)}_{JIK}}^S\right)\pi_S\delta\phi^J\delta\phi^K= \frac 1 2 \left(\Gamma^S_{IJ,K}  + \Gamma^R\Tsr\Gamma^S_{RK}+\alpha {R_{IJK}}^S\right)\pi_S \delta\phi^J\delta\phi^K. \nonumber \\ \label{eq:Talpha}
\eea
It is worth noting that the explicit dependence of $F$ with $\alpha$ arises at the cubic order and not at lower orders, through the `coefficients' ${T^{(\alpha)}_{IKJ}}^S$. Note also that ${T^{(\alpha)}_{IKJ}}^S$ is expected not to be a tensor for $\delta\phi^I$ does not transform covariantly and this applies to any term in the generating function since it mixes non-covariant variables with covariant ones. 

We are interested in deriving ${T^{(\alpha)}_{IJK}}^S\,\pi_S\delta\phi^I\delta\phi^J\delta\phi^K$. Hence it may appears as sufficient to consider ${T^{(\alpha)}_{IJK}}^S$ to be symmetric under $I\leftrightarrow J$, $I\leftrightarrow K$, and $J\leftrightarrow K$ permutations since only these symmetric contributions will remain in the generating function. However, $\Gamma^S_{IJ,K}  + \Gamma^R\Tsr\Gamma^S_{RK}+\alpha {R_{IJK}}^S$ which enters the right-hand-side of the above equality is not symmetric under those permutations (at least for an arbitrary value of $\alpha$). In particular, it is important to keep track of the antisymmetric part of ${T^{(\alpha)}_{IJK}}^S$ by permuting $I$ with either $J$ or $K$. To this end, we adopt a more general perspective. By simple renaming of summed indices on both sides of Eq. (\ref{eq:Talpha}) which only exploits the symmetry under $K\leftrightarrow J$, the equality holds providing that
\bea
	{T^{(\alpha)}_{IJK}}^S+{T^{(\alpha)}_{KIJ}}^S+{T^{(\alpha)}_{JKI}}^S=\frac 1 4 \left[\Gamma^S_{I(J,K)}  + \Gamma^R_{I(J}\Gamma^S_{RK)}+{\alpha} {R_{I(JK)}}^S+{A_{JKI}}^S\right],
\eea
where we now recognize cyclic permutations of $T^{(\alpha)}$ and where ${A_{JKI}}^S$ can be safely added providing it is antisymmetric under $K\leftrightarrow J$ permutation (for instance, it could be  $\beta{R_{JKI}}^S$ with $\beta$ any real number since ${R_{JKI}}^S\,\pi_S\delta\phi^J\delta\phi^K=0$ by antisymmetry of the Riemann tensor permuting $K$ with $J$). An explicit calculation of the right-hand-side yields
\bea
	{T^{(\alpha)}_{IJK}}^S+{T^{(\alpha)}_{KIJ}}^S+{T^{(\alpha)}_{JKI}}^S=\frac 1 4 \left[(1-\alpha)({\widetilde{\Lambda}_{IJK}}^S+{\widetilde{\Lambda}_{KIJ}}^S)+2\alpha{\widetilde{\Lambda}_{JKI}}^S+{A_{JKI}}^S\right], \label{eq:TtoLambda}
\eea
where ${\widetilde{\Lambda}_{IJK}}^S=\Gamma^S_{IJ,K}  + \Gamma^R_{IJ}\Gamma^S_{RK}$. The above admits a solution (up to some parts antisymmetric in $K\leftrightarrow J$ which are of no relevance) providing that $\alpha=1/3$ and which is
\bea
	{T^{(\alpha)}_{IJK}}^S=\frac{1}{6}{\widetilde{\Lambda}_{IJK}}^S.
\eea
It is worth stressing that $\alpha=1/3$ is a necessary condition to find a solution for ${T^{(\alpha)}_{IJK}}^S$ since otherwise, the right-hand-side of Eq. (\ref{eq:TtoLambda}) cannot be expressed using cyclic permutations of an object with three lower indices. This is no surprise that the value $1/3$ appears. Indeed, the generating function has to generate a canonical transformation and it was shown previously that $\alpha=1/3$ is mandatory for the Poisson bracket to be preserved. Hence the construction of the generating function presented above provides us with a nice double-check of the need for fixing the ambiguity parameter to $1/3$.

\section{Action with covariant variables}
\label{app:actionQP}
In this appendix, we show that it is sufficient to implement the canonical transformation relating the covariant variables to the naive one at the linear order in the second-order action. To this end, we implement the canonical transformation up to quadratic order and keeping the linear order in the expanded action using generating function techniques (see Apps. \ref{app:covQP} \& \ref{app:genfunc}). Here, we will omit the $(\alpha)$ superscript since the only viable momentum variable is for $\alpha=1/3$.

Generating function techniques exploit the fact that two actions have the same extremum (hence yields the same equations of motion) providing that their Lagrangian densities differ by a total time-derivative. Since the action with the new variables should have the same extremum as the action derived with the old (and naive) variables, one obtains
\bea
	\ds\int\dd^3x\left[-\dot{\pi}_I\delta\phi^I+\dot{\phi}^I\delta\pi_I+\dot{\delta\phi^I}\delta\pi_I-\mathcal{H}\left(\delta\phi^I,\delta\pi_I\right)\right]&=&\ds\int\dd^3x\left[-\dot{\pi}_IQ^I+\dot{\phi}^IP_I+\dot{Q^I}P_I\right. \nonumber \\
	&&\left.-\mathcal{K}\left(Q^I,P_I\right)+\frac{\dd G}{\dd\tau}\right], \label{eq:actionF}
\eea
where $\mathcal{H}$ is a shorthand notation for the linear and quadratic expansion of $N\mathcal{C}+N^i\mathcal{D}_i$ in $(\delta\phi^I,\delta\pi_I)$, where $\mathcal{K}$ is the new Hamiltonian made of linear and quadratic orders in $(Q^I,P_I)$, and where $G$ is the function generating the canonical transformation. In the above, the background has been already removed since the background degrees of freedom are not subject to the canonical transformation considered here, hence their Lagrangian density is not modified, \ie $\mathcal{K}^{(0)}=\mathcal{H}^{(0)}$. Similarly, contributions from the gravitational perturbations are unchanged which leads to $\mathcal{K}^{G}=\mathcal{H}^{G}$ (note however that one should keep the coupling between scalar fields perturbations and gravitational perturbations).   

Considering a generating function of type 2 as built in App. \ref{app:genfunc}, we plug $\dd G/\dd\tau = -\dot{Q}^IP_I-Q^I\dot{P}_I + (\partial F/\partial \delta\phi^I)\dot{\delta\phi^I}+(\partial F/\partial P_I)\dot{P_I}+\partial F/\partial \tau$ in Eq. (\ref{eq:actionF}). Assuming Eq. (\ref{eq:dFdP}) holds, it leads to
\bea
	\ds\int\dd^3x\left[-\dot{\pi}_I\delta\phi^I+\dot{\phi}^I\delta\pi_I-\mathcal{H}\left(\delta\phi^I,\delta\pi_I\right)\right]=\ds\int\dd^3x\left[-\dot{\pi}_IQ^I+\dot{\phi}^IP_I-\mathcal{K}\left(Q^I,P_I\right)+\frac{\partial F}{\partial\tau}\right].  \nonumber \\ \label{eq:deltaaction}
\eea
 We note that the mapping being already established, it is in fact sufficient to know $\partial F/\partial \tau$ up to quadratic order in the new variables since our analysis is restricted to that order.

Let us now write the new Hamiltonian as  the sum of a linear contribution and a quadratic one and consider each order.

\subsubsection*{Linear action}\label{Linear action}
 At the linear order, there are no difficulties since we can simply replace each occurence of $(\delta\phi^I,\delta\pi_J)$ in Eq. (\ref{eq:S1}) by their expression as functions of $(Q^I, P_J)$ truncated at linear order, \ie replace $\delta\phi^I=Q^I$ and $\delta\pi_J= P_J + \Gamma^K_{IJ}\pi_KQ^J$. Indeed, $F$ being at least quadratic, it does not give any contribution to the new Hamiltonian at the linear order. Hence one has
\beq
 	\ds\int\dd^3x\left[-\dot{\pi}_I\delta_1\phi^I+\dot{\phi}^I\delta_1\pi_I-\delta N\mathcal{C}^{(0)}-N\mathcal{C}^{(1)}(\delta_1\phi^I,\delta_1\pi_I)\right]=\ds\int\dd^3x\left[-\dot{\pi}_IQ^I+\dot{\phi}^IP_I-\mathcal{K}^{(1)}\right],
\eeq
where $\delta_1\phi^I$ means that its expression as a function of the new variables should be restricted to the linear order, and similarly for $\delta_1\pi_I$.
 
 Performing the replacement yields 
 \bea
 	\mathcal{K}^{(1)}=\delta N\mathcal{C}^{(0)}+N\mathcal{C}^{(1)}-\Gamma^K_{IJ}\pi_K\dot{\phi}^IQ^J,
\eea
where 
\bea
	\mathcal{C}^{(1)\,\phi}(Q^I,P_I)&\rightarrow&\left(v\rho_{,I}+\frac{1}{v}G^{KJ}\Gamma^L_{JI}\pi_K\pi_L\right)Q^I+\frac{1}{v}G^{KI}\pi_KP_I, \label{eq:C1QP}
\eea
and where $\rightarrow$ means that we use Eqs. (\ref{mapping phi}) \& (\ref{mapping pi}) restricted to the linear order in the replacement (note that the gravitational part remains unchanged). Combining every term and using the background equations of motion to set $\mathcal{C}^{(0)}=0$ and to express $\dot{\phi}^I$ as a function of $\pi_I$, one arrives at 
\bea
 	\mathcal{K}^{(1)\,\phi}=Nv\rho_{,I}Q^I+\frac{1}{v}G^{KI}\pi_KP_I, \label{eq:scalaralmostcov}
\eea
and the gravitational part is unchanged. The latter can be further simplified by introducing the expression of $\rho_{,I}$ as a function of derivative of the kinetic energy and of the potential energy, and by further noticing that $G^{KL}_{\,\,\,\,,I}\pi_K\pi_L+2G^{KJ}\Gamma^L_{JI}\pi_K\pi_L$ is vanishing since it is proportional to the covariant derivative of the metric. As a result, contributions from derivative in the field space of the kinetic energy of the background drops out and the linear scalar constraint reduces to
\beq
	\mathcal{C}^{(1)\,\phi}=vV_{;I}\,Q^I+\frac{1}{v}G^{KI}\pi_KP_I, \label{eq:scalarcov}
\eeq
where $V_{,I}$ has been replaced by $V_{;I}$ since the potential is a scalar. As such, the linear constraints are now expressed in a way which is  manifestly covariant. We note that $V_{;I}=\rho_{;I}$ for the metric in the field space has a vanishing covariant derivative and the covariant constraints could have been intuitively inferred replacing standard derivatives by covariant ones.

From this, it is straightforward to show that the linear action admits the very same form as Eq. (\ref{first order action}) where $\delta\phi^I$ and $\delta\pi_I$ are replaced by $Q^I$ and $P_I$ and we can once again use the equations of motion Eqs. (\ref{eq:dotphi}) \& (\ref{eq:dotpi}) to see that this is zero as expected.

\subsubsection*{Quadratic action}
The quadratic Hamiltonian will receive three types of contributions. Indeed starting from Eq. (\ref{eq:deltaaction}) and focusing on terms which are strictly quadratic, one arrives at 
\bea
	\ds\int\dd^3x\,\mathcal{K}^{(2)}&=&\ds\int\dd^3x\left[N\mathcal{C}^{(1)}(\delta_2\phi^I,\delta_2\pi_I)+\dot{\pi}_I\,\delta_2\phi^I-\dot{\phi}^I\,\delta_2\pi_I\right. \nonumber \\ 
	&&\left.+\mathcal{H}^{(2)}(\delta_1\phi^I,\delta_1\pi_I)+\frac{\partial F}{\partial\tau}(\delta_1\phi^I,\delta_1\pi_I)\right]
\eea
where $\mathcal{H}^{(2)}=\delta N\mathcal{C}^{(1)}+\delta N^i\mathcal{D}^{(1)}_i+N\mathcal{C}^{(2)}$ now expressed with the new variables (which is quadratic by restricting the canonical transformation to its linear part), where ${\partial F}/{\partial\tau}$ is by construction quadratic, and where $\delta_2\phi_I$ stands for the contribution of $Q^I$ and $P_I$ to $\delta\phi^I$ which are purely quadratic (and similarly for $\delta_2\pi_I$). After doing the replacement in the first line of the right-hand-side of the above equation, one easily sees that $N\mathcal{C}^{(1)}(\delta_2\phi^I,\delta_2\pi_I)+\dot{\pi}_I\,\delta_2\phi^I-\dot{\phi}^I\,\delta_2\pi_I$ takes the exact same form as Eq. (\ref{first order action}) where $\delta\phi^I$ and $\delta\pi_I$ are replaced by $\delta_2\phi^I$ and $\delta_2\pi_I$. Hence this term is vanishing once the equation of motion of the background are implemented and we are left with 
\bea
	\ds\int\dd^3x\,\mathcal{K}^{(2)}&=&\ds\int\dd^3x\left[\mathcal{H}^{(2)}(\delta_1\phi^I,\delta_1\pi_I)+\frac{\partial F}{\partial\tau}(\delta_1\phi^I,\delta_1\pi_I)\right], \label{eq:appKam}
\eea
where it is explicit that only the linear relation between the naive variables and the covariant is needed. This finalizes to prove that despite the action linear in $(\delta\phi^I,\delta\pi_I)$ has quadratic contributions in the covariant variables, these cancels out once the background equations of motion are implemented. (This statement can be extended to higher-orders as shown latter in this appendix.)

Let us now compute $\mathcal{K}^{(2)}$. First, we perform the replacement $(\delta\phi^I,\delta\pi_i)\to(Q^I,P_I)$ in $\mathcal{H}^{(2)}$. On the one hand, the replacement has to be done in the linear constraints. It boils down to
\bea
\mathcal{D}^{(1)\,\phi}&=&\pi_IQ^I, \label{eq:diffeocov}
\eea
for the linear diffeomorphism constraint which is covariant, and to Eq. (\ref{eq:scalarcov}) for the linear scalar constraints $\mathcal{C}^{(1)\,\phi}$ (while the gravitational parts remain unchanged). 

On the other hand, the replacement is also done in the quadratic constraint. For $\mathcal{C}^{(2)\,\phi}$, we arrived at
\bea
    \frac{1}{4v}G\uTsr\left(\delta\pi_I\delta\pi_J^\star+ \mathrm{c.c.}\right) &=& \frac{1}{4v}G\uTsr\left(P_IP_J^\star+ \mathrm{c.c.}\right) \label{replacement double pi} \\
    &&+ \frac{1}{2v}G\uTsr\Gamma^{L}_{JM}\pi_L \left(P_IQ^{M\star}+ \mathrm{c.c.}\right) \nonumber \\
    && + \frac{1}{4v}G\uTsr\Gamma^R_{IS}\Gamma^{L}_{JM}\pi_R\pi_L \left(Q^SQ^{M\star}+ \mathrm{c.c.}\right), \nonumber\\
    \frac{v}{4}\left(\frac{k^2}{v^{2/3}}G_{IJ} + \rho,_{IJ}\right)\left(\delta\phi^I\delta\phi^{J\,\star}+ \mathrm{c.c.}\right) &=&  \frac{v}{4}\left(\frac{k^2}{v^{2/3}}G_{IJ} + \rho,_{IJ}\right)\left(Q^IQ^{J\,\star}+ \mathrm{c.c.}\right),\\
    \frac{1}{2v}G\uTsr,_K\pi_J\left(\delta\phi^K\delta\pi_I^\star + \mathrm{c.c.}\right) &=& \frac{1}{2v}G\uTsr,_K\pi_J\left(Q^KP_I^\star + \mathrm{c.c.}\right) \\
    && +\frac{1}{2v}G\uTsr,_K\Gamma^{L}_{IM}\pi_J\pi_L \left(Q^K Q^{M\star} + \mathrm{c.c.}\right). \nonumber
\eea
For the coupling with the gravitational degrees of freedom, \ie $\mathcal{C}^{(2)\,V}$, it yields 
\bea
    - \frac{\sqrt{3}v^{1/3}}{4}p_{,I}\left(\delta\phi^I \delta\gamma_1^\star+ \mathrm{c.c.}\right) &=& - \frac{\sqrt{3}v^{1/3}}{4}p_{,I} \left(Q^I \delta\gamma_1^\star+ \mathrm{c.c.}\right),  \\
    - \frac{\sqrt{3}}{4}\frac{1}{v^{5/3}}G\uTsr\pi_J\left(\delta\pi_I \delta\gamma_1^\star + \mathrm{c.c.}\right) &=& - \frac{\sqrt{3}}{4}\frac{1}{v^{5/3}}G\uTsr\pi_J\left(P_I \delta\gamma_1^\star + \mathrm{c.c.}\right) \\
    && - \frac{\sqrt{3}}{4}\frac{1}{v^{5/3}}G\uTsr\Gamma^L_{IK}\pi_L \pi_J (Q^K \delta\gamma_1^\star + \mathrm{c.c.}), \nonumber
\eea
and $\mathcal{C}^{(2)\,G}$ remains unchanged. Let us note here that since $\delta\pi_I$ depends on combinations of both $P_K$ and $Q^K$ terms we will have many new terms coming from these. We can clearly see \eg in Eq. (\ref{replacement double pi}) that one single term in the naive variables Hamiltonian leads to three terms in the new Hamiltonian.

Second we will have quadratic terms coming from the partial time derivative of the generating function $F$, \ie 
\bea
    \frac{\partial F}{\partial \tau} &=& \frac 1 2 \left(\partial_t \Gamma^K_{IJ}\pi_K + \Gamma^K_{IJ}\dot \pi_K\right)Q^IQ^J,
\eea
in which $\delta\phi^I$ has been replaced by its expression as a function of $Q^I$, restricting to quadratic orders in $\partial F/\partial\tau$, hence to the linear order in Eq. (\ref{mapping phi main}). Using the background equations of motion, this contribution reads
\bea
    \frac{\partial F}{\partial \tau} &=& \frac N 2 \left(\frac{1}{v}\Gamma^K_{IJ,L}G^{LS}\pi_K\pi_S - v\Gamma^K_{IJ}\,\rho_{,K}\right)Q^IQ^J.
\eea 
Taking the Fourier transform of the above and recasting the integral over $\mathbb{R}^{3-}$ as an integral over $\mathbb{R}^{3+}$, one arrives at
\bea
	\ds\int\dd^3x\,\frac{\partial F}{\partial \tau}=\ds\int_{\mathbb{R}^{3+}}\dd^3k\left[\frac N 2 \left(\frac{1}{v}\Gamma^K_{IJ,L}G^{LS}\pi_K\pi_S - v\Gamma^K_{IJ}\rho_{,K}\right)\left(Q^IQ^{J^\star}+\mathrm{c.c.}\right)\right].
\eea
Since the generating function do not depend on the gravitational degrees of freedom, it will only contribute to $\mathcal{K}^{(2)\,\phi}$.

We can now put all the contributions together to get
\bea
	\mathcal{K}^{(2)\,\phi}&=&\frac{1}{4v}G\uTsr\left(P_IP_J^\star+ \mathrm{c.c.}\right)+\frac{1}{2v}H^{LI}_M\pi_L\left(P_IQ^{M\star}+ \mathrm{c.c.}\right) \\
	&&+\left\{\frac{v}{4}\left(\frac{k^2}{v^{2/3}}G_{IJ}+\rho_{,IJ}-\Gamma^K_{IJ}\rho_{,K}+\frac{1}{v^2}K^{MN}_{IJ}\pi_M\pi_N\right)\right\}\left(Q^IQ^{J\star}+\mathrm{c.c.}\right), \nonumber 
\eea
where
\bea
    H^{LI}_M&=&G^{IL},_M+G\uTsr\Gamma^{L}_{JM}=-G^{JL}\Gamma^I_{JM},
\eea
and
\bea
	K^{MN}_{IJ}&=&G^{KL}\Gamma^M_{KI}\Gamma^{N}_{LJ}+2{G^{KM}},_I\Gamma^{N}_{KJ}+\Gamma^M_{IJ,L}G^{LN}. 
\eea
We note that the second equality for $H^{LI}_M$ is obtained by use of footnote \ref{ftnt}. Let us now focus on the term multiplying $Q^IQ^J$. The contribution of the potential energy reads $V_{,IJ}-\Gamma^{K}_{IJ}V_{,K}$. Since $V$ is a scalar, one has $V_{,I}=V_{;I}$ which is a vector whose covariant derivative is $V_{;IJ}=\partial_IV_{;J}-\Gamma^K_{IJ}V_{;K}$. Hence one arrives at
\bea
	\rho_{,IJ}-\Gamma^K_{IJ}\rho_{,K}=V_{;IJ}+\frac{1}{2v^2}\left({G^{MN}}_{,IJ}-\Gamma^K_{IJ}{G^{MN}}_{,K}\right)\pi_M\pi_N,
\eea
where the second term comes from the contribution of the kinetic energy.  It is combined with $K^{MN}_{IJ}\pi_M\pi_N/v^2$ and by making use of ${G^{KL}},_I=-G^{KJ}\Gamma^L_{JI}-G^{JL}\Gamma^K_{IJ}$ (see footnote \ref{ftnt}), it boils down to  
\bea
	\rho_{,IJ}-\Gamma^K_{IJ}\rho_{,K}+\frac{1}{v^2}K^{MN}_{IJ}\pi_M\pi_N=V_{;IJ}-\frac{1}{v^2}{{R_I}^{MN}}_J\,\pi_M\pi_N,
\eea 

Similarly for the coupling between the gravitational and scalar fields perturbations, we write
\bea
	\mathcal{K}^{(2)\,V}&=&-\frac{\sqrt{3}}{4}v^{1/3}\left(p_{,I}+\frac{1}{v^2}G^{KJ}\Gamma^L_{KI}\pi_J\pi_L\right)\left(Q^I\delta\gamma_1^\star+\mathrm{c.c.}\right)\nonumber \\
	&&-\frac{\sqrt{3}}{4}\frac{1}{v^{5/3}}G^{IJ}\pi_J\left(P_I\delta\gamma_1^\star+\mathrm{c.c.}\right).
\eea
The first line can be simplified using the same strategy as the one used for the linear scalar constraint. First, one expresses $p_{,I}$ as a function of the kinetic and potential energies. Second, one recognizes the covariant derivative of the metric tensor by combining with $(1/v^2)G^{KJ}\Gamma^L_{KI}\pi_J\pi_L$ (see footnote \ref{ftnt}). Hence, the contribution of the kinetic energy drops out and we are left with $-V_{,I}$ which can be further replaced by its covariant derivative. This eventually gives equation \ref{eq:KVcov} \\

\subsection*{Higher orders}
One first notes that the Hamiltonian action expanded at any order in the naive variable reads
\bea
	S=S^{(0)}+S^{(1)}+\ds\sum_{n=2}^\infty S^{(n)},
\eea
where 
\bea
	S^{(0)}&=&\ds\dd\tau\left[\pi_I\dot{\phi}^I-N\mathcal{C}^{(0)}+\mathrm{Grav.}\right], \\
	S^{(1)}&=&\ds\int\dd\tau\int\dd^3x\left[-\left(\dot{\pi}_I-N\left\{\pi_I,\mathcal{C}^{(0)}\right\}\right)\delta\phi^I+\left(\dot{\phi}^I-N\left\{\phi^I,\mathcal{C}^{(0)}\right\}\right)\delta\pi_I\right. \nonumber \\
	&&\left.-\delta N\,\mathcal{C}^{(0)}+\mathrm{Grav.}\right], \\
	\ds\sum_{n=2}^\infty S^{(n)}&=&\ds\int\dd\tau\int\dd^3x\left\{\delta\pi_I\dot{\delta\phi}^I-\sum_{n=1}\left[\delta N\mathcal{C}^{(n)}(\delta\phi^I,\delta\pi_I)+\delta N^i\mathcal{D}^{(n)}_i(\delta\phi^I,\delta\pi_I)\right]\right. \nonumber \\
	&&\left.-N\sum_{n=2}\mathcal{C}^{(n)}(\delta\phi^I,\delta\pi_I)+\mathrm{Grav.}\right\}.
\eea
Note that we made the dependence with gravitational perturbations implicit in order to lighten the expressions. This is not harmful here since we are considering canonical transformations on the scalar field sector only. In the above, the linear action is written by expressing $\int\dd^3x N\mathcal{C}^{(1)}$ using the Poisson brackets of the background variables and a similar writing exist for the gravitational sector.\footnote{Note that this is possible since in the linear action, total spatial derivatives can be removed from $\int\dd^3x N\mathcal{C}^{(1)}$ using integration by parts.  As a result, gradients of perturbative degrees of freedom yield a vanishing  contribution in $\int\dd^3x N\mathcal{C}^{(1)}$ since they appear as total derivative. Also, this ensures that gravitational degrees of freedom which are anisotropic do not contribute to the linear action, since they only arise through gradients, and which is mandatory to be able to rewrite the isotropic part of $\mathcal{C}^{(1)}$ as Poisson brackets of $\mathcal{C}^{(0)}$.}

Let us now consider the general relation between the naive variables and the covariant ones reading
\bea
	\delta\phi^I&=&\ds\sum_{n=1}^\infty\widetilde{\delta_n\phi}^I(Q^I,P_I), \\
	\delta\pi_I&=&\ds\sum_{n=1}^\infty\widetilde{\delta_n\pi}_I(Q^I,P_I),
\eea
where $\delta_n\phi^I$ means that it contains powers of $(Q^I,P_I)$ to the $n$ only (and similarly for $\delta_n\pi_I$). In what follows, quantities denoted with an overtilda means that they are expressed as functions of the covariant variables. The function of type 2 which generates such a transformation is given by $G=-Q^IP_I+F(\delta\phi^I,P_I,t)$ such that 
\bea
	F(\delta\phi^I,P_I,t)=\sum_{n=2}F_n(\delta\phi^I,P_I,t),
\eea
where $F_n$ contains powers of $(\delta\phi^I,P_I)$ to the $n$ only. Note that the sum in $F$ has to start at $n=2$ in the above since one has $\partial F/\partial\delta\phi_I=\delta\pi_I$ and $\partial F/\partial P_I=Q^I$. It is also worth stressing that once expressed using the new variables, one has
\bea
	F(\delta\phi^I,P_I,t)=\sum_{n=2}\widetilde{F}_n(Q^I,P_I,t),
\eea
where importantly, the sum also starts at $n=2$ (details on the relation between the $\widetilde{F}_n$ and the $F_n$ are not needed in what follows). Similarly, performing the replacement of the naive variables as functions of the covariant ones in the expanded constraints yields
\bea
	\mathcal{C}^{(n\geq1)}(\delta\phi^I,\delta\pi_I)&=&\ds\sum_{m=n}^\infty\widetilde{C}^{(n,m)}(Q^I,P_I), \\
	\mathcal{D}^{(n\geq1)}_i(\delta\phi^I,\delta\pi_I)&=&\ds\sum_{m=n}^\infty\widetilde{D}^{(n,m)}_i(Q^I,P_I),
\eea
where it is worth stressing that the sum starts at $m=n$, and where $\mathcal{C}^{(0)}$ remains unchanged. This can be plugged into the expression of the action for $n\geq2$ which then reads
\bea
	\ds\sum_{n=2}^\infty S^{(n)}&=&\ds\int\dd\tau\int\dd^3x\left\{\delta\pi_I\dot{\delta\phi}^I-\sum_{n=1}\left[\delta N\widetilde{\mathcal{C}}^{(n)}+\delta N^i\widetilde{\mathcal{D}}^{(n)}_i\right]-N\sum_{n=2}\widetilde{\mathcal{C}}^{(n)}\right\}.
\eea
As is the case for the generating function, the relation between $(\widetilde{\mathcal{C}}^{(n)},\widetilde{\mathcal{D}}^{(n)}_i)$ and $(\widetilde{C}^{(n,m)},\widetilde{D}^{(n,m)}_i)$ is not needed for what follows.

Let us now write the action with the new variables as
\bea
	S=S^{(0)}+\ds\int\dd\tau\int\dd^3x\left[-\dot\pi_IQ^I+\dot\phi^IP_I-\mathcal{K}^{(1)}+P_I\dot{Q}_I+\ds\sum_{n=2}^\infty\mathcal{K}^{(n)}\right].
\eea
Since the two actions should have the same extremum, they are related one to each other by the total derivative of the generating function [see Eq. (\ref{eq:deltaaction})], hence
\bea
	-\dot\pi_IQ^I+\dot\phi^IP_I-\mathcal{K}^{(1)}&=&-\left(\dot{\pi}_I-N\left\{\pi_I,\mathcal{C}^{(0)}\right\}\right)\delta_1\phi^I \nonumber \\
	&&+\left(\dot{\phi}^I-N\left\{\phi^I,\mathcal{C}^{(0)}\right\}\right)\delta_1\pi_I-\delta N\,\mathcal{C}^{(0)}, \label{eq:Kam1}\\
	\mathcal{K}^{(n\geq2)}&=&-\left(\dot{\pi}_I-N\left\{\pi_I,\mathcal{C}^{(0)}\right\}\right)\widetilde{\delta_n\phi}^I+\left(\dot{\phi}^I-N\left\{\phi^I,\mathcal{C}^{(0)}\right\}\right)\widetilde{\delta_n\pi}_I \nonumber \\
	&&+\sum_{n=1}\left[\delta N\widetilde{\mathcal{C}}^{(n)}+\delta N^i\widetilde{\mathcal{D}}^{(n)}_i\right]+N\sum_{n=2}\widetilde{\mathcal{C}}^{(n)}+\frac{\partial \widetilde{F}_n}{\partial \tau}. \label{eq:Kamn}
\eea
The generating function does not enter at the linear order since it is at least quadratic. By further implementing that the background variables are solutions of the background equations of motion, the right-hand-side of Eq. (\ref{eq:Kam1}) equals zero and one readily recover that the linear action remains vanishing in the covariant variables, as already proved previously. For $n\geq2$, the first line comes from the action linear in $(\delta\phi^I,\delta\pi_I)$ which however leads to higher order contributions in the covariant variables because the latter are non-linearly related to the former. However, these terms multiply the background equations of motion, hence equal zero on the background solution. As a result, the new Hamiltonian at orders higher than one reads
\bea
	\mathcal{K}^{(n\geq2)}&=&\sum_{n=1}\left[\delta N\widetilde{\mathcal{C}}^{(n)}+\delta N^i\widetilde{\mathcal{D}}^{(n)}_i\right]+N\sum_{n=2}\widetilde{\mathcal{C}}^{(n)}+\frac{\partial \widetilde{F}_n}{\partial \tau}.
\eea
showing that it receives no contribution from the action linear in the naive variables, even if it formally contains higher orders in the covariant ones.

\subsubsection*{Off-shell action}
It is also worth stressing that assuming the background equations of motion is needed to further express ${\partial F}/{\partial\tau}$ as a function of the background, phase-space variables. Indeed one has ${\partial F}/{\partial\tau}=({\partial F}/{\partial\phi^I})\dot{\phi}^I+({\partial F}/{\partial\pi_I})\dot{\pi}_I$, because the explicit time-dependence of the generating function comes from $\Gamma^K_{IJ}\pi_K$ which is a function of the background variables. Then, using the background equations of motion yields
\bea
	\frac{\partial F}{\partial\tau}=\frac{N}{v}G^{IJ}\pi_J\,\left(\frac{\partial F}{\partial\phi^I}\right)-N\left(\frac{1}{2v}{G^{KL}}_{,I}\pi_K\pi_L+vV_{,I}\right)\,\left(\frac{\partial F}{\partial\pi_I}\right). 
\eea
We will call the on-shell second-order action, the action assuming the background equations of motion have been fully implemented. Concretely, it means that it is parameterized by the background phase-space variables only without time-derivative of them. It reads 
\bea
	S^{(2)}_\mathrm{on-shell}&=&\ds\int\dd\tau\int\dd^3x\left[P_I\dot{Q}^I-\mathcal{H}^{(2)}(\delta_1\phi^I,\delta_1\pi_I)-\frac{N}{v}G^{IJ}\pi_J\,\frac{\partial F}{\partial\phi^I}(\delta_1\phi^I,\delta_1\pi_I)\right. \nonumber \\
	&&\left.+N\left(\frac{1}{2v}{G^{KL}}_{,I}\pi_K\pi_L+vV_{,I}\right)\,\frac{\partial F}{\partial\pi_I}(\delta_1\phi^I,\delta_1\pi_I)\right],
\eea
where $({\partial F}/{\partial\phi^I})$ and $\left({\partial F}/{\partial\pi_I}\right)$ are restricted to the quadratic order in the covariant variables. (Note that by construction, $\mathcal{H}^{(2)}$ is solely parametrized by the background phase-space variables.)

For the sake of completeness, let us provide with the expression of $S^{(2)}$ where instead, the equations of motion of the background are not assumed to hold. The off-shell second-order action then reads $S^{(2)}_\mathrm{off-shell}=S^{(2)}_\mathrm{on-shell}+\Delta S^{(2)}$ where $\Delta S^{(2)}$ receives two contributions. The first one comes from $N\mathcal{C}^{(1)}(\delta_2\phi^I,\delta_2\pi_I)+\dot{\pi}_I\,\delta_2\phi^I-\dot{\phi}^I\,\delta_2\pi_I$ and reads
\bea
	\Delta S^{(2)}&\supset&\ds\int\dd\tau\int\dd^3x\left\{\frac{1}{2}\left(\dot{\pi}_I+Nv\rho_{,I}\right)\Gamma^I_{KL}Q^KQ^L\right. \\
	&&\left.+\left(\dot\phi^I-\frac{N}{v}G^{MI}\pi_M\right)\left[\Gamma^K_{IL}P^{(\alpha)}_KQ^L+\frac 1 2 \left(\Gamma^S_{IJ,K} - \Gamma^S_{IR}\Gamma^R_{JK} + \Gamma^R\Tsr\Gamma^S_{RK}+\alpha {R_{IJK}}^S\right)\pi_S Q^J Q^K\right]\right\} \nonumber
\eea
The second one comes from the generating function in which we cannot further assume the background equations of motion. It reads
\bea
	\Delta S^{(2)}&\supset&-\frac{1}{2}\ds\int\dd\tau\int\dd^3x\left[\left(\dot\phi^I-\frac{N}{v}G^{MI}\pi_M\right)\Gamma^J_{KL,I}\pi_JQ^KQ^L+\left(\dot{\pi}_I+Nv\rho_{,I}\right)\Gamma^I_{KL}Q^KQ^L\right]. \nonumber \\
\eea
Combining the two, one recognizes the appearance of the Riemann curvature tensor which finally leads to
\bea
	\Delta S^{(2)}&=&\ds\frac{1}{2}\left(\alpha-1\right)\int\dd\tau\int\dd^3x\left(\dot\phi^I-\frac{N}{v}G^{LI}\pi_L\right) {R_{IJK}}^S\,\pi_S Q^J Q^K \nonumber \\
	&&+\int\dd\tau\int\dd^3x\left(\dot\phi^I-\frac{N}{v}G^{LI}\pi_L\right)\Gamma^K_{IL}P^{(\alpha)}_KQ^L. \label{eq:deltaS2}
\eea
Let us note that $\Delta S^{(2)}$ has to be manifestly covariant. Indeed $S^{(2)}_\mathrm{off-shell}$ is covariant since it derives from a covariant action and we showed in the main text that $S^{(2)}_\mathrm{on-shell}$ is manifestly covariant too. Because of that, the second line in Eq. (\ref{eq:deltaS2}) should be zero since it is not manifestly covariant. Indeed, it is vanishing in a locally-flat coordinate system in which the Christoffel symbols are all zero.

\section{Adiabatic and entropic basis} \label{app:AEbasis}
We remind that the adiabatic direction is defined as $e^\sigma_I=\pi_I/\pi_\sigma$ where $\pi_\sigma=\sqrt{G^{IJ}\pi_I\pi_J}$. By using the Hamilton equation for the momenta in the field basis, it is easy to show that 
\beq
	\dot{\pi}_\sigma=-NvV_{;\sigma} \label{eq:eompisigma}
\eeq
 where $V_{;\sigma}=e^I_\sigma V_{;I}$ and where we use $D_\tau\pi_\sigma=\dot{\pi}_\sigma$ for $\pi_\sigma$ is a scalar. On defining $\dot\sigma\equiv e^\sigma_I\dot\phi^I$, the Hamilton equation for $\phi^I$ is projected on the adiabatic direction to give $\dot\sigma=(N/v)\pi_\sigma$. Plugging this in Eq. (\ref{eq:eompisigma}) yields the Klein-Gordon equation
 \beq
 	\ddot{\sigma}+\left(\frac{\dot{v}}{v}-\frac{\dot{N}}{N}\right)\dot\sigma+N^2V_{;\sigma}=0.
\eeq
Note also that by definition of the adiabatic direction, the covariant time-derivative consists in projecting the covariant derivatives on the adiabatic direction, \ie 
 \beq
 	D_\tau=\dot\phi^JD_J=\frac{N}{v}\pi_\sigma G^{IJ}e^\sigma_I D_J \label{eq:timediffadiabatic}
\eeq
 where the second equality is obtained from the Hamilton equation for $\phi^I$.
 
 The turning-rate covector, $\omega_I$, is defined as $N\omega_I\equiv D_\tau e^\sigma_I$. One obtains from the background equations of motion
\bea
	\omega_I=-\left(\frac{v}{\pi_\sigma}\right)\perp_{KI}\,G^{KL}V_{,L}, \label{eq:turnratevector}
\eea
where $\perp_{KI}=G_{KI}-e^\sigma_K e^\sigma_I$ is the operator projecting on the subspace of entropic modes; hence $e^I_\sigma\perp_{KI}=0$. The first entropic direction is thus defined as the unit vector aligned with $\omega_I$, \ie we write $\omega_I=\omega_1e^1_I$ with $\omega_1$ the norm of $\omega_I$. Projecting Eq. (\ref{eq:turnratevector}) on the first entropic direction leads to
\bea
	\frac{\pi_\sigma}{v}\omega_1+V_{;1}=0 \label{eq:eomV1}
\eea
where $V_{;1}=e^I_1V_{;I}$. One can further make use of the background equations of motion, \ie $\pi_\sigma=v\sqrt{\dot{\theta}/N}$ and $\dot\theta=N\epsilon_1\theta^2/(2\Mp^2)$, to obtain
\bea
	\sqrt{\frac{\epsilon_1}{2}}\omega_1\theta+\Mp V_{;1}=0.
\eea
By construction, the remaining entropic directions are orthogonal to the first one, hence to the turning-rate covector. As a result, projecting Eq. (\ref{eq:turnratevector}) on the remaining entropic directions yields $V_{;s}=0$ for $s\geq2$ and where $V_{,s}=e^I_sV_{;s}$.

The second derivative of the potential can be related to the second entropic direction. On the one hand, one can start from the definition of the first entropic direction, \ie $\omega_I=\omega_1e^1_I$, and make use of Eqs. (\ref{eq:dotAEbasis}) \& (\ref{eq:omegamatrix}) to arrive at
\beq
	D_\tau\omega_I=-N\omega_1^2e^\sigma_I+\dot{\omega}_1e^1_I+N\omega_1\omega_2 e^{2}_I, \label{eq:dtomegaE}
\eeq
where we use that $\omega_1$ is a scalar. It shows that the second entropic direction is related to the covariant time-derivative of the turning-rate covector. On the other hand, one can use the background equations of motion and Eq. (\ref{eq:turnratevector}) to derive
\beq
	D_\tau\omega_I=-Ne^s_I\left(V_{;\sigma s}+\frac{v^2}{\pi_\sigma^2}V_{;\sigma}V_{;s}-\frac{3}{2\Mp^2}\frac{v\theta}{\pi_\sigma} V_{;s}\right)+\frac{Nv}{\pi_\sigma}\omega_1\left(e^\sigma_IV_{;1}+e^1_IV_{;\sigma}\right), \label{eq:dtomegaV}
\eeq
where $V_{;ab}=e^I_a e^J_b V_{;IJ}$. Because $V_{;s}=0$ for $s\geq 2$, it yields $e^{I}_sD_\tau\omega_I=-NV_{;\sigma s}$ for $s\geq2$. As a result of the above expressions, one derives the following identities by projecting Eqs. (\ref{eq:dtomegaE}) \& (\ref{eq:dtomegaV}) on the basis vectors $e^I_a$, \ie
\bea
	\omega_1^2&=&-\frac{v}{\pi_\sigma}\omega_1V_{;1}, \label{eq:omega12} \\
	\dot{\omega}_1&=&-N\left(V_{;\sigma 1}+\frac{v^2}{\pi_\sigma^2}V_{;\sigma}V_{;1}-\frac{3}{2\Mp^2}\frac{v\theta}{\pi_\sigma} V_{;1}\right)+\frac{Nv}{\pi_\sigma}\omega_1V_{;\sigma}, \label{eq:dotomega1}\\
	\omega_1\omega_2&=& -V_{;\sigma 2}, \label{eq:omega1omega2}
\eea
and $V_{;\sigma s}=0$ for $s\geq3$ (note that we made use of $V_{;s}=0$ for $s\geq2$). Eq. (\ref{eq:omega12}) is no more than Eq. (\ref{eq:eomV1}) and as a second consistency check, it is shown that Eq. (\ref{eq:dotomega1}) is the time-derivative of Eq. (\ref{eq:omega12}).\footnote{Alternatively, the above results can be obtained starting from
\beq
	e^I_aD_Ie^J_bD_JV=V_{;ab}+\left(e^I_aD_Ie^J_b\right)D_JV.
\eeq
By fixing $a=\sigma$ and using Eqs. (\ref{eq:dotAEbasis}) \& (\ref{eq:timediffadiabatic}), the above yields
\beq
	\frac{N\pi_\sigma}{v}V_{;\sigma b}=D_\tau V_{;b}-N{\Omega_b}^cV_{;c}. 
\eeq
Setting $b=1$ and making use of Eq. (\ref{eq:eomV1}) boils down to Eq. (\ref{eq:dotomega1}). Then, setting $b=2$ and using $V_{;c}=0$ for $c\geq2$ gives
\beq
	\frac{\pi_\sigma}{v}V_{;\sigma 2}=\omega_2V_{;1}.
\eeq
Combined with Eq. (\ref{eq:eomV1}) it gives $V_{;\sigma 2}=-\omega_1\omega_2$. Finally, by setting $b\geq3$, one recovers that $V_{;\sigma b}=0$ for $b\geq3$ since $V_{;b}=0$ for $b\geq2$ which leads to ${\Omega_b}^cV_{;c}\propto\omega_{b-1}V_{;b-1}-\omega_bV_{;b+1}=0$.}  

In principle, the above can be continued to higher derivatives of the turning-rate covector. Indeed, using Eqs. (\ref{eq:dtomegaE}), (\ref{eq:dotAEbasis}) \& (\ref{eq:omegamatrix}), one can express $D^2_\tau\omega_I$ as a linear combinations of $e^{\sigma}_I$ and of the $e^{s}_I$'s up to $s=3$. In parallel, $D^2_\tau\omega_I$ can be expressed as a function of the third derivative of the potential starting from Eq. (\ref{eq:dtomegaV}) and using the background equations of motion. Hence, third derivatives of the potential can be related to $\ddot{\omega}_1$, $\dot{\omega}_2$ and $\omega_3$.

\section{Perturbed background equations}\label{app:separate universe}

The equations of motion of perturbations in the separate universe approach are obtained by simply linearising the background equations of motion. Using the definitions of $\overline{\delta\gamma_1}$ and $\overline{\delta\pi_1}$ in linearising Eqs. (\ref{eq:dotgamma}) \& (\ref{eq:dotpig}), we get :
\bea
	&&\left\{\begin{array}{rcl}
	\dot{\overline{\delta\gamma}}_1&=&\ds-\frac{\sqrt{3}}{\Mp^2}v^{2/3}\theta\overline{\delta N}-\frac{N}{\Mp^2}\left(2v^{1/3}\overline{\delta\pi}_1+\frac{\theta}{2}\overline{\delta\gamma}_1\right) \\
	\dot{\overline{\delta\pi}}_1&=&\ds\frac{v^{1/3}}{2\sqrt{3}}\left(\rho+3p\right)\overline{\delta N}-N\left[\frac{1}{6v^{1/3}}\left(5\rho+3p\right)\overline{\delta\gamma}_1-\frac{\theta}{2\Mp^2}\overline{\delta\pi}_1\right]  \\
	&&\ds-\frac{\sqrt{3}}{2}v^{1/3}N\left(V_{;I}\overline{Q}^I-\frac{G^{IJ}\pi_J}{v^2}\overline{P}_I\right),
	\end{array}\right. \label{eq:SU1} 
\eea
where we use $V_{,I}=V_{;I}$. The dynamics of the fluctuations of the scalar fields are derived linearizing Eqs. (\ref{eq:dotphi}) \& (\ref{eq:dotpi}). It gives for the configuration variables 
\bea
	\dot{\overline{Q}}^I&=&\ds\frac{1}{v}G^{IJ}\pi_J \overline{\delta N} \label{eq:dotOVQ} \\
	&&+ N\left[\frac{1}{v}G^{IJ}\overline{P}_J+\frac{1}{v}\left({G^{IJ}}_{,K}+G^{IL}\Gamma^J_{KL}\right)\pi_J\overline{Q}^K-\frac{\sqrt{3}}{2v^{5/3}}G^{IJ}\pi_J\overline{\delta\gamma}_1\right]. \nonumber
\eea
The second term in the square brackets of the second line can be combined with $\dot{\overline{Q}}^I$ which leads to the covariant derivative (see footnote \ref{ftnt}). It boils down to
\bea
	D_\tau {\overline{Q}}^I&=&\ds\frac{1}{v}G^{IJ}\pi_J \overline{\delta N} + N\left[\frac{1}{v}G^{IJ}\overline{P}_J-\frac{\sqrt{3}}{2v^{5/3}}G^{IJ}\pi_J\overline{\delta\gamma}_1\right]. \label{eq:SUQ}
\eea
The Hamilton equation for the momenta is
\bea
	\dot{\overline{P}}_I+\frac{\dd}{\dd\tau}\left(\Gamma^K_{IJ}\pi_K\overline{Q}^J\right)&=&\ds-\left(\frac{1}{2v}{G^{KL}}_{,I}\pi_K\pi_L+vV_{,I}\right)\overline{\delta N}-\frac{N}{v}{G^{KL}}_{,I}\pi_K \overline{P}_L \\
	&&\ds-N\left(\frac{1}{v}{G^{KL}}_{,I}\Gamma^M_{LJ}\pi_K\pi_M+\frac{1}{2v}{G^{KL}}_{,IJ}\pi_K\pi_L+vV_{,IJ}\right)\overline{Q}^J \nonumber \\
	&&+N\frac{\sqrt{3}}{2}v^{1/3}\left(\frac{1}{2v^2}{G^{KL}}_{,I}\pi_K\pi_L-V_{,I}\right)\overline{\delta\gamma}_1.\nonumber 
\eea
The second term in left-hand-side is expressed as functions of the phase-space variables and the Lagrange multiplier using the background equations of motion and Eq. (\ref{eq:dotOVQ}). This is further combined with the right-hand-side and making multiple use of footnote \ref{ftnt}  to express first and second derivatives of the metric as functions of the Christoffel symbols and their first derivatives, one arrives at 
\bea
	D_\tau  \overline{P}_I&=&\ds-vV_{;I}\overline{\delta N} -N\left[v\left(V_{;IJ}-\frac{1}{v^2}R_I{}^{KL}{}_J\pi_K\pi_L\right)\overline{Q}^J+\frac{\sqrt{3}}{2}v^{1/3}V_{;I}\overline{\delta\gamma}_1\right]. \label{eq:SUP}
\eea
These dynamical equations have to be solved under the scalar constraint at first order which is obtained by linearizing Eq. (\ref{eq:C0}), \ie 
\bea
	\overline{\mathcal{C}}^{(1)}&=&-\frac{\sqrt{3}}{2}{v^{1/3}}\left(\rho-\frac{\theta^2}{4\Mp^2}\right)\overline{\delta\gamma}_1-\frac{\sqrt{3}}{\Mp^2}v^{2/3}\theta\,\overline{\delta\pi}_1 \\
	&&+\left[\frac{1}{2v}\left({G^{IJ}}_{,K}+2G^{LJ}\Gamma^I_{LK}\right)\pi_I\pi_J+vV_{,K}\right]\overline{Q}^K+\frac{1}{v}G^{IJ}\pi_I\overline{P}_J. \nonumber
\eea
One can further use the background constraint, Eq. (\ref{eq:C0}), to simplify the coefficients multiplying $\overline{\delta\gamma}_1$ and using footnote \ref{ftnt} to show that $\left({G^{IJ}}_{,K}+2G^{LJ}\Gamma^I_{LK}\right)\pi_I\pi_J=0$, leading to:
\beq
	\overline{\mathcal{C}}^{(1)}=vV_{;I}\,\overline{Q}^I+\frac{1}{v}G^{IJ}\pi_J\overline{P}_I-\frac{v^{1/3}}{2\sqrt{3}}\left(\rho+3p\right)\overline{\delta\gamma}_1-\frac{\sqrt{3}}{\Mp^2}v^{2/3}\theta\,\overline{\delta\pi}_1.
\eeq

\bibliographystyle{JHEP}
\bibliography{ref}

\end{document}